\documentclass[a4paper,11pt]{article}
\pdfoutput=1 
\usepackage{physics}

\usepackage{jheppub} 
\usepackage[T1]{fontenc} 
\usepackage{slashed}
\newcommand{\be}{\begin{equation}}
\newcommand{\ee}{\end{equation}}

\usepackage{subcaption}
\usepackage{ucs}
\usepackage{hyperref}
\usepackage[utf8x]{inputenc}
\usepackage{amsmath,bm}
\usepackage{amsfonts}
\usepackage{comment}
\usepackage{amssymb}
\usepackage{array}
\usepackage{epsfig}
\usepackage{slashed}
\usepackage{lipsum}
\usepackage{hyperref}
\hypersetup{
	colorlinks=true,
	linkcolor=blue,
	filecolor=red,
	urlcolor=blue,
	citecolor=red
} 

\title{\boldmath Higher spin 3-point functions in 3d CFT using spinor-helicity variables}
\author[a]{Sachin Jain,}
\author[a]{ Renjan Rajan John,}
\author[a]{Abhishek Mehta,}
\author[b]{Amin A.~Nizami,}
\author[a]{Adithya Suresh}

\affiliation[a]{Indian Institute of Science Education and Research, Homi Bhabha Road, Pashan, Pune 411 008, India}
\affiliation[b]{Department of Physics, Ashoka University, India}

\emailAdd{sachin.jain@iiserpune.ac.in}
\emailAdd{renjan.john@acads.iiserpune.ac.in}
\emailAdd{aan27cam@gmail.com}
\emailAdd{\{abhishek.mehta,s.adithya\}@students.iiserpune.ac.in}

\abstract{In this paper we use the spinor-helicity formalism to calculate  3-point functions involving scalar operators and spin-$s$ conserved currents in general 3d CFTs. In spinor-helicity variables we notice that the parity-even and the parity-odd parts of a correlator are related. Upon converting spinor-helicity answers to momentum space, we show that correlators involving spin-$s$ currents can be expressed in terms of some simple conformally invariant conserved structures. This in particular allows us to understand and separate out contact terms systematically, especially for the parity-odd case. We also reproduce some of the correlators using  weight-shifting operators.}

\begin{document}
\maketitle
\raggedbottom
\flushbottom

\section{Introduction}

The study of CFT correlation functions in momentum space was initiated systematically in \cite{Coriano:2013jba,Bzowski:2013sza}. It is important for a variety of reasons. It has wide ranging applications in cosmology \cite{Mata:2012bx,Ghosh:2014kba,Kundu:2014gxa,Arkani-Hamed:2015bza,Maldacena:2011nz, Arkani-Hamed:2018kmz,Sleight:2019mgd,Sleight:2019hfp,Baumann:2019oyu,Baumann:2020dch} - in particular in computing cosmological correlators,  and condensed matter physics  \cite{Huh:2013vga, Chowdhury:2012km} - especially in studying quantum phase transitions.  Recent works on aspects of momentum space CFTs include \cite{Bzowski:2013sza, Coriano:2013jba,Bonora:2015nqa,Bonora:2015odi,Bonora:2016ida,sissathesis,Bzowski:2015pba,Bzowski:2017poo,Coriano:2018bbe,Bzowski:2018fql,Gillioz:2018mto,Coriano:2018tgn,Albayrak:2018tam,Farrow:2018yni,Isono:2018rrb, Isono:2019wex,Isono:2019ihz,Maglio:2019grh,Gillioz:2019lgs,Bzowski:2019kwd,Bautista:2019qxj,Albayrak:2019yve,Coriano:2019nkw,Albayrak:2019asr,Lipstein:2019mpu,Gillioz:2020mdd,Gillioz:2020wgw,Albayrak:2020isk,Bzowski:2020kfw,Jain:2020rmw,Jain:2020puw,Coriano:2020ccb,Albayrak:2020bso,Albayrak:2020fyp,Armstrong:2020woi,Serino:2020pyu,Skvortsov:2018uru,Jain:2021wyn,Jain:2021qcl,Coriano:2018zdo,Coriano:2020ees,Gillioz:2019iye}. Via holography, momentum space CFT correlators are related to flat space scattering amplitudes \cite{Gary:2009ae, Gary:2009mi, Komatsu:2020sag, Penedones:2010ue, Raju:2012zr, Fitzpatrick:2011hu}. Thus CFT correlators in momentum space, besides enabling a connection between the conformal and S-matrix bootstrap, also reveal interesting structures such as double copy and colour-kinematics duality relations which are hard to discern without working in momentum space \cite{Farrow:2018yni, Lipstein:2019mpu,Jain:2021qcl}.

 3d CFT correlators for conserved currents in position space are quite well explored. A detailed position space analysis of higher spin $CFT_3$ correlators was performed in \cite{Giombi:2011rz}. 
For the position space 3-point correlators of conserved currents of the form $\langle J_{s_1}J_{s_2}J_{s_3}\rangle$, where $s_i\ge 1$, it was shown that the correlators have two parity-even structures and one parity-odd structure.
When the sum of spins is even, the parity-even contributions arise from three-point correlators in the free-boson and free-fermion theories. When the sum of spins is odd,  the parity-even contributions arise from correlators of non-abelian currents in the free theory of multiple scalars or fermions. The parity-odd structure is not generated by a free theory. When the spins of the conserved currents satisfy the triangle inequality, i.e. $s_i\le s_{i+1}+s_{i+2}$ for $i=1,2,3$ (modulo 3), there exists one parity-odd structure. When the inequality is not satisfied the odd structure is zero. A proof of this using an integral representation of the correlators was presented in \cite{Maldacena:2011jn}. It was also noted in \cite{Giombi:2011rz, Maldacena:2011jn} that a correlator of the form $\langle J_{s}J_{s}O_{\Delta}\rangle$, where $O_{\Delta}$ is a scalar operator with scaling dimension $\Delta$, contains only two structures, one parity-odd and another parity-even. Also, a correlator of the form $\langle J_{s}O_{\Delta_1}O_{\Delta_2}\rangle$ is non-zero only for $\Delta_1=\Delta_2$ and only has one parity-even structure.
 In \cite{Giombi:2011rz} it was shown that in position space, one can write down arbitrary CFT 3-point higher spin correlators as multinomials of just a few simple conformal invariants. 

A similar exhaustive analysis of CFT correlators in momentum space or spinor-helicity variables has not been done yet. 
A helicity basis was used in \cite{Caron-Huot:2021kjy} for studying higher spin 3d CFT correlators. Spinor-helicity variables have been used earlier in the study of CFTs, see for example \cite{Maldacena:2011nz, Mata:2012bx, Baumann:2020dch}.
Our goal in this paper is to study  3-point correlators of scalar operators and conserved spin-$s$ currents in $d=3$ CFTs in spinor-helicity variables as well as in momentum space.


In \cite{Bzowski:2013sza}, parity-even 3-point momentum space correlators involving scalar operator and conserved currents up to spin-2 (stress tensor) were computed by solving the conformal Ward identities (CWI). These correlators were later obtained using weight-shifting operators in \cite{Baumann:2019oyu,Baumann:2020dch}. In \cite{Jain:2021wyn}, we explored  momentum space $CFT_3$ parity-odd  3-point correlators such as $\langle JJO_\Delta\rangle$  by solving CWIs directly in momentum space. We also calculated parity-odd correlators of the form $\langle TTO_\Delta \rangle$ and $\langle JJJ \rangle$ using weight-shifting operators, as the  direct use of CWIs became very complex and inefficient. More complex higher spin correlators were not computed directly using momentum space CWIs \footnote{See \cite{Maldacena:2011nz} for a calculation of $\langle TTT\rangle_{\text{odd}}$ based on $dS_4$ tree-level Feynman diagram.}. 

The main difficulty in calculating complicated correlators such as  $\langle TTT\rangle$ is that 
 there is a high degree of degeneracy in the tensor structures in 3d, both in the parity-even and the parity-odd sector, which makes it difficult to choose an appropriate basis to write an ansatz for the correlator. In the parity-odd sector one has to deal with the additional complication of non-trivial Schouten identities, which makes it difficult to solve for $\langle TTT\rangle_{\text{odd}}$ directly. The problem becomes even more complicated if we want to calculate a correlator involving higher spin conserved currents ($J_s$ with $s > 2$) both for the parity-even and parity-odd case.  In this paper, we overcome this problem by working in the spinor-helicity formalism where the degeneracy is automatically taken care of. We solve the CWIs in these variables and then convert the results back to momentum space. In this way we obtain the momentum space expressions for all correlators of the form $\langle J_{s_1}J_{s_2} J_{s_3}\rangle$ with spins satisfying the triangle inequality. We make contact with the counting of structures in \cite{Giombi:2011rz} by showing that correlators of conserved spinning operators in momentum space  have 2 parity-even and 1 parity-odd structure up to contact terms. 

In our analysis we split the correlation function into two pieces, homogeneous and non-homogeneous parts.
 We show that for the correlation functions we consider the parity-odd contribution to the non-homogeneous piece is always a contact term. Interestingly, spinor-helicity variables reveal that parity-even and parity-odd contributions to homogeneous pieces are completely identical, although they look completely different in momentum space as well as position space. Moreover, for divergent correlation functions which require completely different regularization and renormalization for parity-even and parity-odd parts, the relation between these parts in spinor-helicity variables holds even after renormalization. Upon converting spinor-helicity answers to momentum space, we see that the results for correlators involving spin-$s$ currents can be expressed in terms of some simple conformally invariant conserved structures. In certain cases (such as $\langle TTO_4\rangle$), the correlators are divergent in momentum space and require a careful application of the renormalization procedure, but in spinor-helicity variables they turn out to give directly the finite part without any renormalization. We also verify some of the results using weight-shifting operators.
 

The plan of the rest of the paper is as follows. In Section \ref{CCSHV} we introduce the basic idea of expressing conformal correlators in terms of spinor-helicity variables and discuss the preliminary case of 2-point functions. We also discuss some general features of our 3-point function analysis. Section \ref{sectionSHresults} has the results of various 3-point correlators of spinning conserved currents and scalar operators in spinor-helicity variables. In Section \ref{shmomentumspaceresults} we translate these results to momentum space after carefully taking the degeneracies into account and Section \ref{renor11} has a discussion of the renormalisation of some of these correlators which have divergences. In Section \ref{wso} some of these momentum space results are re-derived using weight-shifting operators acting on seed correlators. Section \ref{INV} contains a discussion of momentum space higher-spin conserved current correlators expressed in terms of 3-point momentum space invariants. In Section \ref{sio} we make some important observations, including the connection between the parity-even and parity-odd parts of a correlator. We conclude in Section \ref{fdir} with a brief summary and a discussion on future directions of study. At the end we have a number of appendices supplementing the main text and providing various technical details. Appendix \ref{shn} outlines our spinor-helicity notation.  In Appendix \ref{hnhvstl} we describe in detail our terminology of homogeneous and non-homogeneous contributions to a correlator and discuss how they differ from the usual splitting of a correlation function into transverse and longitudinal pieces. Appendix \ref{CWI} has the technical details of solutions of various conformal Ward identities quoted in Section \ref{sectionSHresults}. Appendix \ref{idtrk} contains useful triple-$K$ integral identities and Appendix \ref{higherspin-correlator} lists the momentum space form of various 3-point correlators of conserved currents. Finally Appendix \ref{sdro} contains the required details of some weight-shifting operators which are used in Section \ref{wso}.


\section{Conformal correlators in spinor-helicity variables}
\label{CCSHV}
Momentum space expressions for parity preserving two and three-point conformal correlators of spinning operators were obtained in \cite{Coriano:2013jba,Bzowski:2013sza,Bzowski:2015pba,Bzowski:2017poo,Bzowski:2018fql,Coriano:2018bbe} by solving momentum space conformal Ward identities. Recently in \cite{Jain:2021wyn} we derived momentum space expressions for parity-odd correlators using two different techniques. The first one, following \cite{Bzowski:2013sza}, involved solving conformal Ward identities directly in momentum space. The second one, following \cite{Baumann:2019oyu}, involved using the technique of spin-raising and weight-shifting operators in momentum space. In \cite{,Jain:2020rmw,Jain:2020puw} following the position space analysis in  \cite{Maldacena:2011jn, Maldacena:2012sf} it was shown that one could make use of momentum space higher-spin equations arising from Ward identities associated to (weakly broken) higher spin symmetry to compute spinning correlators including the parity-odd ones. 

 The analysis of parity-odd correlators was restricted to correlators such as $\langle JJO\rangle$, $\langle TTO\rangle$ and $\langle JJJ\rangle$ due to various technical difficulties. One of the main obstacles was to identify the correct basis of tensor structures to work with, due to various non-trivial Schouten identities and other degeneracies in three-dimensions. 

In this section, we compute 3-point CFT correlators in  spinor-helicity variables. It turns out that solving for CFT correlators in spinor-helicity variables is a lot simpler than doing so in momentum space. The reader may wish to refer Appendix \ref{shn} at this point to get familiar with our notation and convention regarding spinor-helicity variables.

We start with an ansatz for the correlator in spinor-helicity variables. To do so, we use the fact that a Lorentz transformation of the momentum $\vec{k}$ corresponds to a scale transformation of the spinors. Therefore, a Lorentz-covariant structure in spinor-helicity variables is a structure that has the correct scaling based on the helicities of the operators. An operator $O$ with helicity $h$ transforms in the following way under a scale transformation of spinors :
\begin{align}
O^h(t\lambda, t^{-1}\bar{\lambda}) &= t^{-2h}O(\lambda, \bar{\lambda})
\end{align}
Therefore, the ansatz for a general correlator is given by
\begin{align}\label{generalansatz}
\langle O^{h_1}(k_1)O^{h_2}(k_2)O^{h_3}(k_3) \rangle&=(c_1\,F_1(k_1, k_2, k_3) +i\,c_2\,F_2(k_1, k_2, k_3) )\nonumber\\[5pt]
&\hspace{1cm}\langle 12 \rangle^{h_3-h_1-h_2} \langle 23 \rangle^{h_1-h_2-h_3} \langle 31 \rangle^{h_2-h_3-h_1}
\end{align}
where $F_1(k_1, k_2, k_3)$ and $F_2(k_1, k_2, k_3)$ are form-factors that we will determine by imposing dilatation and special conformal invariance. 
For parity-even correlators $c_2=0$ and for parity-odd correlators $c_1=0$, and for the latter the '$i$' ensures that the correlator changes sign under conjugation, since conjugation corresponds to a parity transformation for spinors. 

\subsection{Conformal generators}
\label{SectionCG}
The conformal Ward identities are differential equations determined by the action of the special conformal generator on a conformal correlator. The special conformal generator in spinor-helicity variables takes the form  \cite{Witten:2003nn} :
\begin{align}
\widetilde{K}^{\kappa}=2\sum_{i=1}^n\left(\sigma^{\kappa}\right)_{\alpha}^{\;\; \beta}\frac{\partial^2}{\partial \lambda _{i\alpha}\partial \bar{\lambda}_i^{\beta}}
\end{align}
The action of $\widetilde{K}$ on a scalar with $\Delta=2$ is given by \cite{Baumann:2020dch} :
\begin{align}
\widetilde{K}^{\kappa}O_2 = -K^{\kappa}O_2
\end{align}
where 
\begin{align}\label{Kkappascalar}
K^{\kappa}=-2 \partial_{k_{\kappa}}-2 k^{\alpha} \partial_{k^{\alpha}} \partial_{k_{\kappa}}+k^{\kappa} \partial_{k^{\alpha}} \partial_{k_{\alpha}}
\end{align}
The action of $\widetilde{K}$ on a scalar with $\Delta\ne 2$ is given by \cite{Baumann:2020dch} :
\begin{align}
\label{Kkappageneralscalar}
\widetilde{K}^{\kappa}\left(\frac{O_{\Delta}}{k^{\Delta-2}}\right)= -\frac{1}{k^{\Delta-2}}K^{\kappa}O_{\Delta}+\frac{O_{\Delta}}{k^{\Delta}}k^{\kappa}(\Delta-1)(\Delta-2)
\end{align}
Similarly, the action of $\widetilde{K}$  on spin-one and spin-two conserved currents is as follows \cite{Baumann:2020dch} :
\begin{align}\label{Kkappaspinning}
\begin{aligned}
\widetilde{K}^{\kappa} J^{\pm} &=\left(-z_{\pm}^{\alpha} K^{\kappa}+2 z_{\pm}^{\kappa} \frac{k^{\alpha}}{k^{2}}\right) J_{\alpha} \\
\widetilde{K}^{\kappa}\left(\frac{T^{\pm}}{k}\right) &=\left(-\frac{1}{k} z_{\pm}^{(\alpha} z_{\pm}^{\beta)} K^{\kappa}+12 z_{\pm}^{\kappa} \frac{z_{\pm}^{(\alpha} k^{\beta)}}{k^{3}}\right) T_{\alpha \beta}
\end{aligned}
\end{align}
where  $J^+=z^{+}_{\mu}J^\mu$ and  $T^+=z^{+}_{\mu}z^{+}_{\nu}T^{\mu\nu}$.
In \eqref{Kkappascalar} and \eqref{Kkappaspinning}, $K^{\kappa}$ corresponds to the special conformal generator in momentum space with $\Delta=2$. Its action on a conformally invariant correlator is zero. Therefore, the action of $\widetilde{K}^{\kappa}$ on a correlator in which all the operators have $\Delta=2$ will just have a part proportional to the R.H.S. of the Ward-Takahashi identity of the correlator. When the correlator has operators with scaling dimensions other than 2, it is convenient to divide them by appropriate powers of $k$ so that the insertion has $\Delta=2$. For a derivation,  see \cite{Mata:2012bx}.

\subsection{Two-point functions}
In this section we present the expressions for a few two-point correlators in spinor-helicity variables. These will later turn out to be useful when dealing with transverse Ward identities associated to spinning three-point correlators. 
For conserved currents of generic integer spin $s$ we have the following two-point functions:
\begin{align}\label{JsJstwopointfn}
\begin{split}
\langle J^{s-}(k_1) J^{s-}(k_2) \rangle &=\left(c_{J_s}+i\,c'_{J_s}\right)\frac{\langle 12 \rangle^{2s}}{2s k_2},\\[5 pt]
\langle J^{s+}(k_1) J^{s+}(k_2) \rangle &=\left(c_{J_s}-i\,c'_{J_s}\right)\frac{\langle {\bar 1}{\bar 2} \rangle^{2s}}{2s k_2},
\end{split}
\begin{split}
\langle J^{s+}(k_1) J^{s-}(k_2) \rangle &=\left(c_{J_s}+i\,c'_{J_s}\right)\frac{\langle \bar{1}2 \rangle^{2s}}{2s k_2}\\[5 pt]
\langle J^{s-}(k_1) J^{s+}(k_2) \rangle &=\left(c_{J_s}-i\,c'_{J_s}\right)\frac{\langle  1{\bar 2} \rangle^{2s}}{2s k_2}
\end{split}
\end{align}
where $c_{J_s}$ and $c'_{J_s}$ are the two-point function coefficients of the spin-$s$ current for the even and odd cases respectively.

\subsection{Three-point functions: General discussion}
We will now consider three-point functions with spinning operator insertions. The parity odd sector of a few correlators such as $\langle JJO\rangle$, $\langle JJJ\rangle$, and $\langle TTO\rangle$ have been studied in momentum space by solving conformal Ward identities, using spin-raising and weight-shifting operators and using higher spin equations \cite{Jain:2020rmw,Jain:2020puw,Jain:2021wyn}. In extending our analysis to more complicated three-point correlators we faced some difficulties as described in the beginning of this section. However, working in spinor-helicity variables, we are able to circumvent this problem and get expressions for more complicated 3-point correlators as described in detail below. We will first introduce the terminology of homogeneous and non-homogeneous solutions to conformal Ward identities which we will use throughout this paper.
\subsubsection{Homogeneous and non-homogeneous solutions}\label{hnh}
The action of the special conformal generator in spinor-helicity variables on a generic 3-point correlator takes the following form :
\begin{align}
\widetilde K^\kappa\left\langle\frac{J_{s_1}}{k_1^{s_{1}-1}}\frac{J_{s_2}}{k_2^{s_{2}-1}}\frac{J_{s_3}}{k_3^{s_{3}-1}}\right\rangle=\text{transverse Ward identity terms}
\end{align}
where the R.H.S. contains contact-term contributions and is expressible in terms of 2-point functions. The explicit form of the generator $\widetilde K^\kappa$ is given in Section \ref{SectionCG}.

Being a linear differential equation, the general solution of the above is expressible as the sum of homogeneous and non-homogeneous solutions :
\begin{align}
\langle J_{s_1} J_{s_2}J_{s_3}\rangle=\langle J_{s_1} J_{s_2}J_{s_3}\rangle_{\bf{h}}+\langle J_{s_1} J_{s_2}J_{s_3}\rangle_{\bf{nh}}
\end{align}
where $\langle J_{s_1} J_{s_2}J_{s_3}\rangle_{\bf{h}}$ solves:
\begin{align}
\widetilde K^\kappa\left\langle\frac{J_{s_1}}{k_1^{s_{1}-1}}\frac{J_{s_2}}{k_2^{s_{2}-1}}\frac{J_{s_3}}{k_3^{s_{3}-1}}\right\rangle_{\bf{h}}=0
\end{align}
and $\langle J_{s_1} J_{s_2}J_{s_3}\rangle_{\bf{nh}}$ is a solution of :
\begin{align}\label{nhpiece}
\widetilde K^\kappa\left\langle\frac{J_{s_1}}{k_1^{s_{1}-1}}\frac{J_{s_2}}{k_2^{s_{2}-1}}\frac{J_{s_3}}{k_3^{s_{3}-1}}\right\rangle_{\bf{nh}}=\text{transverse Ward identity terms}
\end{align}
This distinction will be important to keep in mind since the homogeneous and non-homogeneous parts have different structures and properties. One way to distinguish between the two kinds of solutions in the final answer will be that the non-homogeneous solution depends on the coefficient of the two-point function. Another way is to make use of the transverse Ward identities :
\begin{align}
   \langle k_1\cdot J_{s_1}(k_1) J_{s_2}(k_2)J_{s_3}(k_3)\rangle_{\bf{h}} &=0\notag\\
   \langle k_1\cdot J_{s_1}(k_1) J_{s_2}(k_2)J_{s_3}(k_3)\rangle_{\bf{nh}}&=\rm{WT~identity~terms}.
\end{align}
 In other words, while the homogeneous solution is completely transverse, the non-homogeneous solution gets contribution from both transverse as well as local (or longitudinal) terms.
 
Since the 3-point correlators can be parity-violating, it will be useful to break up the homogeneous and non-homogeneous parts further into parity-even and parity-odd contributions:
 \begin{align}\label{jssph1}
    \langle J_{s_1} J_{s_2} J_{s_3} \rangle &= \langle J_{s_1} J_{s_2} J_{s_3} \rangle_{\bf{h}}+ \langle J_{s_1} J_{s_2} J_{s_3} \rangle_{\bf{nh}}\nonumber\\
    \langle J_{s_1} J_{s_2} J_{s_3} \rangle_{\bf{h}} &= \langle J_{s_1} J_{s_2} J_{s_3} \rangle_{\bf{h},\text{even}}+\langle J_{s_1} J_{s_2} J_{s_3} \rangle_{\bf{h},\text{odd}}\nonumber\\
     \langle J_{s_1} J_{s_2} J_{s_3} \rangle_{\bf{nh}} &= \langle J_{s_1} J_{s_2} J_{s_3} \rangle_{\bf{nh},\text{even}}+\langle J_{s_1} J_{s_2} J_{s_3} \rangle_{\bf{nh},\text{odd}}
\end{align}
 For a detailed discussion on the homogeneous and non-homogeneous contributions to three-point correlators and their distinction from transverse and longitudinal contributions see appendix \ref{hnhvstl}.
\subsubsection{Degeneracy structure}
\label{sectiondeg}
In three dimensions, there exist degeneracies in tensor structures which complicate the analysis of correlators. The existence of degeneracy is tied to the simple fact that not more than three vectors can be linearly independent in three dimensions.

The basic problem is that the different tensor structures in the ansatz for a correlator become linearly dependent due to degeneracies. This affects the analysis of both parity-even and parity-odd correlators. For the parity-odd correlator, Schouten identities, which relate various tensor structures involving Levi-Civita tensors, are an additional source of complication. The main problem is that while solving the conformal Ward identity, one needs to identify the correct independent set of tensor structures to be able to write down differential equations for the form-factors. However, this process becomes very complicated for correlators involving spin-2 or higher spin operators.

An example of such an identity in three dimensions is :
\begin{align}\label{jjoS1}
\epsilon^{z_1 z_2 k_1}(k_1\cdot k_2)+\epsilon^{z_1 k_1 k_2}k_1\cdot z_2-\epsilon^{z_1 z_2 k_2}k_1^2-\epsilon^{z_2 k_1 k_2}k_1\cdot z_1 &=0,
\end{align}
where we have used the notation\footnote{We will often use this notation in this paper.} $\epsilon^{z_2 k_1 k_2}=\epsilon_{\mu\nu\rho}z_2^{\mu} k_1^{\nu} k_2^{\rho}.$
The structures that appear in the above equation arise in the ansatz for various parity-odd correlators such as $\ev{JJO}_{\text{odd}}$. The above equation then implies that a term with $\epsilon^{z_1 k_1 k_2}$ in the ansatz can be eliminated in favour of other structures \footnote{See \cite{Jain:2021wyn} for details of the complete momentum space analysis of $\ev{JJO}_{\text{odd}}$.}. This, while essential to be taken into account, makes cumbersome the correct ansatz with a minimal basis of independent structures.

Other than Schouten identities, there are identities such as \cite{Bzowski:2013sza} :
\be\label{degen}
\delta^{\mu \nu}= \frac{4}{J^2}\Big( k_i^2 k_j^{\mu} k_j^{\nu} + k_j^2 k_i^{\mu} k_i^{\nu}-\vec{k}_i . \vec{k}_j(k_i^{\mu} k_j^{\nu}+k_j^{\mu} k_i^{\nu})+n^{\mu}n^{\nu}  \Big)
\ee
where $n^{\mu}=\epsilon^{\mu \nu \rho}k_{\nu}k_{\rho}$ and $i \ne j=1,2,3$. We also have \cite{Bzowski:2013sza} :
\begin{align}\label{eid}
\begin{array}{l}
\Pi_{\alpha \beta}^{\mu \nu}\left(k_{j}\right) n^{\alpha} n^{\beta}=-k_{j}^{2} \Pi_{\alpha \beta}^{\mu \nu}\left(k_{j}\right) k_{(j+1) \bmod 3}^{\alpha} k_{(j+1) \bmod 3}^{\beta} \quad\quad\quad\quad j =1,2,3
\end{array}
\end{align}
Another example of a degeneracy is \cite{Bzowski:2017poo} :
\begin{align}\label{dgev}
    &\Pi_{\mu_1\nu_1\beta_1}^{\alpha_1}(k_1) \Pi_{\mu_2\nu_2\beta_2}^{\alpha_2}(k_2)4!\delta^{\beta_1}_{\left[\alpha_1\right.}\delta^{\beta_2}_{\alpha_2}k_{1\alpha_2}k_{\left.2\alpha_4\right]}k_1^{\alpha_3}k_2^{\alpha_4}\notag\\[5 pt]
    &\hspace{.5cm}=\Pi_{\mu_1\nu_1\alpha_1\beta_1}(k_1) \Pi_{\mu_2\nu_2\alpha_2\beta_2}(k_2)\bigg[k_2^{\alpha_1}k_2^{\beta_1}k_3^{\alpha_2}k_3^{\beta_2}\bigg.\notag\\[5 pt]
    &\hspace{1cm}\bigg. -(k_1^2+k_2^2-k_3^2)\delta^{\beta_1\beta_2}k_2^{\alpha_1}k_3^{\alpha_2}-\frac{J^2}{4}\delta^{\alpha_1\alpha_2}\delta^{\beta_1\beta_2}\bigg]=0
\end{align}
These also allow certain basis structures to be expressed in terms of others. Both parity-even and parity-odd degeneracies complicate the analysis when computing correlation functions. 

One of the advantages of working with spinor-helicity variables is that the degeneracies become trivial in these variables. For example, the left hand side of both \eqref{jjoS1} and \eqref{dgev} become identically zero in spinor-helicity variables. One can check that all the Schouten identities and other identities relating various tensor structures also become trivial in spinor-helicity variables.



\section{Three-point functions : Explicit solutions in spinor-helicity variables}
\label{sectionSHresults}
In this section we focus on determining $CFT_3$ 3-point correlators in spinor-helicity variables. In particular, we compute correlators of the form $\langle J_s O_{\Delta} O_{\Delta}\rangle,\langle J_s J_s O_{\Delta}\rangle,\langle J_s J_s J_s\rangle$ and $\langle J_{s_1} J_s J_s\rangle$ where $J_s$ is a symmetric, traceless, spin-$s$ conserved current  with scaling dimension $\Delta=s+1$, and $O_{\Delta}$ is a scalar operator with scaling dimension $\Delta$. In three dimensions, 3-point correlators involving only spinning operators are always finite, whereas those involving a scalar operator require renormalization for large enough values of $\Delta$.

We will observe that splitting the correlator into homogeneous and non-homogeneous parts in the sense explained in Section \ref{hnh} is useful. As we demonstrate, whenever there exists a homogeneous parity-even solution to the conformal Ward identity in spinor-helicity variables, there also exists a homogeneous parity-odd solution and the two are identical up-to some signs. Interestingly, in the case of divergent correlators, the parity-odd and the parity-even correlators continue to match even after renormalization, although the renormalization procedure for the two differs. Further, it turns out that the non-homogeneous part is always parity-even. Any parity-odd contribution to the non-homogeneous part is always a contact term. After the first example in which we present all the details, in each case we will give the correlator ansatz and then write down the form-factors as solution of the CWI's, relegating the details to Appendix \ref{CWI}.

\subsection*{Notation}
 A spin $s$ current has various helicity components such as 
$J_s^{-\cdots -},J_s^{-\cdots +-},\cdots, J_s^{+\cdots+}$. Due to tracelessness, mixed helicity components vanish. Hence the only nontrivial helicity components are 
$J_s^{-\cdots-}$ and $ J_s^{+\cdots +}$ which we denote by $J_s^{-}$ and $J_s^{+}$, respectively.

\subsection{$\langle J_s O_{\Delta} O_{\Delta}\rangle$}
\label{jsododshsection}
In this section, we calculate correlators of the form $\langle J_s O_{\Delta}O_{\Delta} \rangle$. The Ward-Takahashi (WT) identity when the spinning operator is either a spin-one conserved current or the stress-tensor (i.e. when $s=1$ or $s=2$) is given by the following \cite{Bzowski:2013sza, Baumann:2020dch}:
\begin{align}\label{jToowi}
k_{1\mu}\langle J^{\mu}O_{\Delta}O_{\Delta} \rangle &= \langle O_{\Delta}(k_3)O_{\Delta}(-k_3) \rangle -\langle O_{\Delta}(k_2)O_{\Delta}(-k_2) \rangle\nonumber\\
k_{1\mu}z_{1\nu}\langle T^{\mu\nu}O_{\Delta}O_{\Delta}\rangle &= (k_2 \cdot z_1)\left(\langle O_{\Delta}(k_3)O_{\Delta}(-k_3) \rangle -\langle O_{\Delta}(k_2)O_{\Delta}(-k_2) \rangle\right)
\end{align}
where in the second equation we have contracted both sides of the WT identity with null transverse polarization vectors.
It is straightforward to generalise the WT identity to arbitrary spin-$s$ conserved currents by matching the spin and scaling dimensions on both sides of the identity. This gives the following :
\begin{align}\label{jsoowi}
z_{1\mu_2}\cdots z_{1\mu_s}k_{1\mu_1}\langle J^{\mu_1 \cdots \mu_s} O_{\Delta} O_{\Delta} \rangle = (k_2 \cdot z_1)^{s-1}(\langle O_{\Delta}(k_3)O_{\Delta}(-k_3) \rangle-\langle O_{\Delta}(k_2)O_{\Delta}(-k_2) \rangle)
\end{align}
We will see that the homogeneous part of the correlator is zero. The non-homogeneous part has the scalar two-point function on the right hand side\footnote{A correlator comprising one conserved current and two scalar operators with different scaling dimensions also vanishes, i.e.
\begin{align}
\langle J_s O_{\Delta_1}O_{\Delta_2} \rangle &= 0~~~~~\rm{for~~~ \Delta_1 \ne \Delta_2}
\end{align}}. 
Consequently, the odd part of the correlator goes to zero as there is no parity-odd scalar two-point function. Thus this correlator has only a parity-even non-homogeneous part.

As noted in Section \ref{SectionCG}, when the correlator involves operators with scaling dimensions other than 2, it is convenient to divide the insertions by appropriate powers of the corresponding momenta $k$ such that they have $\Delta=2$. The correlator itself is obtained at the end by restoring the powers of $k$. Keeping this in mind, we start with the following ansatz for the correlator :
\begin{align}
\label{jooansatz}
\left\langle \frac{J_s^-}{k_1^{s-1}} \frac{O_{\Delta}}{k_2^{\Delta-2}}\frac{O_{\Delta}}{k_3^{\Delta-2}} \right\rangle = {F}(k_1, k_2, k_3)\langle 12 \rangle^{s} \langle \bar{2}1\rangle^{s}
\end{align}
The action of the generator of special conformal transformations $\widetilde{K}$ is then given by (see Section \ref{SectionCG}) :
\begin{align}\label{jseqk}
\widetilde{K}^{\kappa}\left\langle \frac{J_s^-}{k_1^{s-1}}\frac{O_{\Delta}}{k_2^{\Delta-2}}\frac{O_{\Delta}}{k_3^{\Delta-2}} \right\rangle \nonumber&=\frac{2z_1^{-\kappa}c_O}{k_1^{2s-1} k_2^{\Delta-2}k_3^{\Delta-2}}(k_3^{2\Delta-3}-k_2^{2\Delta-3})\notag\\
&\hspace{.5cm}+(\Delta-1)(\Delta-2)\left\langle J_s^- \frac{O_{\Delta}}{k_2^{\Delta-2}}\frac{O_{\Delta}}{k_3^{\Delta-2}}\right\rangle \left(\frac{k_2^{\kappa}}{k_2^{2}} -\frac{k_3^{\kappa}}{k_3^{2}}\right)
\end{align}
Contracting \eqref{jseqk} with  $b_{\kappa}= (\sigma^{\kappa})_{\beta}^{\;\;\alpha}\lambda_{1\alpha}\lambda_1^{\beta}$, $b_{\kappa}= (\sigma^{\kappa})_{\beta}^{\;\;\alpha}(\lambda_{1\alpha}\lambda_2^{\beta}+\lambda_{2\alpha}\lambda^{\beta}_1)$ and $b_{\kappa}=(\sigma^{\kappa})_{\beta}^{\;\;\alpha}\lambda_{2\alpha}\lambda_2^{\beta}$ gives the following :
\begin{align}
\label{jsooeqn1}
&\frac{\partial^2 {F}}{\partial k_2^2}-\frac{\partial^2 {F}}{\partial k_3^2}= -\frac{{F}}{k_2^{2}k_3^{2}}(\Delta-1)(\Delta-2)(k_2^2-k_3^2)\\[10 pt]
\label{jsooeqn2}
&\frac{k_1}{2}\left(\frac{\partial^2 {F}}{\partial k_3^2}-\frac{\partial^2 {F}}{\partial k_1^2}\right)+\frac{k_2}{2}\left(\frac{\partial^2 {F}}{\partial k_2^2}-\frac{\partial^2 {F}}{\partial k_3^2}\right)-s\frac{\partial {F}}{\partial k_1}\notag\\[5pt]
&\hspace{4cm}=\frac{2(\Delta-1)(\Delta-2){F}}{k_2^{2}k_3^{2}}k_2(k_2^2+k_3^2-k_1k_2)\\[3pt]
\label{jsooeqn3}
&\frac{1}{4}(k_1-k_2+k_3)(-k_1+k_2+k_3)\left(\frac{\partial^2 {F}}{\partial k_1^2}-\frac{\partial^2 {F}}{\partial k_3^2}\right)+s^2 {F}+s k_2\left(\frac{\partial {F}}{\partial k_1}+\frac{\partial {F}}{\partial k_2}\right)\notag\\[5 pt]
&\hspace{1.5cm}=c_O\frac{k_3^{2\Delta-3}-k_2^{2\Delta-3}}{k_1^3}+\frac{{F}}{k_3^{2}}(\Delta-1)(\Delta-2)(k_1-k_2+k_3)(-k_1+k_2+k_3)
\end{align}
Finally, the dilatation Ward identity is given by :
\begin{align}\label{jsoodil}
\left(\sum_{i=1}^3 k_i \frac{\partial {F}}{\partial k_i}\right)+2s{F}=0
\end{align}
The above differential equations \eqref{jsooeqn1},\eqref{jsooeqn2}, \eqref{jsooeqn3} and \eqref{jsoodil} can be solved to get :
\begin{align}
 F =  c_O k_2^{-\Delta+2}k_3^{-\Delta+2} I_{\frac{1}{2}+s\{\frac{1}{2}-s,\Delta-\frac{3}{2},\Delta-\frac{3}{2}\}}
\end{align}
where the triple-$K$ integral \cite{Bzowski:2013sza} which occurs in the RHS of this equation is defined in \eqref{triplekint}. After taking the momentum factors in the denominator of the LHS of \eqref{jooansatz} to the RHS and using the above result for the form factor we obtain the correlator :
\begin{align}\label{Jsoofnl}
\langle J_s^- O_{\Delta}O_{\Delta} \rangle= c_O k_1^{s-1} I_{\frac{1}{2}+s\{\frac{1}{2}-s,\Delta-\frac{3}{2},\Delta-\frac{3}{2}\}}\langle 12 \rangle^s \langle\bar{2}1\rangle^s
\end{align}
When $c_O=0$, there is no non-trivial solution to the differential equations and one has :
\begin{equation}
    \langle J_sO_{\Delta}O_{\Delta} \rangle _{{\bf{h}}} =0.
\end{equation}

\subsection{$\langle J_s J_s O_{\Delta}\rangle$}
In this section, we compute correlators of the form $\langle J_s J_s O_{\Delta}\rangle$ for general spin $s$. As discussed in Section \ref{hnh}, we separate out the correlator into homogeneous and non-homogeneous parts :
\begin{equation}
  \langle J_s J_s O_{\Delta}\rangle = \langle J_s J_s O_{\Delta}\rangle_{{\bf{h}}}+ \langle J_s J_s O_{\Delta}\rangle_{{\bf{nh}}}  
\end{equation}
 The correlator $\langle J_s J_s O_{\Delta}\rangle$ is completely transverse :
\begin{equation}\label{Wtjsjso}
    \langle {k_1}\cdot J_s(k_1) J_s(k_2) O_{\Delta}(k_3) \rangle = \langle J_s(k_1) {k_2}\cdot J_s(k_2)  O_{\Delta}(k_3) \rangle=0
\end{equation}
where $k\cdot J_s(k)=k_{\mu_1} J^{\mu_1 \mu_2 ....\mu_s}(k).$ 
This implies that the non-homogeneous part of the correlator is zero :
\begin{equation}
  \langle J_s J_s O_{\Delta}\rangle_{{\bf{nh}}}  =0.  
\end{equation}
We will now compute the explicit form of the correlators for arbitrary $\Delta$. We find that for $\Delta \ge 4$, there is a divergence and we need to regularize and renormalize to obtain finite correlators.

We consider the following ansatz for the correlator :
\eqref{generalansatz} :
\begin{align}\label{ssoansatzsh}
\left\langle \frac{J^{s-}(k_1)}{k_1^{s-1}}\frac{J^{s-}(k_2)}{k_2^{s-1}}\frac{O_{\Delta}(k_3)}{k_3^{\Delta-2}} \right\rangle &=(c_1\,{F}_1(k_1, k_2, k_3)+i\,c_2\,{F}_2(k_1, k_2, k_3))\,\langle 12 \rangle^{2s}\nonumber\\
\left\langle \frac{J^{s+}(k_1)}{k_1^{s-1}}\frac{J^{s+}(k_2)}{k_2^{s-1}}\frac{O_{\Delta}(k_3)}{k_3^{\Delta-2}} \right\rangle &=(c_1\,{F}_1(k_1, k_2, k_3)-i\,c_2\,{F}_2(k_1, k_2, k_3))\,\langle {\bar {1}}{\bar {2}} \rangle^{2s}\nonumber\\
\left\langle \frac{J^{s-}(k_1)}{k_1^{s-1}}\frac{J^{s+}(k_2)}{k_2^{s-1}}\frac{O_{\Delta}(k_3)}{k_3^{\Delta-2}} \right\rangle &=(d_1\,{G}_1(k_1, k_2, k_3)+i\,d_2\,{G}_2(k_1, k_2, k_3))\,\langle 1{\bar{2}} \rangle^{2s}
\end{align}
It is interesting to note that the conformal Ward identity gives identical equations for the parity-odd and the parity-even parts. The details of these equations and their solution are provided in Appendix \ref{appJsJsOdetails} where we also discuss examples for special values of $\Delta$ and $s$. Here we give the final form of the solution :
\begin{align}
    \label{ssoformfactor}
 F_1(k_1, k_2, k_3)&= F_2(k_1, k_2, k_3)=k_3^{2-\Delta}I_{(\frac{1}{2}+2s)\{\frac{1}{2},\frac{1}{2},\Delta-\frac{3}{2}\}}\nonumber\\
     G_1(k_1, k_2, k_3)&= G_2(k_1, k_2, k_3)=0.
\end{align}
Substituting the form-factor in the ansatz \eqref{ssoansatzsh} we obtain 
\begin{align}
\begin{split}
\langle J_s^- J_s^- O_{\Delta} \rangle&=\langle J_s^- J_s^- O_{\Delta} \rangle_{\text{even}}+\langle J_s^- J_s^- O_{\Delta} \rangle_{\text{odd}}=\left(c_1+ i c_2\right)\,\left(k_1 k_2\right)^{s-1} I_{(\frac{1}{2}+2s)\{\frac{1}{2},\frac{1}{2},\Delta-\frac{3}{2}\}} \langle 12 \rangle^{2s}\\[5 pt]
\langle J_s^+ J_s^+ O_{\Delta} \rangle&=\langle J_s^+ J_s^+ O_{\Delta} \rangle_{\text{even}}+\langle J_s^+ J_s^+ O_{\Delta} \rangle_{\text{odd}}=\left(c_1 - i c_2\right)\,\left(k_1 k_2\right)^{s-1} I_{(\frac{1}{2}+2s)\{\frac{1}{2},\frac{1}{2},\Delta-\frac{3}{2}\}} \langle \bar 1\bar 2 \rangle^{2s}\\[5 pt]
\langle J_s^- J_s^+ O_{\Delta} \rangle &= 0\label{jjofinalanswersh}
\end{split}
\end{align}
For $\Delta \ge 4$, the above triple-$K$ integrals and thereby the correlators are divergent. A detailed study of the renormalization of these correlators will be carried out in Section \ref{renor11}. We will see that the relationship between the parity-even and the parity-odd parts of a correlator in spinor-helicity variables continues to hold even after renormalization.


\subsection{$\langle J_s J_s J_s\rangle$}
In this subsection we concentrate on the three point function of a general spin $s$ conserved current $J_s$ \footnote{If $s$ is odd then we need to consider a  non-abelian current to have a non-trivial correlator}. Since the correlator $\langle J_s J_s J_s\rangle$ satisfies a nontrivial transverse WT identity it has both the homogeneous as well as the non-homogeneous contributions. 

Let us split the correlator into the odd and even contributions :
$$\langle J_s J_s J_s\rangle=\langle J_s J_s J_s\rangle_{\text{even}}+\langle J_s J_s J_s\rangle_{\text{odd}}.$$ It will turn out that $\langle J_s J_s J_s\rangle_{\text{even}}$ has both the homogeneous and the non-homogeneous contributions whereas $\langle J_s J_s J_s\rangle_{\text{odd}}$ has a non-trivial homogeneous part but the non-homogeneous part is always a contact term. 

\subsubsection{$\langle JJJ \rangle$}
Let us start our analysis with the 3-point function of the spin-1 current $J_{\mu}$. As noted earlier, for this correlator to be non-zero, the currents have to be non-abelian. The WT identity is given by \cite{Bzowski:2013sza,Bzowski:2017poo,Baumann:2020dch} :
\begin{align}
\label{WTjjj}
\begin{split}
k_{1\mu}\langle J^{\mu a}(k_1) J^{\nu b}(k_2) J^{\rho c}(k_2) \rangle &= \left(f^{adc}\langle J^{\rho d}(k_2)J^{\nu b}(-k_2) \rangle-f^{abd}\langle J^{\nu d}(k_3)J^{\rho c}(-k_3)\rangle\right)\\[5 pt]
&\hspace{-1cm}+\bigg[\left(\frac{k_2^{\nu}}{ k_2^2}f^{abd}k_{2\alpha}\langle J^{\alpha d}(k_3)J^{\rho c}(-k_3)\rangle\right)+\left((k_2, \nu) \leftrightarrow (k_3, \rho)\right)\bigg]
\end{split}
\end{align}
Let us consider the following ansatz for the two helicity components of the correlator \footnote{We will suppress the color indices which amounts to suppressing an overall factor of $f^{abc}$.} :
\begin{align}\label{jjjevenansatzab}
\langle J^-(k_1) J^-(k_2)J^-(k_3) \rangle &=\left(F_1(k_1, k_2, k_3)+ i F_2(k_1, k_2, k_3)\right)\langle 12 \rangle \langle 23 \rangle \langle 31 \rangle\\
\langle J^-(k_1) J^-(k_2)J^+(k_3) \rangle &=\left(G_1(k_1, k_2, k_3)+ i G_2(k_1, k_2, k_3)\right)\langle 12 \rangle \langle 2\bar{3} \rangle \langle \bar{3}1 \rangle
\end{align}
The solutions of the conformal Ward identity  are given by (see Appendix \ref{CWIjjj})
\begin{align}
F_1(k_1, k_2, k_3)&= \frac{c_1 }{E^3}+\frac{c_J}{k_1 k_2 k_3}\\[5 pt]
G_1(k_1, k_2, k_3)&=\frac{c_2 }{(k_1+k_2-k_3)^3}+ c_J \frac{E-2k_3}{E(k_1 k_2 k_3)}\\[5 pt]
F_2(k_1, k_2, k_3)&= \frac{c'_1}{E^3}+\frac{c'_J}{k_1 k_2 k_3}\\[5 pt]
G_2(k_1, k_2, k_3)&=\frac{c_2'}{(k_1+k_2-k_3)^3}+\frac{ c'_J}{k_1 k_2 k_3}\label{F1G1F2G2JJJ}
\end{align} where $c_J$ and  $c_J'$ are the parity-even and parity-odd coefficients of the two-point function of conserved currents (see \eqref{JsJstwopointfn}). 
The terms proportional to $c_1, c_1'$ and $c_2,c_2'$ are the homogeneous solutions to the differential equations and those  proportional to $c_J,c_J'$ are the non-homogeneous solutions. Since $G(k_1, k_2, k_3)$ and ${\widetilde G}(k_1, k_2, k_3) $ both have an un-physical pole when $k_1+k_2=k_3$, we set the coefficients of these terms to zero, i.e. $c_2=c_2'=0.$

\subsection*{Summary of the solution}
Taking into account both the parity-even and the parity-odd contributions, we obtain :
\begin{align}
\label{jjjshal}
\langle J^-(k_1) J^-(k_2)J^-(k_3) \rangle&=\left(\frac{c_1+i c'_1}{E^3}+\frac{c_J+ i c'_J }{k_1 k_2 k_3}\right)\langle 12 \rangle \langle 23 \rangle \langle 31 \rangle\\
\langle J^-(k_1) J^-(k_2)J^+(k_3) \rangle &=\,\frac{ 1}{k_1 k_2 k_3}\left((c_J+i c_J')-c_J\frac{2 k_3}{E}\right)\langle 12 \rangle \langle 2\bar{3} \rangle \langle \bar{3}1 \rangle
\end{align}
In the next section, we will convert these expressions into the momentum space and see that the non-homogeneous contribution to the parity-odd correlator (term proportional to $c_J'$) becomes a contact term.

\subsubsection{$\langle TTT \rangle$}
Let us now consider the correlator with three insertions of the stress-tensor operator.  
The transverse Ward identity satisfied by the correlator is given by \cite{Bzowski:2013sza,Bzowski:2017poo,Baumann:2020dch}:
\begin{align}\label{wtidttt}
\begin{split}
z_{1\mu}k_{1\nu}&\langle T^{\mu\nu}(k_1) T(k_2) T(k_3) \rangle \\
&= -(z_1 \cdot k_2) \langle T(k_1 + k_2) T(k_3) \rangle +2(z_1 \cdot z_2)k_{2\mu}z_{\nu}\langle T^{\mu\nu}(k_1+k_2) T(k_3) \rangle\\[5 pt]
&\hspace{.3cm}-(z_1 \cdot k_3)\langle  T(k_1+k_3) T(k_2) \rangle + 2(z_1 \cdot z_3)k_{3\mu}z_{3\nu}\langle T^{\mu\nu}(k_1+k_3) T(k_2) \rangle\\[5 pt]
&\hspace{.3cm}+(k_1 \cdot z_2)z_{1\mu}z_{2\nu}\langle T^{\mu\nu}(k_1+k_2) T(k_3) \rangle+(z_1 \cdot z_2)k_{1\mu}z_{2\nu}\langle T^{\mu\nu}(k_1+k_2) T(k_3) \rangle\\[5 pt]
&\hspace{.3cm}+(k_1 \cdot z_3)z_{1\mu}z_{3\nu}\langle T^{\mu\nu}(k_1+k_3) T(k_2) \rangle+(z_1 \cdot z_3)k_{1\mu}z_{3\nu}\langle T^{\mu\nu}(k_1+k_3) T(k_2) \rangle
\end{split}
\end{align}
where $T(k)\equiv z_{\mu}z_{\nu}T^{\mu\nu}(k)$. Thus the correlator can have both homogeneous and non-homogeneous solutions for the parity-even  and parity-odd correlation functions. The parity-even solution was already discussed in \cite{Bzowski:2013sza}. The parity-odd homogeneous contribution was computed in \cite{Maldacena:2011nz} using Feynman diagram computations in $dS_4$. Here we reproduce the same result by solving the conformal Ward identities. We also get a nontrivial non-homogeneous contribution to the parity-odd correlator which in momentum space turns out to be a contact term. 
\subsubsection*{{\bf{$\langle TTT \rangle_{\text{even}}$}}}
The parity-even contribution to the correlator $\langle TTT \rangle_{\text{even}}$ is given by \cite{Bzowski:2013sza,Baumann:2020dch} 
\begin{align}\label{evenTTTso}
\langle T^-T^-T^- \rangle_{\text{even}} &= \left(c_1 \frac{c_{123}}{E^6}+c_T\frac{E^3-E b_{123}-c_{123}}{c_{123}^2}\right) \langle 12 \rangle^2 \langle 23 \rangle^2 \langle 31 \rangle^2\notag\\[5 pt]
\langle T^-T^-T^+ \rangle_{\text{even}} &= c_T \frac{(E-2k_3)^2(E^3-E b_{123}-c_{123})}{E^2 c_{123}^2} \langle 12 \rangle^2 \langle 2\bar{3} \rangle^2 \langle \bar{3}1 \rangle^2
\end{align}
where $b_{123} = (k_1 k_2+k_2 k_3+k_3 k_1)$, $c_{123} = k_1 k_2 k_3$, and $c_T$ comes from the parity-even two point function of the stress tensor \eqref{JsJstwopointfn}.
The terms proportional to $c_1$ and $c_T$ are the homogeneous  and the non-homogeneous contributions to the correlator respectively.
\subsubsection*{$\langle TTT \rangle_{\text{odd}}$}
The parity-odd part of the correlator can be solved analogously (see Appendix \ref{CWItttapp} for details). The answer is given by 
\begin{align}
\label{tttoddsphresult}
\left\langle \frac{T^-}{k_1}\frac{T^-}{k_2} \frac{T^-}{k_3} \right\rangle_{odd} &=i\left(c_1'\frac{1}{E^6}+c_T'\frac{E^3-E\,b_{123}-c_{123}}{c_{123}^3}\right)\langle 12 \rangle^2 \langle 23 \rangle^2 \langle 31 \rangle^2\\[6 pt]
\left\langle \frac{T^-}{k_1} \frac{T^-}{k_2} \frac{T^+}{k_3} \right\rangle_{odd} &=i \left(c_T'\frac{(E-2k_3)^2-(E-2k_3)(b_{123}-2k_3\,a_{12})+c_{123}}{c_{123}^3}\right)\langle 12 \rangle^2 \langle 2\bar{3} \rangle^2 \langle \bar{3}1 \rangle^2
\end{align}
where  $a_{12}=k_1+k_2$, $b_{123}=k_1k_2+k_2k_3+k_1k_3$ and $c_{123}=k_1\,k_2\,k_3$ and 
$c_T'$ arises in the parity-odd two point function of the stress tensor \eqref{JsJstwopointfn}.
The terms proportional to $c_1'$ and $c_T'$ are the homogeneous  and the non-homogeneous contributions to the correlator respectively.

\subsection*{Summary of the solution}
Taking into account both the parity-even and the parity-odd contributions, we obtain :
\begin{align}
\label{TTTshal}
\langle T^-(k_1) T^-(k_2)T^-(k_3) \rangle&=\left[\left(c_1+i c_1'\right) \frac{c_{123}}{E^6}+\left( c_T+ i c_T'\right)\frac{E^3-E b_{123}-c_{123}}{c_{123}^2}\right] \langle 12 \rangle^2 \langle 23 \rangle^2 \langle 31 \rangle^2\notag\\[5 pt]\\
\langle T^-(k_1) T^-(k_2)T^+(k_3) \rangle &=\Big[c_T \frac{(E-2k_3)^2(E^3-E b_{123}-c_{123})}{E^2 c_{123}^2} + \notag\\
&i c_T' \frac{(E-2k_3)^2-(E-2k_3)(b_{123}-2k_3\,a_{12})+c_{123}}{c_{123}^3}\Big]\langle 12 \rangle^2 \langle 2\bar{3} \rangle^2 \langle \bar{3}1 \rangle^2 
\end{align} 
The other helicity components of the correlator can be obtained by complex conjugating the above results.

In the next section, we will show that the non-homogeneous contribution to the parity odd correlator (the term proportional to $c_T'$ in \eqref{TTTshal}) is a contact term. Thus the only non-trivial contribution to the non-homogeneous piece in the correlator comes from the parity-even part.

\subsubsection{$\langle J_s J_s J_s \rangle$ for general spin: The homogeneous part}
In this subsection we generalize the above discussion to the three-point function $\langle J_s J_s J_s \rangle$ of arbitrary spin $s$ conserved current. We first split the correlator into homogeneous and non-homogeneous pieces, and then separate them into their parity-even and parity-odd parts as indicated in \eqref{jssph1}. The non-homogeneous piece $\langle J_s J_s J_s \rangle_{\bf{nh}}$ requires us to know the WT identity which for general spin is quite complicated. However, on general grounds we can argue that the parity-odd part of this term, $\langle J_s J_s J_s \rangle_{\bf{nh},\text{odd}}$, is a contact term for general spin $s$. We refer the reader to Section \ref{contact} for details. In the rest of this subsection, we focus on obtaining the homogeneous part of the correlator which does not require the WT identity.
 
\subsubsection*{Homogeneous solution}
For the homogeneous solution, the parity-even and the parity-odd contributions are again the same in spinor-helicity variables.
We start with the following ansatz for $\langle J_s J_s J_s \rangle_{\textbf{h}}$ :
\begin{align}\label{jsjsjsHH}
\left\langle \frac{J_s^-}{k^{s-1}_1}\frac{J_s^-}{k^{s-1}_2} \frac{J_s^-}{k^{s-1}_3} \right\rangle_{\textbf{h}} &=\,F(k_1,k _2, k_3)\langle 12 \rangle^s \langle 23 \rangle^s \langle 31 \rangle^s\\[6 pt]
\left\langle \frac{J_s^-}{k_1^{s-1}} \frac{J_s^-}{k_2^{s-1}} \frac{J_s^+}{k_3^{s-1}} \right\rangle_{\textbf{h}} &=\,G(k_1,k _2, k_3)\langle 12 \rangle^s \langle 2\bar{3} \rangle^s \langle \bar{3}1 \rangle^s.
\end{align}
Since we are focusing on the homogeneous part, the action of the conformal generator is given by :
\begin{align}\label{Kkappajsjsjs}
\widetilde{K}^{\kappa} \left\langle \frac{J_s^-}{k_1^{s-1}}\frac{J_s^-}{k_2^{s-1}} \frac{J_s^-}{k_3^{s-1}}  \right\rangle_{\textbf{h}} &=0 ,~~
\widetilde{K}^{\kappa} \left\langle \frac{J_s^-}{k_1^{s-1}} \frac{J_s^-}{k_2^{s-1}} \frac{J_s^+}{k_3^{s-1}} \right\rangle_{\textbf{h}} =0.
\end{align}
The action of $\widetilde{K}^{\kappa}$ on the ansatz, after dotting with $b_{\kappa}=(\sigma^{\kappa})_{\beta}^{\;\;\alpha}(\lambda_{2\alpha}\lambda^{\beta}_3+\lambda_{3\alpha}\lambda_2^{\beta})$, becomes 
\begin{align}
2s\left(\frac{\partial F}{\partial k_2}-\frac{\partial F}{\partial k_3}\right)+k_3\left(\frac{\partial^2 F}{\partial k_2^2}-\frac{\partial^2 F}{\partial k_3^2}\right)-k_2\left(\frac{\partial^2 F}{\partial k_1^2}-\frac{\partial^2 F}{\partial k_2^2}\right)&= 0\\[5 pt]
2s\left(\frac{\partial G}{\partial k_2}+\frac{\partial G}{\partial k_3}\right)-k_3\left(\frac{\partial^2 G}{\partial k_2^2}-\frac{\partial^2 G}{\partial k_3^2}\right)-k_2\left(\frac{\partial^2 G}{\partial k_1^2}-\frac{\partial^2 G}{\partial k_2^2}\right)&= 0
\end{align}
The dilatation Ward identity is given by :
\begin{align}
&\left(\sum_{i=1}^3 k_i \frac{\partial F}{\partial k_i}\right)+3sF=0,~~\left(\sum_{i=1}^3 k_i \frac{\partial G}{\partial k_i}\right)+3sG=0
\end{align}
The solutions for $F$ and $G$ are then given (see appendix \ref{CWI}) by
\begin{align}
\label{tttsh}
F(k_1, k_2, k_3) &= \frac{c_1}{E^{3s}}\\[5 pt]
G(k_1, k_2, k_3) &= 0
\end{align}
\subsection*{Summary of result}
Considering the parity-even and the parity-odd contributions we obtain the homogeneous part of the correlator :
\begin{align}\label{jsjsjsHH}
\left\langle J_s^- J_s^- J_s^- \right\rangle_{\textbf{h}} &=(c_1+ i c_2)k^{s-1}_1 k^{s-1}_2 k^{s-1}_3\frac{1}{E^{3s}}\langle 12 \rangle^s \langle 23 \rangle^s \langle 31 \rangle^s,~~~
\left\langle  J_s^- J_s^- J_s^+\right\rangle_{\textbf{h}} =0
\end{align}
The other helicity components can be obtained by a simple complex conjugation.
\subsection{$\langle J_{s_1}J_s J_s \rangle$}
In this subsection we generalize the above discussion to three-point correlators of the kind $\langle J_{s_1} J_s J_s \rangle$ for arbitrary spins $s$ and $s_1$. We again find it convenient to split the correlator into various parts as in \eqref{jssph1}.
The WT identity for $ \langle J_{s_1} J_s J_s \rangle$ for general spins $s$ and $s_1$ is quite complicated. However, as discussed in Section \ref{contact}, we can argue that the parity-odd contribution to the non-homogeneous part, $\langle J_s J_s J_s \rangle_{\bf{nh},\text{odd}}$, is a contact term. In the following we will calculate the homogeneous and the non-homogeneous contribution to the correlator $\langle TJJ \rangle$. For general spins $s$ and $s_1$, we present only the homogeneous solution.
 
\subsubsection{$\langle TJJ \rangle$}
Let us consider the correlator with a single insertion of the stress-tensor operator and two insertions of the spin-one current operator. The transverse WT identity is given by : \cite{Bzowski:2013sza,Bzowski:2017poo}
\begin{align}\label{tjjoddwi}
\begin{split}
k_{1\mu}\langle T^{\mu\nu}(k_1)J^{\rho}(k_2)J^{\sigma}(k_3) \rangle &= k_{3\mu}\delta^{\nu\sigma} \langle J^{\mu}(k_1+k_3)J^{\rho}(k_2) \rangle+k_{2\mu}\delta^{\nu\rho}\langle J^{\mu}(k_1+k_2)J^{\sigma}(k_3) \rangle\\[5 pt]
&\hspace{.5cm}-k_{3\nu}\langle J^{\sigma}(k_1+k_3)J^{\rho}(k_2) \rangle-k_{2\nu}\langle J^{\rho}(k_1+k_2) J^{\sigma}(k_3) \rangle\\
k_{2\rho}\langle T^{\mu\nu}(k_1)J^{\rho}(k_2)J^{\sigma}(k_3) \rangle &=0.
\end{split}
\end{align}
Since the WT identity is not trivial, the correlator can have both homogeneous and non-homogeneous solutions for the parity-even and the parity-odd correlation functions. The parity-even solution was already discussed in \cite{Bzowski:2013sza,Bzowski:2017poo}. 
\subsubsection*{{\bf{$\langle TJJ \rangle_{\text{even}}$}}}

The even part of the correlator was calculated in \cite{Bzowski:2013sza,Bzowski:2017poo} in momentum space and it is straightforward to convert that into spinor-helicity variables :
\begin{align}\label{evenTTTso}
    \langle T^- J^- J^- \rangle &= c_1 \frac{k_1}{E^4}\langle 12 \rangle^2 \langle 13 \rangle^2\notag\\
    \langle T^+ J^- J^- \rangle &= 0\notag\\
    \langle T^-J^- J^+ \rangle &= c_J \frac{E+k_1}{k_1^2 E^2}\langle 12 \rangle^2 \langle 1\bar{3}\rangle^2
\end{align}
where the term proportional to $c_1$ is the homogeneous term and the term proportional to $c_J$ is the non-homogeneous term.

\subsubsection*{{\bf{$\langle TJJ \rangle_{\text{odd}}$}}}
Let us now consider the parity-odd contribution to the correlator. In the parity-odd case the transverse WT identity 
\eqref{tjjoddwi} gives :
\begin{align}\label{wtjt1}
k_{1\mu}\langle T^{\mu\nu}(k_1)J^{\rho}(k_2)J^{\sigma}(k_3) \rangle_{\text{odd}}&= c_J'\left(k_{2\nu}\epsilon^{\rho\sigma k_3} +k_{3\nu}\epsilon^{\sigma\rho k_2}-\delta^{\nu\rho}\epsilon^{\sigma k_3 k_2}-\delta^{\nu\sigma}\epsilon^{\rho k_2 k_3}\right)
\end{align}
where we have used  $\langle J^{\alpha}(k_1+k_2) J^{\beta}(k_3) \rangle = -c_J' \epsilon^{\alpha \beta k_3}.$ 
Interestingly it turns out that the R.H.S. of the above equation vanishes upon using a Schouten identity. Thus, in addition to the trivial transverse WT identities w.r.t $k_2^\rho$ and $k_3^\sigma$, we have the following trivial transverse WT identity :
\begin{align}\label{tjjoddwifnl}
k_{1\mu}\langle T^{\mu\nu}(k_1)J^{\rho}(k_2)J^{\sigma}(k_3) \rangle_{\text{odd}} &= 0
\end{align}
This immediately implies that the parity odd part of the non-homogeneous part of the correlator is zero :
\begin{equation}
   \langle T^{\mu\nu}(k_1)J^{\rho}(k_2)J^{\sigma}(k_3) \rangle_{\bf{nh},\text{odd}} =0.
\end{equation}
We now turn our attention to computing the homogeneous contribution. 
Let us start with the following ansatz for $\langle TJJ \rangle_{\text{odd}}$ : 
\begin{align}\label{tjjansatz}
\left\langle \frac{T^-}{k_1}J^-J^- \right\rangle_{\text{odd}}&= i\,F(k_1, k_2, k_3)\langle 12 \rangle^2 \langle 13 \rangle^2,\notag\\
\left\langle \frac{T^-}{k_1}J^-J^+ \right\rangle_{\text{odd}} &=i\,G(k_1, k_2, k_3)\langle 12 \rangle^2 \langle 1\bar{3} \rangle^2\notag\\
\left\langle \frac{T^+}{k_1} J^- J^- \right\rangle_{\text{odd}}&=i\,H(k_1, k_2, k_3)\langle \bar{1}2 \rangle^2 \langle \bar{1}3 \rangle^2
\end{align}
The solutions to the resulting CWIs are given (see appendix \ref{CWI}) by :
\begin{align}
\label{tjjoddformfactorssh}
F(k_1, k_2, k_3)&=\frac{c'_1}{E^4},~~
G(k_1, k_2, k_3) =\frac{c'_2}{E^4 (k_2+k_3-k_1)^2},~~~H(k_1, k_2, k_3)=0.
\end{align}
Since the solution for $G$ has an unphysical pole, we set its coefficient $c'_2=0$. Substituting the above form-factors in the ansatz \eqref{tjjansatz}, we obtain the following:
\begin{align}
\label{tjjsh}
\langle T^-J^-J^- \rangle_{\text{odd}} &= i\,c'_1 \frac{k_1}{E^4}\langle 12 \rangle^2 \langle 13 \rangle^2\\[5 pt]
\langle T^+J^+J^+ \rangle_{\text{odd}} &= -i\,c'_1 \frac{k_1}{E^4}\langle \bar 1\bar 2 \rangle^2 \langle \bar 1\bar 3 \rangle^2
\end{align} 
The other helicity components of the correlator are zero.
\par 
\subsubsection*{Summary of Homogeneous contribution to {\bf{$\langle TJJ \rangle$}}}
Adding up the contribution coming from the parity-even and parity-odd parts we obtain
\begin{align}
\label{tjjshal}
\langle T^-J^-J^- \rangle_{\bf{h}} &= \left(c_1+i\,c_1'\right) \frac{k_1}{E^4}\langle 12 \rangle^2 \langle 13 \rangle^2\\[5 pt]
\langle T^+J^+J^+ \rangle_{\bf{h}} &= \left(c_1-i\,c_1'\right)\frac{k_1}{E^4}\langle \bar 1\bar 2 \rangle^2 \langle \bar 1\bar 3 \rangle^2
\end{align} with all other components being zero.
\subsubsection*{Summary of non-homogeneous contribution to {\bf{$\langle TJJ \rangle$}}}
As discussed above, the parity-odd contribution to the non-homogeneous part of the correlator vanishes. Thus from \eqref{evenTTTso} we have the following for the non-homogeneous part of the correlator :
\begin{align}\label{evenTTTsonha}
    \langle T^-J^- J^+ \rangle_{\bf{nh}} &= c_J \frac{E+k_1}{k_1^2 E^2}\langle 12 \rangle^2 \langle 1\bar{3}\rangle^2
\end{align} with all other components zero except the one obtained from complex conjugating \eqref{evenTTTsonha}.

\subsubsection{$\langle J_{s_1} J_{s} J_s \rangle$ for general spin: The homogeneous part}
\label{SectionJs1JsJs}
\subsubsection*{Homogeneous solution}
We start with the following ansatz for $\langle J_{s_1} J_s J_s \rangle$ :
\begin{align}
\label{jsjsjsgeneralformfactors}
\left\langle \frac{J^-_{s_1}}{k_1^{s_1-1}}\frac{J^-_s}{k_2^{s-1}} \frac{J^-_{s}}{k_3^{s-1}} \right\rangle_{\textbf{h}} &=F(k_1,k _2, k_3)\langle 12 \rangle^{s_1} \langle 23 \rangle^{2s-s_1} \langle 31 \rangle^{s_1}\notag\\[6 pt]
\left\langle \frac{J^-_{s_1}}{k_1^{s_1-1}}\frac{J^-_{s}}{k_2^{s-1}} \frac{J^+_{s}}{k_3^{s-1}} \right\rangle_{\textbf{h}} &=G(k_1,k _2, k_3)\langle 12 \rangle^{s_1} \langle 2\bar{3} \rangle^{2s-s_1} \langle \bar{3}1 \rangle^{s_1}\notag\\
\left\langle \frac{J^+_{s_1}}{k_1^{s_1-1}}\frac{J^-_{s}}{k_2^{s-1}} \frac{J^-_{s}}{k_3^{s-1}} \right\rangle_{\textbf{h}} &=H(k_1,k _2, k_3)\langle 12 \rangle^{s_1} \langle 2\bar{3} \rangle^{2s-s_1} \langle \bar{3}1 \rangle^{s_1}
\end{align}
In our analysis, we assume that $2s>s_1.$ The action of the conformal generator on the homogeneous part is given by :
\begin{align}\label{Kkappajs1jsjs}
\widetilde{K}^{\kappa} \left\langle \frac{J^-_{s_1}}{k_1^{s_1-1}}\frac{J_s^-}{k_2^{s-1}} \frac{J_s^-}{k_3^{s-1}} \right \rangle_{\textbf{h}} =
\widetilde{K}^{\kappa} \left\langle \frac{J^+_{s_1}}{k_1^{s_1-1}}\frac{J_s^-}{k_2^{s-1}} \frac{J_s^-}{k_3^{s-1}} \right \rangle_{\textbf{h}} =
\widetilde{K}^{\kappa} \left\langle \frac{J^-_{s_1}}{k_1^{s_1-1}}\frac{J_s^-}{k_2^{s-1}} \frac{J_s^+}{k_3^{s-1}} \right \rangle_{\textbf{h}} =0.
\end{align}
The solutions for $F$, $G$ and $H$ are given (see appendix \ref{CWI}) by:
\begin{align}
\label{js1jsjsshsh}
F(k_1, k_2, k_3) = \frac{1}{E^{2s+s_1}},~~
G(k_1, k_2, k_3) = 0,~~
H(k_1, k_2, k_3) = 0.
\end{align}
 We will now summarise the results for the homogeneous solution.
\subsubsection*{Summary of Homogeneous contribution to {\bf{$\langle J_{s_1} J_s J_s \rangle$}}}
\begin{align}\label{s1ssJ}
&\langle J_{s_1}^- J_s^- J_s^-\rangle_{\textbf{h}} = \left(c_1+ i c'_1\right)\frac{k_1^{s_1-1}k_2^{s-1}k_3^{s-1}}{E^{2s+s_1}}\langle 12 \rangle^{s_1}\langle 23 \rangle^{2s-s_1}\langle 31 \rangle^{s_1}\\[5 pt]
&\langle J_{s_1}^+ J_s^- J_s^-\rangle_{\textbf{h}} = 0\\[5 pt]
&\langle J_{s_1}^- J_s^- J_s^+\rangle_{\textbf{h}} = 0
\end{align} while other components can be obtained by complex conjugating the result in \eqref{s1ssJ}.

\section{Conformal correlators in momentum space}
\label{shmomentumspaceresults}
In this section we present the results for higher spin $CFT_3$ correlators in momentum space. As explained in Section \ref{sectiondeg} a direct computation of parity-odd correlators in momentum space becomes complicated due to the large amount of degeneracy in 3d. Rather than solving the CWIs directly in momentum space, we convert our expressions for the correlators in spinor-helicity variables obtained in the previous section to momentum space. The simplest way to do this is to write down the ansatz for the correlator in momentum space and convert it to spinor-helicity variables. One can then compare it to the explicit results in spinor-helicity variables and solve for the form factors. For correlators involving higher spins, this procedure also becomes complicated, and in such cases we make use of transverse polarization tensors to represent the answers.
\subsection{Two point function}
Two-point functions of various conserved currents are as follows :
\begin{align}
\begin{split}
    \langle J^{\mu}(k)J^{\nu}(-k) \rangle_{\text{odd}} &= c'_J \epsilon^{\mu\nu k}\\[5 pt]
    \langle T^{\mu\nu}(k)T^{\rho\sigma}(-k) \rangle_{\text{odd}} &= c'_T \Delta^{\mu\nu\rho\sigma}(k)k^2
    \end{split}
    \begin{split}
         \langle J^{\mu}(k)J^{\nu}(-k) \rangle_{\text{even}} &= c_J \pi^{\mu\nu}(k)k\\[5 pt]
          \langle T^{\mu\nu}(k)T^{\rho\sigma}(-k) \rangle_{\text{even}} &= c_T\Pi^{\mu\nu\rho\sigma}(k)k^3
    \end{split}
\end{align}
where 
\begin{align}
    \Delta^{\mu \nu \rho \sigma}(k)&=\epsilon^{\mu \rho k} \pi^{\nu \sigma}(k)+\epsilon^{\mu \sigma k} \pi^{\nu \rho}(k)+\epsilon^{\nu \sigma k} \pi^{\mu \rho}(k)+\epsilon^{\nu \rho k} \pi^{\mu \sigma}(k)\\[5 pt]
    \Pi^{\mu\nu\rho\sigma}(k) &= \frac{1}{2}\left(\pi^{\mu\rho}(k)\pi^{\nu\sigma}(k)+\pi^{\mu\sigma}(k)\pi^{\nu\rho}(k)-\pi^{\mu\nu}(k)\pi^{\rho\sigma}(k)\right)\\[5 pt]
    \pi^{\mu\nu}(k) &= \delta^{\mu\nu}-\frac{k^{\mu}k^{\nu}}{k^2}
\end{align}
For arbitrary spin $s$, we have the following expression for the two-point function after contracting with polarization vectors :
\begin{align}\label{jsjs2pt}
\begin{split}
    \langle J_s(k)J_s(-k) \rangle_{\text{odd}} &= c'_s\epsilon^{z_1 z_2 k}\,(z_1 \cdot z_2)^{s-1}k^{2(s-1)}\\[5 pt]
        \langle J_s(k)J_s(-k) \rangle_{\text{even}} &= c_s (z_1 \cdot z_2)^s k^{2s-1}
    \end{split}
\end{align}
From \eqref{jsjs2pt}, it is clear the the parity-odd two-point function for a spin-$s$ current is a contact term as it is analytic in $k^2$.
\subsection{Three point function}
 The parity-even sector of momentum space 3-point correlators involving spin 1 and spin 2 conserved currents and scalar operators of arbitrary scaling dimensions was obtained by solving conformal Ward identities in  \cite{Bzowski:2013sza,Bzowski:2015pba,Bzowski:2017poo,Bzowski:2018fql}. The parity odd sector of $CFT_3$ was studied for specific correlators using momentum space Ward identities and weight-shifting operators in \cite{Jain:2021qcl}. In this section we convert the spinor-helicity expressions in the previous section to momentum space expressions and obtain parity-even as well as parity-odd three point correlators comprising generic spin $s$ conserved currents and scalar operators of arbitrary scaling dimensions. 
\subsubsection{$\langle J_s O_{\Delta} O_{\Delta} \rangle$}
In this section we give the momentum space expression for correlators of the form $\langle J_s O_{\Delta}O_{\Delta} \rangle$. As discussed earlier, the parity-odd part is zero. The parity-even part is straightforward to write down from the spinor-helicity expressions. For general spin-$s$, it is given by :
\begin{align}
    \langle J_s O_{\Delta} O_{\Delta} \rangle_{\text{even}} &= c_1 k_1^{2s-1}I_{\frac{1}{2}+s,\{\frac{1}{2}-s,\Delta-\frac{3}{2},\Delta-\frac{3}{2}\}}(k_2 \cdot z_1)^s
\end{align}
Let us now consider the correlator for some specific values of $s$ and $\Delta$.
\subsubsection*{$\langle J_s O_2 O_2 \rangle$}
For $\Delta=2$, we have
\begin{align}
    \langle J_s O_2 O_2 \rangle_{\text{even}} &= c_1 k_1^{2s-1}I_{\frac{1}{2}+s,\{\frac{1}{2}-s,\frac{1}{2},\frac{1}{2}\}}(k_2 \cdot z_1)^s
\end{align}
For $s=1$, $s=2$ and $s=3$, we have
\begin{align}\label{jso2o2momentum}
    \begin{split}
        \langle J O_2 O_2 \rangle_{\text{even}} &= c_1\frac{1}{(k_1+k_2+k_3)}(k_2 \cdot z_1)\\[5 pt]
        \langle T O_2 O_2 \rangle_{\text{even}} &= c_1\frac{2k_1+k_2+k_3}{(k_1+k_2+k_3)^2}(k_2 \cdot z_1)^2\\[5 pt]
        \langle J_3 O_2 O_2 \rangle_{\text{even}} &= c_1\frac{8k_1^2+9k_1(k_2+k_3)+3(k_2+k_3)^2}{(k_1+k_2+k_3)^3}(k_2 \cdot z_1)^3
    \end{split}
\end{align}
\subsubsection*{$\langle T O_3 O_3 \rangle$}
For $s=2$ and $\Delta=3$, we have
\begin{align}\label{TO3O3}
    \langle T O_3 O_3 \rangle_{\text{even}} &= c_1\frac{k_1^3+k_2^3+k_3^3+2(k_1^2+k_2k_3)(k_2+k_3)+2k_1(k_2^2+k_2k_3+k_3^2)}{(k_1+k_2+k_3)^2}(k_2 \cdot z_1)^2
\end{align}
Let us now consider the three-point correlator with two insertions of the spin-$s$ conserved current and a scalar operator of scaling dimension $\Delta$.
\subsubsection{$\langle J_sJ_sO_{\Delta} \rangle$}
\label{jsjsodeltamomentum}
In this section we determine the momentum space expression for correlators of the kind $\langle J_sJ_sO_{\Delta} \rangle$. We first discuss the $s=1$ and $s=2$ cases in detail. We then present the final result for the general spin case expressed in terms of a transverse polarization tensor.
\subsubsection*{$\langle JJO_{\Delta} \rangle$}

The correlator is purely transverse and the even part of it takes the following form in momentum space \cite{Bzowski:2018fql}:
%
\begin{align}
\label{jjodeltaevenmomentum}
\langle J^\mu(k_1)J^\nu(k_2)O_{\Delta}\rangle_{\text{even}}=A_1(k_1,k_2,k_3)\pi^\mu_{\alpha}(k_1)\pi^\nu_{\beta}(k_2)\left[k_2^{\alpha}k_3^{\beta}-\chi\delta^{\alpha\beta}\right]
\end{align}
where the form factor $A(k_1,k_2,k_3)$ is given by :
\begin{align}
\label{tripkjjo}
A_1(k_1,k_2,k_3)=I_{\frac 52,\{\frac 12,\frac 12,\Delta-\frac 32\}}
\end{align}
and
\begin{align}
\label{chidef}
\chi=\frac{1}{2}(k_1+k_2+k_3)(k_1+k_2-k_3)
\end{align}
As an example let us consider the case when the scaling dimension of the scalar operator is $\Delta=2$. In this case, after evaluating the integral \eqref{tripkjjo} we obtain the form factor in the ansatz \eqref{jjodeltaevenmomentum} to take the form :
%
\begin{align}
    A_1(k_1,k_2,k_3)=\frac{1}{(k_1+k_2+k_3)^2}
\end{align}
Let us now consider the parity-odd sector. In \cite{Jain:2021qcl} we computed $\langle JJO_{\Delta} \rangle_{\text{odd}}$ by imposing conformal invariance and obtained :
%
\begin{align}
\label{JJOansatz1}
\langle J^\mu(k_1)\,J^\nu(k_2)\,O(k_3)\rangle_{\text{odd}}=\pi^{\mu}_{\alpha}(k_1)\pi^{\nu}_{\beta}(k_2)\left[ A(k_1,k_2,k_3) \epsilon^{\alpha k_1 k_2} k_1^{\beta}+ B(k_1,k_2,k_3) \epsilon^{\beta k_1 k_2}k_2^{\alpha}\right]
\end{align}
where 
\begin{align}
    A(k_1,k_2,k_3)=\frac{k_2^2(I_1k_1^2+I_2k_1\cdot k_2)}{k_1^2k_2^2-(k_1\cdot k_2)^2}\cr
    B(k_1,k_2,k_3)=\frac{k_1^2(I_2k_2^2+I_1k_1\cdot k_2)}{k_1^2k_2^2-(k_1\cdot k_2)^2}
\end{align}
where $I_1$ and $I_2$ are the following two triple-$K$ integrals respectively :
\begin{align}
I_1&=c_1\,I_{\frac{5}{2}\{-\frac{1}{2},\frac{1}{2},\Delta-\frac 32\}} \\
I_2&=-c_1\,I_{\frac{5}{2}\{\frac{1}{2},-\frac{1}{2},\Delta-\frac 32\}} 
\end{align}
We further verified our results by computing the correlator for $\Delta=1,\ldots,5$ using weight-shifting operators and matching with the results obtained above.

Again, when $\Delta=2$, after evaluating the above triple-$K$ integrals we obtain the form factors in \eqref{JJOansatz1} to be :
\begin{align}\label{JJOff}
    A(k_1,k_2,k_3)&=-\frac{k_2}{(k_1+k_2+k_3)^2\left((k_1-k_2)^2-k_3^2\right)}\cr
     B(k_1,k_2,k_3)&=\frac{k_1}{(k_1+k_2+k_3)^2\left((k_1-k_2)^2-k_3^2\right)}
    \end{align}
Note that, as expected there is a total energy singularity when $E=k_1+k_2+k_3 \rightarrow 0$. It seems from the above expression that there is also an apparent collinear divergence when any two momentum vectors are proportional to each other. In this case, momentum conservation implies that all 3 momenta are along a line and it is easy to check that the $((k_1-k_2)^2 -k_3^2)$ factor in the denominator above vanishes. However, in this case the numerator of the full correlator also vanishes appropriately,
leaving the correlator finite. Hence as expected the correlator has no singularities other than the $E\rightarrow 0$ singularity \footnote{There is also an alternative form of this correlator (see eq. 5.16 of \cite{Jain:2021wyn}) using which it is easy to see that there are no collinear divergences.}.

\subsubsection*{$\langle TTO_{\Delta}\rangle$}
The transverse and traceless part of the correlator in three dimensions is given by \cite{Bzowski:2018fql}:
\begin{align}
 \langle T_{\mu_1 \nu_1}T_{\mu_2 \nu_2}O_{\Delta} \rangle_{\text{even}}=-2k_1^2k_2^2A_1\Pi_{\mu_1\nu_1\alpha\beta_1}(k_1)\Pi_{\mu_2\nu_2\alpha\beta_2}(k_2)(k_2^{\beta_1}k_3^{\beta_2}-\chi\,\delta^{\beta_1\beta_2})   
\end{align}
where
\begin{align}
\label{parityeventtoformfactor}
    A_1(k_1,k_2,k_3)=c_1\,I_{\frac 92,\{\frac 12,\frac 12,\Delta-\frac{3}{2}\}}
\end{align}
and $\chi$ is defined in \eqref{chidef}. For the case when $\Delta=2$ one obtains the following :
\begin{align}
A_1(k_1,k_2,k_3)=\frac{1}{(k_1+k_2+k_3)^4}
\end{align}
Let us now consider the parity odd contribution to the correlator. In \cite{Jain:2021wyn} we obtained the momentum space expressions for $\langle TTO_1 \rangle_{\text{odd}}$ and $\langle TTO_2 \rangle_{\text{odd}}$ using spin-raising and weight-shifting operators. We will now use the expressions we obtained in spinor-helicity variables  for $\langle TTO_{\Delta}\rangle$, for a generic $\Delta$, to obtain a momentum space expression for the same.

We start with the following ansatz in momentum space :
\begin{align}
 \langle T_{\mu_1 \nu_1}T_{\mu_2 \nu_2}O_{\Delta} \rangle_{\text{odd}} = \Pi_{\mu_1 \nu_1}^{\alpha_1 \beta_1}(k_1) \Pi_{\mu_2 \nu_2}^{\alpha_2 \beta_2}\left(A_1\epsilon^{\mu_1\mu_2 k_1}\delta^{\nu_1 \nu_2}+A_2 \epsilon^{\mu_1 \mu_2 k_2}\delta^{\nu_1 \nu_2}\right)
\end{align}
Symmetry considerations tell us that :
\begin{align}
    A_2= -A_1(k_1 \leftrightarrow k_2)
\end{align}
Dotting with transverse, null polarization vectors, we get
\begin{align}
 \langle TTO_{\Delta} \rangle_{\text{odd}} = A_1 \epsilon^{k_1 z_1 z_2} z_1 \cdot z_2-A_1(k_1 \leftrightarrow k_2)\epsilon^{k_2 z_1 z_2}z_1 \cdot z_2
\end{align}
Converting into spinor-helicity variables we get :
\begin{align}\label{ttodeleqns}
\begin{split}
\langle T^- T^+ O_{\Delta} \rangle _{\text{odd}}&=\frac{A_1 k_1 -A_1(k_1 \leftrightarrow k_2) k_2 }{4 k_1^2 k_2^2}\langle 1 \bar{2} \rangle^4\\[5 pt]
\langle T^- T^- O_{\Delta} \rangle_{\text{odd}} &= \frac{A_1 k_1 + A_1(k_1 \leftrightarrow k_2) k_2 }{4 k_1^2 k_2^2}\langle 1 2 \rangle^4
\end{split}
\end{align}
Comparing \eqref{ttodeleqns} and \eqref{ttodeltasph}, we get the following for the form factor :
\begin{align}
 A_1 = 2c_1 k_1^2 k_2^3I_{\frac{9}{2}\{\frac{1}{2},\frac{1}{2},\Delta-\frac{3}{2}\}}
\end{align}
which matches the form factor \eqref{parityeventtoformfactor} that appears in the even part of the same correlator.
Let us now come to the case where the spinning operator in the correlator is a generic spin $s$ conserved current.
\subsubsection*{$\langle J_sJ_sO_{\Delta}\rangle$}
As mentioned in the introduction to this section, for correlators involving higher spin operators, it is convenient to introduce transverse polarization tensors. It is straightforward to write down the momentum space expression for these correlators from their expression in spinor-helicity variables. This can be done upon observing that 
\begin{align}
    \begin{split}
        \left[k_2\,\epsilon^{ k_1 z_1 z_2}- k_1\,\epsilon^{k_2 z_1  z_2}\right] \mapsto i\langle 12 \rangle^2\\[5 pt]
        \left[(\vec{z}_1\cdot \vec{k}_2)( \vec{z}_2\cdot \vec{k}_1) +\frac{1}{2}E (E-2k_3)\vec{z}_1\cdot \vec{z}_2 \right] \mapsto \langle 12 \rangle^2
    \end{split}
\end{align}
We then have 
\begin{align}
\langle J_sJ_sO_{\Delta}\rangle_{\text{even}}&= (k_1 k_2)^{s-1}I_{\frac{1}{2}+2s\{\frac{1}{2},\frac{1}{2},\Delta-\frac{3}{2}\}}\left[2(\vec{z}_1\cdot \vec{k}_2)( \vec{z}_2\cdot \vec{k}_1) +E (E-2k_3)\vec{z}_1\cdot \vec{z}_2 \right]^{s}\\[5 pt]
\langle J_sJ_sO_{\Delta}\rangle_{\text{odd}}&=(k_1 k_2)^{s-1}I_{\frac{1}{2}+2s\{\frac{1}{2},\frac{1}{2},\Delta-\frac{3}{2}\}}\left[k_2\,\epsilon^{ k_1 z_1 z_2}- k_1\,\epsilon^{k_2 z_1  z_2}\right]\cr
&\hspace{.5cm}\times \left[2(\vec{z}_1\cdot \vec{k}_2)( \vec{z}_2\cdot \vec{k}_1) +E (E-2k_3)\vec{z}_1\cdot \vec{z}_2 \right]^{s-1}
\end{align}
Let us now look at this correlator for specific values of the scaling dimension of the scalar operator.
\subsubsection*{$\langle J_sJ_sO_{1}\rangle$}
For $\Delta=1$, the correlator is given by :
\begin{align}
\langle J_sJ_sO_{1}\rangle_{\text{odd}}&=(k_1 k_2)^{s-1}\frac{1}{k_3 E^{2s}}\left[k_2\,\epsilon^{ k_1 z_1 z_2}- k_1\,\epsilon^{k_2 z_1  z_2}\right]\cr
&\hspace{.5cm}\times \left[2(\vec{z}_1\cdot \vec{k}_2)( \vec{z}_2\cdot \vec{k}_1) +E (E-2k_3)\vec{z}_1\cdot \vec{z}_2 \right]^{s-1}
\end{align}
\subsubsection*{$\langle J_sJ_sO_{2}\rangle$}
For $\Delta=2$, the correlator is given by :
\begin{align}
\langle J_sJ_sO_{2}\rangle_{\text{odd}}&=(k_1 k_2)^{s-1}\frac{1}{E^{2s}}\left[k_2\,\epsilon^{ k_1 z_1 z_2}- k_1\,\epsilon^{k_2 z_1  z_2}\right]\cr
&\hspace{.5cm}\times \left[2(\vec{z}_1\cdot \vec{k}_2)( \vec{z}_2\cdot \vec{k}_1) +E (E-2k_3)\vec{z}_1\cdot \vec{z}_2 \right]^{s-1}
\end{align}
\subsubsection*{$\langle J_sJ_sO_{3}\rangle$}
For $\Delta=3$, the correlator is given by :
\begin{align}
\langle J_sJ_sO_{3}\rangle_{\text{odd}}&=(k_1 k_2)^{s-1}\frac{(E+(2s-1)k_3)}{E^{2s}}\left[k_2\,\epsilon^{ k_1 z_1 z_2}- k_1\,\epsilon^{k_2 z_1  z_2}\right]\cr
&\hspace{.5cm}\times \left[2(\vec{z}_1\cdot \vec{k}_2)( \vec{z}_2\cdot \vec{k}_1) +E (E-2k_3)\vec{z}_1\cdot \vec{z}_2 \right]^{s-1}
\end{align}
\subsubsection{$\langle J_sJ_sJ_s\rangle$}
\label{jsjsjsmomentumsp}
We will now determine the correlator $\langle J_sJ_s J_s\rangle$ in momentum space. We first discuss the $s=1$ and $s=2$ cases in detail. For the case of general spin, we present only the final result expressed in terms of transverse polarization tensor. For this, we restrict our attention to the homogeneous part. For the parity-odd part of the correlator, the non-homogeneous contribution is always a contact term.

\subsubsection*{$\langle JJJ\rangle$}
Unlike $\langle J_sJ_sO_{\Delta}\rangle$ the correlator $\langle JJJ\rangle$ is not purely transverse and it has a local term given by \cite{Bzowski:2017poo}:
\begin{align}
\label{jjjlocal}
\begin{split}
\langle J^{\mu a} J^{\nu b} J^{\rho c} \rangle_{\text{local}} &= \bigg[\frac{k_1^{\mu}}{k_1^2}\left(f^{adc}\langle J^{\rho d}(k_2)J^{\nu b}(-k_2) \rangle-f^{abd}\langle J^{\nu d}(k_3)J^{\rho c}(-k_3)\rangle\right)\bigg.\\[5 pt]
&\hspace{.5cm}\bigg.+\left((k_1, \mu) \leftrightarrow (k_2, \nu)\right)+\left((k_1, \mu) \leftrightarrow (k_3, \rho)\right)\bigg]+\bigg[\left(\frac{k_1^{\mu}\,k_2^{\nu}}{k_1^2 k_2^2}f^{abd}k_{2\alpha}\langle J^{\alpha d}(k_3)J^{\rho c}(-k_3)\rangle\right)\bigg.\\[5 pt]
&\hspace{.5cm}\bigg.+\left((k_1, \mu) \leftrightarrow (k_3, \rho)\right)+\left((k_2, \nu) \leftrightarrow (k_3, \rho)\right)\bigg]
\end{split}
\end{align}
The transverse part of the even part of the correlator denoted by $\langle j^{\mu_1a_1}j^{\mu_2a_2}j^{\mu_3a_3}\rangle_{\text{even}}$ was computed in \cite{Bzowski:2017poo} :
\begin{align}
  \langle j^{\mu_1a_1}j^{\mu_2a_2}j^{\mu_3a_3}\rangle_{\text{even}}=\pi^{\mu_1}_{\alpha_1}(k_1)\pi^{\mu_2}_{\alpha_2}(k_2)\pi^{\mu_3}_{\alpha_3}(k_3)[&A_1k_2^{\alpha_1}k_3^{\alpha_2}k_1^{\alpha_3}+A_2\delta^{\alpha_1\alpha_2}k_1^{\alpha_3}\notag\\
 &\hspace{-2.5cm} +A_2(k_3,k_1,k_2)\delta^{\alpha_1\alpha_3}k_3^{\alpha_2}
  +A_2(k_2,k_3,k_1)\delta^{\alpha_2\alpha_3}k_2^{\alpha_1}] 
\end{align}
where the form factors are given by :
\begin{align}
    A_1&=\frac{2c_1}{(k_1+k_2+k_3)^3}\cr
    A_2&=c_1\frac{k_3}{(k_1+k_2+k_3)^2}-\frac{2c_{J}}{(k_1+k_2+k_3)}
\end{align}
Here and in the following we suppress the color indices for brevity. After dotting with transverse, null polarization vectors, the correlator can be separated into homogeneous and non-homogeneous parts as follows :
\begin{align}
    \begin{split}
        \langle JJJ \rangle_{\text{even}, \bf{h}} &= \frac{c_1}{(k_1+k_2+k_3)^2}\left[\frac{2(k_2 \cdot z_1)(k_3 \cdot z_2)(k_1 \cdot z_3)}{(k_1+k_2+k_3)}+\left(k_3(z_1 \cdot z_2)(k_1 \cdot z_3)+\text{cyclic perm.}\right)\right]\\[5 pt]
        \langle JJJ \rangle_{\text{even}, \bf{nh}} &= -\frac{2c_J}{(k_1+k_2+k_3)}\left((z_1 \cdot z_2)(k_1 \cdot z_3) +\text{cyclic perm.}\right)
        \end{split}
\end{align}
In \cite{Jain:2021wyn} we computed the odd part of $\langle JJJ\rangle$ using the action of spin-raising and weight-shifting operators on a scalar seed correlator. 
The correlator is given by the sum of its local terms \eqref{jjjlocal} and transverse parts. The latter is given by :
\begin{equation}
 \langle j^{\mu a} j^{\nu b} j^{\rho c} \rangle_{\text{odd}}=\pi^{\mu}_{\alpha}(k_1)\pi^{\nu}_{\beta}(k_2)\pi^{\rho}_{\gamma}(k_3)X^{\alpha\beta\gamma}
\end{equation}
where
\begin{align}\label{jjjansatz}
\begin{split}
X^{\alpha\beta\gamma}&=A_1\epsilon^{k_1 k_2 \alpha}k_1^{\gamma}k_3^{\beta}+A_2\epsilon^{k_1 k_2 \alpha}\delta^{\beta\gamma}+A_3\epsilon^{k_1 \alpha \beta}k_1^{\gamma}+A_4\epsilon^{k_1 \alpha \gamma}k_3^{\beta}+\text{cyclic perm.}
\end{split}
\end{align}
where
\begin{align}
A_1&=-\frac{2}{k_1(k_1+k_2+k_3)^3},\quad\quad
A_2=-\frac{1}{(k_1+k_2+k_3)^2}\nonumber\\[5 pt]
A_3&=\frac{k_1+k_2+2\,k_3}{k_1(k_1+k_2+k_3)^2},\quad\quad\,\,\,\,
A_4=\frac{k_1+2\,k_2+k_3}{k_1(k_1+k_2+k_3)^2}
\end{align}
After dotting with transverse polarization vectors, the correlator can be rewritten in the following form using Schouten identities :
  \begin{align}\label{JJJhn1}
\langle JJJ\rangle_{\text{odd,\bf{h}}}&=\frac{c'_{1}}{E^3}\left[\left\{(\vec{k}_1 \cdot \vec{z}_3)\left(\epsilon^{k_3 z_1 z_2}k_1-\epsilon^{k_1 z_1 z_2}k_3\right)+(\vec{k}_3 \cdot \vec{z}_2)\left(\epsilon^{k_1 z_1 z_3}k_2-\epsilon^{k_2 z_1 z_3}k_1\right)\right.\right.\nonumber\\[5 pt]
&\hspace{1.5cm}\left.\left.-(\vec{z}_2 \cdot \vec{z}_3)\epsilon^{k_1 k_2 z_1}E+\frac{k_1}{2} \epsilon^{z_1 z_2 z_3}E(E-2k_1)\right\}+\text{cyclic perm}\right]\nonumber\\[5 pt]
\langle JJJ\rangle_{\text{odd,\bf{nh}}}&=c'_{J} \epsilon^{z_1z_2z_3}
\end{align}
We see that the non-homogeneous part of the parity-odd part of the correlator is a contact term. This term can be explained from the $dS_4$ perspective by considering the three-point tree-level amplitude arising from the interaction term $F\widetilde F$.

In the rest of this section we obtain the momentum space expressions for the correlators $\langle TJJ \rangle$ and $\langle TTT \rangle$.  As described in Section \ref{sectiondeg}, a direct computation of these correlators by solving the conformal Ward identities in momentum space is quite difficult. 

Our strategy to circumventing the difficulties follows our approach in Section \ref{jsjsodeltamomentum} to obtain $\langle TTO_\Delta\rangle$. We start with an ansatz for the correlator in momentum space and then convert it into spinor-helicity variables. We then compare it with the explicit results obtained in the previous section. This gives us algebraic equations involving the momentum space form-factors which can then be solved. 

\subsubsection*{$\langle TTT \rangle$}
The momentum space expression for the even part of the correlator $\langle TTT \rangle$ was obtained in \cite{Bzowski:2017poo} and it was shown to have two structures. We will now obtain the momentum space expression for the odd part of the correlator. 

In \eqref{tttoddsphresult} we obtained the following result for the parity odd part of the correlator $\langle TTT \rangle$ in spinor-helicity variables :  
\begin{align}\label{tttsph1}
\begin{split}
\langle T^-T^-T^- \rangle_{odd} &= \left(c'_1\frac{c_{123}}{E^6}+c'_T\frac{E^3-Eb_{123}-c_{123}}{c_{123}^2}\right)\langle 12 \rangle^2 \langle 23 \rangle^2 \langle 31 \rangle^2\\[5 pt]
\langle T^-T^-T^+ \rangle_{odd} &= c'_T\frac{(E-2k_3)^3-(E-2k_3)(b_{123}-2k_3 a_{12})+c_{123}}{c_{123}^2}\langle 12 \rangle^2 \langle 2\bar{3} \rangle^2 \langle \bar{3}1 \rangle^2
\end{split}
\end{align}
Let us consider the following ansatz for the transverse part of the correlator : 
\begin{align}\label{tttoda}
\begin{split}
\langle T^{\mu_1\nu_1}T^{\mu_2 \nu_2}T^{\mu_3\nu_3} \rangle_{\text{odd}}&=\Pi^{\mu_1\nu_1}_{\alpha_1\beta_1}(k_1)\Pi^{\mu_2\nu_2}_{\alpha_2\beta_2}(k_2)\Pi^{\mu_3\nu_3}_{\alpha_3\beta_3}(k_3)\bigg(A_1 \epsilon^{k_1 k_2 \alpha_1}\epsilon^{k_1 k_2 \alpha_2}\epsilon^{k_1 k_2 \alpha_3}k_1^{\beta_3}k_2^{\beta_1}k_3^{\beta_2}\bigg.\\[5 pt]
&\bigg.+A_2 \epsilon^{k_1 k_2 \alpha_3}k_1^{\beta_3}k_3^{\alpha_2}k_3^{\beta_2}k_2^{\alpha_1}k_2^{\beta_1}+A_2 (k_2 \leftrightarrow k_3) \epsilon^{k_2 k_3 \alpha_2}k_3^{\beta_2}k_1^{\alpha_3}k_1^{\beta_3}k_2^{\alpha_1}k_2^{\beta_1}\bigg.\\[5 pt]
&\bigg.+A_2 (k_1 \leftrightarrow k_3) \epsilon^{k_1 k_2 \alpha_1}k_2^{\beta_1}k_3^{\alpha_2}k_3^{\beta_2}k_1^{\alpha_3}k_1^{\beta_3}\bigg)
\end{split}
\end{align}
One could have started with a more general ansatz with many more tensor structures than exhibited by \eqref{tttoda}. However, it turns out that there are several Schouten identities that relate those tensor structures and upon using them one ends up with the minimal (and complete) ansatz  in \eqref{tttoda} \footnote{Schouten identities that turn out to be useful are given in Appendix C of \cite{Jain:2021wyn}.}.

Contracting with null and transverse polarization vectors, we get
\begin{align}\label{TTTmomspace}
\begin{split}
\langle TTT \rangle_{\text{odd}}&= A_1 \epsilon^{k_3 k_1 z_1}\epsilon^{k_1 k_2 z_2}\epsilon^{k_2 k_3 z_3}(k_2 \cdot z_1)(k_3 \cdot z_2)(k_1 \cdot z_3)+ A_2 \epsilon^{k_2 k_3 z_3}(k_1 \cdot z_3)(k_3 \cdot z_2)^2(k_2 \cdot z_1)^2\\[5 pt]
&+ A_2 (k_2 \leftrightarrow k_3)\epsilon^{k_1 k_2 z_2}(k_1 \cdot z_3)^2(k_3 \cdot z_2)(k_2 \cdot z_1)^2+ A_2 (k_1 \leftrightarrow k_3)\epsilon^{k_3 k_1 z_1}(k_1 \cdot z_3)^2(k_3 \cdot z_2)^2(k_2 \cdot z_1)
\end{split}
\end{align}
Converting this into spinor-helicity variables, we get
\begin{align}\label{tttsph2}
\begin{split}
\langle T^-T^-T^+ \rangle &= \langle 12 \rangle^2 \langle 2\bar{3} \rangle^2 \langle \bar{3}1 \rangle^2\frac{J^4 \left(A_1 c_{123}+A_2 k_3-A_2(k_2 \leftrightarrow k_3)k_2-A_2(k_1 \leftrightarrow k_3)k_1\right)}{(E-2k_3)^2 c_{123}^2}\\[5 pt]
\langle T^-T^-T^- \rangle &= -\langle 12 \rangle^2 \langle 23 \rangle^2 \langle 31 \rangle^2\frac{J^4 \left(A_1 c_{123}+A_2 k_3+A_2(k_2 \leftrightarrow k_3)k_2+A_2(k_1 \leftrightarrow k_3)k_1\right)}{E^2 c_{123}^2}
\end{split}
\end{align}
 Comparing \eqref{tttsph1} and \eqref{tttsph2} and solving for the momentum space form factors we get : 
\begin{align}\label{sola1a2}
A_1 &=c'_1\frac{c_{123}^2}{2J^4 E^4}-c'_T \frac{12(k_1^2+k_2^2+k_3^2)}{J^4}\\[5 pt]
A_2 &= -c'_1\frac{b_{12}c^2_{123}}{2J^4 E^4}+c'_T\frac{\left(k_3^4+7k_3^2(k_1^2+k_2^2)+4(k_1^4+4k_1^2 k_2^2+k_2^4)\right)}{J^4}
\end{align}
Naively it might look like there are two contributions to the parity-odd correlation function. However, as will be shown below the term proportional to $c'_T$ is a contact term. 
\subsection*{Contact term in parity-odd $\langle TTT \rangle$} 
The fact that the term proportional to $c_T$ is a contact term can be seen more explicitly by switching to a basis where the factor of $J^4$ in the denominator disappears.
One such basis is given by\footnote{This choice of basis is not unique. One can find several other bases in which un-physical poles do not appear. To do this, we start with the most general ansatz for $\langle TTT \rangle$ containing all possible tensor structures. We then solve for the form-factors in this most general ansatz by relating it to the known answer for the correlator. Not all the form-factors in the ansatz are fixed this way and those that are not fixed can be set to zero. Which of them are set to zero is a choice made while solving and different choices lead to different bases.}:
\begin{align}\label{tttnewbasis}
\begin{split}
\langle TTT \rangle &= \bigg[B_1 \epsilon^{k_1 z_1 z_2} (z_1 \cdot z_2) (k_1 \cdot z_3)^2 - B_1 (k_1 \leftrightarrow k_2)\epsilon^{k_2 z_1 z_2} (z_1 \cdot z_2) (k_1 \cdot z_3)^2\bigg.\\[5 pt]
&\bigg.+B_2 \epsilon^{k_1 z_1 z_2} (z_1 \cdot z_3) (z_2 \cdot z_3)-B_2(k_1 \leftrightarrow k_2)\epsilon^{k_2 z_1 z_2} (z_1 \cdot z_3) (z_2 \cdot z_3)\bigg] + \text{cyclic perm.}
\end{split}
\end{align}
The ansatz in \eqref{tttnewbasis} is related to \eqref{TTTmomspace} by Schouten identities. Converting \eqref{tttnewbasis} into spinor-helicity variables and comparing with \eqref{tttsph2}, we can solve for $B_1$ and $B_2$. We get the solutions for the terms proportional to $c'_T$ to be :
\begin{align}
\begin{split}
B_1=c'_T \frac{1}{24},
\end{split}
\begin{split}
B_2 = c'_T \frac{1}{12}\left(k_1^2 + \frac{7}{4} k_2^2+ \frac{7}{4} k_3^2\right)
\end{split}
\end{align}
The fact that $B_1$ is a constant and $B_2$ is dependent on $k^2$ implies that if we evaluate $\langle TTT \rangle$ in the basis \eqref{tttnewbasis} and convert it to position space, we will get delta functions or derivatives on delta functions which are nothing but contact terms.
Since the odd non-homogeneous part is a contact term, the full $\langle TTT \rangle$ correlator has only 3 non-trivial contributions, 2 parity-even and 1  parity-odd conformally invariant structures. This agrees with the analysis of \cite{Maldacena:2011jn}. 

From the $dS_4$ perspective this contribution to the correlator can be understood to arise from the $W\widetilde W$ interaction. The $W\widetilde W$  interaction also reproduces the  parity-odd two-point function of the stress tensor. 

\subsubsection*{$\langle J_s J_s J_s \rangle$}
For general spin $s$ it is easy to write down the homogeneous part of $\langle J_s J_s J_s \rangle$ in momentum space using the transverse polarization.
The answer is given by :
\begin{align}
   \langle J_sJ_sJ_s\rangle_{\text{even},\bf{h}}
&=(k_1 k_2 k_3)^{s-1}\left[\frac{1}{E^3} \Big\{2\,(\vec{z}_1\cdot \vec{k}_2) \, (\vec{z}_2\cdot \vec{k}_3) \, (\vec{z}_3\cdot \vec{k}_1)+E \{k_3\, (\vec{z}_1\cdot \vec{z}_2) \, (\vec{z}_3\cdot \vec{k}_1)+ \text{cyclic}\}\Big\}\right]^s\cr
\langle J_sJ_sJ_s\rangle_{\text{odd},\bf{h}}
&=(k_1 k_2 k_3)^{s-1}\frac{1}{E^3}\left[\left\{(\vec{k}_1 \cdot \vec{z}_3)\left(\epsilon^{k_3 z_1 z_2}k_1-\epsilon^{k_1 z_1 z_2}k_3\right)+(\vec{k}_3 \cdot \vec{z}_2)\left(\epsilon^{k_1 z_1 z_3}k_2-\epsilon^{k_2 z_1 z_3}k_1\right)\right.\right.\nonumber\\[5 pt]
&\hspace{1.5cm}\left.\left.-(\vec{z}_2 \cdot \vec{z}_3)\epsilon^{k_1 k_2 z_1}E+\frac{k_1}{2} \epsilon^{z_1 z_2 z_3}E(E-2k_1)\right\}+\text{cyclic perm}\right]\cr
&\hspace{.5cm}\times\left[\frac{1}{E^3} \Big\{2\,(\vec{z}_1\cdot \vec{k}_2) \, (\vec{z}_2\cdot \vec{k}_3) \, (\vec{z}_3\cdot \vec{k}_1)+E \{k_3\, (\vec{z}_1\cdot \vec{z}_2) \, (\vec{z}_3\cdot \vec{k}_1)+ \text{cyclic}\}\Big\}\right]^{s-1}
\end{align}
The parity-odd contribution to the non-homogeneous piece is just a contact term. 

\subsubsection{$\langle J_{2s}J_sJ_s \rangle$}
We will now look at correlators of the form $\langle J_{2s}J_sJ_s \rangle$. We focus the discussion on the $\langle TJJ \rangle$ correlator and also give the results for the $\langle J_4 TT \rangle$ correlator.
\subsubsection*{$\langle TJJ \rangle$}
We saw in \eqref{tjjoddwi} and \eqref{tjjoddwifnl} that the odd part of the correlator $\langle TJJ \rangle$ satisfies trivial transverse Ward identities \footnote{As a result the non-homogeneous part of this parity-odd correlator is zero. One can understand this from the $dS_4$ perspective by the following argument. The only interaction term that could possibly have contributed to this correlator is $F\widetilde F$. However since this term is independent of the metric the contribution from it to $\langle TJJ\rangle$ is zero. In fact there is no interaction term that one can have from the gravity side that contributes to the non-homogeneous parity-odd part of $\langle TJJ\rangle$.}.
We also note that in three-dimensions the trace Ward identity obeyed by this correlator is trivial. Taking these into account we write down the following ansatz for the correlator in momentum space :
\begin{align}
\langle T^{\mu_1\nu_1}(k_1)J^{\mu_2}(k_2)J^{\mu_3}(k_3)\rangle_{\text{odd}}&=\Pi^{\mu_1\nu_1}_{\alpha_1\beta_1}(k_1)\pi^{\mu_2}_{\alpha_2}(k_2)\pi^{\mu_3}_{\alpha_3}(k_3)\left(A_1 k_2^{\alpha_1}k_3^{\alpha_2}\epsilon^{\beta_1\alpha_3k_1}\right.\nonumber\\[4 pt]
&\hspace{2.5cm}\left.+A_2k_2^{\alpha_1}k_3^{\alpha_2}\epsilon^{\beta_1\alpha_3 k_3}
+A_3\delta^{\alpha_1\alpha_2}\epsilon^{\beta_1\alpha_3 k_1}+A_4\delta^{\alpha_1\alpha_2}\epsilon^{\beta_1\alpha_3 k_3}\right)
\end{align}
Let us now contract this with polarization vectors and this gives :
\begin{align}\label{jjta1a2}
\langle T J J\rangle_{\text{odd}}=A_{1}(k_2 \cdot z_1)(k_3 \cdot z_2)\epsilon^{k_1 z_1 z_3}+&A_2 (k_2 \cdot z_1)(k_3 \cdot z_2)\epsilon^{k_3 z_1 z_3}\nonumber\\[4 pt]
&+A_3 (z_1 \cdot z_2)\epsilon^{k_1 z_1 z_3}+A_4 (z_1 \cdot z_2)\epsilon^{k_3 z_1 z_3}
\end{align}
We now convert this expression into spinor-helicity variables to obtain :
\begin{align}\label{tjjsph2}
\begin{split}
\left\langle T^{-} J^{-} J^{-}\right\rangle_{\text{odd}}&=\frac{\langle 12\rangle^{2}\langle 13\rangle^{2}}{8 k_{1}^{2} k_{2} k_{3}}\left(2A_3 k_1-2A_4 k_3+(k_1-k_2-k_3)(k_1-k_2+k_3)(A_1k_1-A_2k_3)\right) \\[5 pt]
\left\langle T^{+} J^{-} J^{-}\right\rangle_{\text{odd}}&=\frac{\langle\overline{1} 2\rangle^{2}\langle\overline{1} 3\rangle^{2}}{8 k_{1}^{2} k_{2} k_{3}}\left(2A_3 k_1+2A_4 k_3+(k_1+k_2+k_3)(k_1+k_2-k_3)(A_1k_1+A_2k_3)\right)  \\[5 pt]
\left\langle T^{-} J^{-} J^{+}\right\rangle_{\text{odd}}&=\frac{\langle 12\rangle^{2}\langle 1 \overline{3}\rangle^{2}}{8 k_{1}^{2} k_{2} k_{3}}\left(2A_3 k_1+2A_4 k_3+(k_1-k_2-k_3)(k_1-k_2+k_3)(A_1k_1+A_2k_3)\right) \\[5 pt]
\left\langle T^{-} J^{+} J^{-}\right\rangle_{\text{odd}}&=\frac{\langle 1\overline{2}\rangle^{2}\langle 1 3\rangle^{2}}{8 k_{1}^{2} k_{2} k_{3}}\left(2A_3 k_1-2A_4 k_3+(k_1+k_2-k_3)(k_1+k_2+k_3)(A_1k_1-A_2k_3)\right) 
\end{split}
\end{align}
We obtained the following explicit results for these correlators in \eqref{tjjsh} :
\begin{align}\label{tjjsph1}
\begin{split}
\langle T^-J^-J^- \rangle_{\text{odd}} &= c'_1 \frac{k_1}{E^4}\langle 12 \rangle^2 \langle 13 \rangle^2\\
\langle T^-J^-J^+ \rangle_{\text{odd}}  &= 0
\end{split}
\begin{split}
\langle T^+J^-J^- \rangle_{\text{odd}} &= 0\\
\langle T^-J^+J^- \rangle_{\text{odd}} &=0
\end{split}
\end{align}
%
Comparing \eqref{tjjsph1} and \eqref{tjjsph2}, we get the following solutions for the form factors :
\begin{align}\label{a1a2a3}
\begin{split}
A_1 &= -c'_1 \frac{k_1 k_3}{E^4}\\
A_2 &=c'_1\frac{k_1^2}{E^4}
\end{split}
\begin{split}
A_3 &= c'_1\frac{k_1 k_3(k_1+k_2-k_3)}{2E^3}\\
A_4 &=- c'_1 \frac{k_1^2(k_1+k_2-k_3)}{2E^3}
\end{split}
\end{align}
Plugging back the solution \eqref{a1a2a3} in \eqref{jjta1a2} we obtain :
    \begin{align}\label{jjta1a3}
       \langle TJJ \rangle_{\text{odd}} = \frac{k_1}{4E^4} \left(-2(k_3 \cdot z_2)(k_2 \cdot z_1)+E(E-2k_3)(z_1 \cdot z_2)\right)\left(k_1 \epsilon^{z_1 z_3 k_3}-k_3\epsilon^{k_1 z_1 z_3}\right)
    \end{align}
    which matches the result in \cite{Jain:2021qcl} obtained by computing a tree level $dS_4$ amplitude.
   \par 
The solution \eqref{jjta1a3} is not manifestly symmetric under a $(2 \leftrightarrow 3)$ exchange in this basis. However, we can use Schouten identities to convert the ansatz \eqref{jjta1a2} to the following form where it is manifestly symmetric under a $(2 \leftrightarrow 3)$ exchange :
\begin{align}
\label{jjta1a4} 
\langle T J J\rangle_{\text{odd}}= B_1 \epsilon^{k_1 k_2 z_1}(k_1 \cdot z_1)&(k_1 \cdot z_3)(k_3 \cdot z_2)+B_2 \epsilon^{k_1 k_2 z_1}(k_2 \cdot z_1)^2(k_1 \cdot z_3)\notag\\
&\hspace{-1cm}+B_3 \epsilon^{k_1 k_2 z_3}(k_2 \cdot z_1)^2(k_3 \cdot z_2)+B_4\epsilon^{k_1 k_2 z_1}(k_2 \cdot z_1)(z_2 \cdot z_3)
\end{align}
 The relation between the form-factors in the two bases \eqref{jjta1a2} and \eqref{jjta1a4}  is given by :
\begin{align}
    \begin{split}
        B_1 &= \frac{16}{J^4}\bigg(4A_4((k_1^2-k_2^2)^2+2(k_1^2+k_2^2)k_3^2-3k_3^4)-2(A_1(k_1^2-k_2^2)+(A_1-2A_2)k_3^2)\bigg.\nonumber\\[5 pt]
        \hspace{.5cm}&\bigg.\times((k_1^2-k_2^2)^2-2(k_1^2+k_2^2)k_3^2+k_3^4)+4A_3(-3k_1^4+(k_2^2-k_3^2)^2+2k_1^2(k_2^2+k_3^2)\bigg)\\[5 pt]
        B_2 &= -\frac{128}{J^4}\bigg(k_1^2(-2A_4 k_3^2+A_3(k_1^2+k_3^2-k_2^2)\bigg)\\[5 pt]
        B_3 &= -\frac{16}{J^4}\bigg(-8A_3 k_1^2(k_1^2+k_2^2-k_3^2)+4A_4(k_1^4-2(k_2^2-k_3^2)^2)-8(2A_1 k_1^2-A_2(k_1^2-k_2^2+k_3^2))J^2\bigg)\\[5 pt]
        B_4 &= \frac{8}{J^2}\bigg(-2A_3 k_1^2+A_4(k_1^2-k_2^2+k_3^2)\bigg)
    \end{split}
    \end{align}
    
    For the case $s_1=4$ and $s=2$, the momentum space expression that we get after converting the answer in spinor-helicity variables given in Section \ref{SectionJs1JsJs} is the following :
\begin{align}
    \langle J_4 T T \rangle_{\text{odd}}= &c'_{1}\frac{k_1^3 k_2 k_3}{E^8} \bigg[\left(2 (k_2 \cdot z_1)(k_3 \cdot z_2)-\left(z_{1} \cdot z_{2}\right)(E-2k_3)E\right) \left( k_1\epsilon^{z_{1} z_{3} k_{3}}-  k_3  \epsilon^{z_{1} z_{3} k_{1}}\right)\bigg] \nonumber\\[5 pt]
     &\times\bigg[\left((k_3 \cdot z_2)(k_2 \cdot z_1)-\frac{1}{2}E(E-2k_3)(z_1 \cdot z_2)\right)
   \left((k_1 \cdot z_3)(k_2 \cdot z_1)-\frac{1}{2}E(E-2k_2)(z_1 \cdot z_3)\right)\bigg]
\end{align}
The parity-odd contribution to the non-homogeneous part is again a contact term.


\section{Renormalisation}\label{renor11}
In Sections \ref{sectionSHresults} and \ref{shmomentumspaceresults} we saw that CFT correlators in spinor-helicity variables and momentum space are given by triple-$K$ integrals of the kind :
\begin{align}\label{triplekint}
I_{\alpha\{\beta_1,\beta_2,\beta_3\}}(k_1,k_2,k_3)=\int_0^\infty dx\,x^\alpha\,\prod_{j=1}^3\,k_j^{\beta_j}\,K_{\beta_j}(k_j\,x)
\end{align}
where $K_\nu$ is a modified Bessel function of the second kind and $\alpha$ and $\beta_i$ are discrete parameters that depend on the dimension of space and the conformal dimensions of the operators. 
The integral is convergent except when the following equality is satisfied for any (or many) choice of signs \cite{Bzowski:2013sza,Bzowski:2015pba,Bzowski:2017poo,Bzowski:2018fql} :
\begin{align}
\label{divergencecondition}
\alpha+1\pm\beta_1\pm\beta_2\pm\beta_3=-2n,\quad\quad n\in\mathbb Z_{\ge 0}
\end{align}
When the integral is divergent, one regulates it by shifting the dimension of the space and the conformal dimensions \cite{Bzowski:2013sza,Bzowski:2015pba,Bzowski:2017poo,Bzowski:2018fql} :
\begin{align}
d&\rightarrow\widetilde d=d+2u\epsilon\cr
\Delta_i&\rightarrow\widetilde\Delta_i=\Delta_i+(u+v_i)\epsilon
\end{align}
where $u$ and $v_i$ are four real parameters. This results in a  shift in the discrete parameters of the triple-$K$ integral indicated as follows :
\begin{align}
I_{\alpha\{\beta_1,\beta_2,\beta_3\}}\rightarrow I_{\widetilde\alpha\{\widetilde\beta_1,\widetilde\beta_2,\widetilde\beta_3\}}=I_{\alpha+u\epsilon\{\beta_1+v_1\epsilon,\beta_2+v_2\epsilon,\beta_3+v_3\epsilon\}}
\end{align}
Here we note that when one deals with parity odd contributions to a correlator one has to set $u=0$ as the Levi-Civita tensors are defined in the original dimensions and not in the shifted dimensions and hence one cannot use dimensional regularisation.

For cases where the divergence condition \eqref{divergencecondition} is satisfied for the choice of signs $(---)$ or $(--+)$ or its permutations, one gets rid of the singularities in the regularised correlator by adding suitable counter-terms to the CFT action. For cases where the equality \eqref{divergencecondition} is satisfied for the choice of signs $(-++)$ and its permutations or $(+++)$ there are no appropriate counter-terms that one can add to the action. These correspond to cases where it is the triple-$K$ representation of the correlator itself that is singular\cite{Bzowski:2015pba,Bzowski:2017poo,Bzowski:2018fql}.

In the following we will show using the example of the $\langle JJO_\Delta\rangle$ correlator that  in the spinor-helicity variables the relation between the parity-even and the parity-odd parts continues to hold even after renormalisation.
To renormalise the correlators, we first convert our answers in the spinor-helicity variables to momentum space expressions, cure the divergences and obtain finite answers in momentum space. We then convert the resulting correlators back to spinor-helicity variables.  
\subsection{$\langle JJO_\Delta\rangle$}
The momentum space expression for the even part of the correlator was given in \eqref{jjodeltaevenmomentum}.
%
%
%
It can be checked that the discrete parameters in the triple-$K$ integral satisfy the divergence condition \eqref{divergencecondition} when $\Delta\ge 4$. When $\Delta=4$, the divergence condition \eqref{divergencecondition} is satisfied for the choice of signs given by $(---)$. 
%
 A convenient scheme of regularisation is choosing $v_3\ne 0$ and $u=v_1=v_2=0$ \cite{Bzowski:2018fql}.
 The divergence terms are removed by adding the counter-term with three sources \cite{Bzowski:2018fql}:
\begin{align}
S_{ct}=a(\epsilon)\int d^{3}x\,\mu^{v_3\epsilon}\,F_{\mu\nu}F^{\mu\nu}\phi
\end{align}
where $\mu$ is the renormalisation scale. After removing the divergence by choosing an appropriate $a(\epsilon)$ such that the singular term in the regularised correlator is canceled, we get the following finite renormalised form factor  :
\begin{equation}\label{jjo4form}
A_1^{\text{reno}}(k_1, k_2, k_3)=3\,c_1\text{log}\left(\frac{k_1+k_2+k_3}{\mu}\right)-c_1\frac{k_3^2+3k_3(k_1+k_2+k_3)}{(k_1+k_2+k_3)^2}
\end{equation}
In spinor-helicity variables, the renormalised correlator takes the following form :
\begin{align}
\langle J^{-}J^{-}O_{4}\rangle_{\text{even}}=A_1^{\text{reno}}(k_1,k_2,k_3)\langle 12\rangle^4
\end{align}
Let us now discuss the parity odd part of $\langle JJO_\Delta\rangle$. The momentum space expression for the correlator takes the form in \eqref{JJOansatz1}. 
%
%
For $\Delta=4$, the two triple-$K$ integrals are singular for the choice of signs $(+--)$ and $(-+-)$ respectively. To remove the singularity, we add the following parity-odd counter-term with two sources and one operator : 
\begin{equation}
\label{ctjjotype1}
S_{ct}=a(\epsilon)\int{d^{3}x\; \mu^{-\epsilon}\,\epsilon^{\mu\nu\lambda}\,F_{\mu\nu}\,J_{\lambda}\,\phi}
\end{equation}
 After removing the divergences, the resulting form factor is given by :
\begin{equation}\label{jjo4form}
B^{\text{reno}}_1(k_1, k_2, k_3)=c_1\frac{3}{k_1}\text{log}\left(\frac{k_1+k_2+k_3}{\mu}\right)-c_1\frac{k_3^2+3k_3(k_1+k_2+k_3)}{k_1(k_1+k_2+k_3)^2}
\end{equation}
Note that the form factor $B^{\text{reno}}_1(k_1,k_2,k_3)$ is related to the one in the even case $A^{\text{reno}}_1(k_1,k_2,k_3)$ by the following simple relation :
\begin{align}
B_1^{\text{reno}}(k_1, k_2, k_3)=\frac{1}{k_1}A_1^{\text{reno}}(k_1, k_2, k_3)
\end{align}
In spinor-helicity variables, the correlator again takes the same form as in the parity even case :
\begin{align}
\langle J^{-}J^{-}O_{4}\rangle_{\text{odd}}=i\,A^{\text{reno}}_1(k_1,k_2,k_3)\langle 12\rangle^4
\end{align}
Thus we have illustrated following the case of $\langle JJO_4\rangle$ that the parity-even and the parity-odd parts of the correlator are given by the same form factor even after renormalisation.

\subsection{$\langle TTO_\Delta\rangle$}
Let us now consider the $\langle TTO_\Delta\rangle$ correlator.
The transverse and traceless part of the even part of the correlator is given by \cite{Bzowski:2018fql} :
\begin{align}
\label{ttonewansatz}
&\langle T_{\mu_1\nu_1}(k_1)T_{\mu_2\nu_2}(k_2)O(k_3) \rangle_{\text{even}}\cr
&\hspace{.4cm}=\Pi_{\mu_1\nu_1\alpha_1\beta_1}(k_1)\Pi_{\mu_2\nu_2\alpha_2\beta_2}(k_2)\left[A_1\,k_2^{\alpha_1}k_2^{\beta_1}k_3^{\alpha_2}k_3^{\beta_2}+A_2\delta^{\alpha_1\alpha_2}k_2^{\beta_1}k_3^{\beta_2}+A_3\delta^{\alpha_1\alpha_2}\delta^{\beta_1\beta_2}\right]\cr
\end{align}
In $d=3$, the solutions of the primary Ward identities were obtained to be \cite{Bzowski:2018fql} :
\begin{align}
A_1&=c_1 I_{\frac 92\{\frac 32,\frac 32,\Delta-\frac 32\}}\cr
A_2&=4c_1 I_{\frac 72\{\frac 32,\frac 32,\Delta-\frac 12\}}+c_2 I_{\frac 52\{\frac 32,\frac 32,\Delta-\frac 32\}}\cr
A_3&=2c_1 I_{\frac 52\{\frac 32,\frac 32,\Delta+\frac 12\}}+c_2 I_{\frac 32\{\frac 32,\frac 32,\Delta-\frac 12\}}+c_3 I_{\frac 12\{\frac 32,\frac 32,\Delta-\frac 32\}}
\end{align}
One can easily check that for $\Delta=1$, $\Delta=2$, $\Delta=3$ a subset of the triple-$K$ integrals that appear in the form-factors above are divergent. A convenient regularisation scheme to work with is $u=v_1=v_2$ and $v_3\ne u$. 
For $\Delta=1,2,3$ one does not have counter-terms to remove the divergences. It turns out that the divergences that appear in these cases are exactly cancelled by the primary constants determined by the secondary Ward identities \cite{Bzowski:2018fql}. 
\subsubsection{$\langle TTO_4\rangle$}
$\langle TTO_4\rangle$ deserves special discussion as this is the first case where a type-$A$ anomaly could occur \cite{Bzowski:2018fql}. It was noticed in \cite{Bzowski:2018fql} that in the regularised correlator the pole in the regulator $\epsilon$ multiplies a degenerate combination of form factors in the numerator and hence 
the divergent form factors amount to a finite anomalous contribution to the correlator. Thus counter-terms are not essential to remove such divergences. It was also shown in \cite{Bzowski:2018fql} that the divergences in the regularised correlator and the anomaly can be removed using an appropriate counter-term with suitable coefficients. The form-factors of the renormalised correlator takes the following form (we present only the scheme independent part) :
\begin{align}
A_1&=\frac{c_1}{E^4}\mathcal E_1\cr
A_2&=\frac{c_1}{E^3}(-\mathcal E_1(k_1+k_2-k_3)+2\mathcal E_2 k_1k_2)\cr
A_3&=\frac{c_1(k_1+k_2-k_3)}{4E^2}(\mathcal E_1(k_1+k_2-k_3)-4\mathcal E_2 k_1k_2)
\end{align}
where
\begin{align}
\mathcal E_1&=(k_1+k_2)^2((k_1+k_2)^2+12k_1k_2)+16(k_1+k_2)((k_1+k_2)^2+3k_1k_2)k_3\cr
&\hspace{1cm}+6(7(k_1+k_2)^2+10k_1k_2)k_3^2+32(k_1+k_2)k_3^3+5k_3^4\cr
\mathcal E_2&=(k_1+k_2)^3+15(k_1+k_2)^2k_3+27(k_1+k_2)k_3^2+5k_3^3
\end{align}

We now convert this result in momentum space to the spinor-helicity variables and obtain :
\begin{align}
\langle T^{-}T^{-}O_4\rangle=k_1\,k_2\frac{(k_1+k_2)^2+4(k_1+k_2)k_3+5k_3^2}{(k_1+k_2+k_3)^4}\langle 12\rangle^4
\end{align}
This precisely matches the result that we obtained for the correlator by directly solving the conformal Ward identities in spinor-helicity variables \eqref{jjofinalanswersh}.
%
For $\Delta=4$ (or more generally $\Delta\le 5$) the  triple-$K$ integral in \eqref{jjofinalanswersh} is convergent and we get finite results for the correlator without any renormalisation. For $\Delta\ge 6$ the above triple-$K$ integral is singular and one needs to regularise and renormalise appropriately. 

%


\section{Weight-shifting operators}
\label{wso}
In this section we obtain correlators in momentum space using weight-shifting operators, following \cite{Baumann:2020dch} and \cite{Jain:2021wyn}. 
 The operators that we will primarily use are given by
\begin{align}
   H_{12}&=2\left(\vec{z}_{1} \cdot \vec{K}_{12}\right)\left(\vec{z}_{2} \cdot \vec{K}_{12}\right)-\left(\vec{z}_{1} \cdot \vec{z}_{2}\right) K_{12}^{2}\\[5 pt]
   \widetilde{D}_{12}&=\epsilon^{z_{1} z_{2} k_{1}} W_{12}^{--}+\epsilon^{z_{1} k_{1} K_{12}^{-}}\left(\vec{z}_{2} \cdot \vec{K}_{12}^{-}\right)+\left(2-\Delta_{1}\right) \epsilon^{z_{1} z_{2} K_{12}^{-}}
\end{align}
and their permutations, where
\begin{align}
   K_{12}^{-\mu} &=\partial_{k_{1\mu}}-\partial_{k_{2\mu}}\\[5 pt]
   W_{12}^{--} &= \frac{1}{2}\vec{K}_{12}^{-}\cdot \vec{K}_{12}^-
\end{align}
\subsection{$\langle JJO_{\Delta} \rangle_{\text{odd}}$}
In \cite{Jain:2021wyn} we obtained $\langle JJO_{\Delta} \rangle_{\text{odd}}$ using weight-shifting operators was discussed earlier . 
The seed correlator, in terms of triple-$K$ integrals is given by :
\begin{align}
    \langle O_2 O_3 O_{\Delta} \rangle = c_1I_{\frac{1}{2},\{\frac{1}{2},\frac{3}{2},\Delta-\frac{3}{2}\}}
\end{align}
The correlator $\langle JJO_{\Delta} \rangle_{\text{odd}}$ is then obtained by the action of the parity-odd operator $\widetilde{D}_{12}$ which raises the spin of the operators at points 1 and 2 and lowers the weight of the operator at point 2 :
\begin{align}
    \langle JJO_{\Delta} \rangle_{\text{odd}} &= \widetilde{D}_{12}\langle O_2 O_3 O_{\Delta} \rangle\nonumber \\[5 pt]
    &= c_1 I_{\frac{5}{2},\{\frac{1}{2},\frac{3}{2},\Delta-\frac{3}{2}\}}\left(\epsilon^{k_1 z_1 z_2}k_2-\epsilon^{k_2 z_1 z_2}k_1\right)
\end{align}
after using appropriate Schouten identities.
\par
It was seen that this reproduces the correct answer even for the case of $\Delta=4$ and $\Delta=5$ where $\langle JJO_{\Delta} \rangle_{\text{odd}}$ is divergent. As is explained below, it is better to start without renormalizing the seed correlator. The target correlator is renormalized at the end, in case it has divergences. 

\subsubsection{Subtleties associated with divergences}
\label{subtlety}
There are some subtleties to note when the correlators are divergent as observed in \cite{Jain:2021wyn}. When the target and seed correlators both have the same kind of divergence, one can renormalize the seed correlator appropriately and use weight-shifting operators to get the correct result for the target correlator. When the target correlator is not divergent but the seed correlator is, this method does not always work. For example, it does not work when the seed correlator has a non-local divergence while the target correlator is not divergent. To illustrate this, consider the following sequence that also reproduces $\langle JJO_3 \rangle_{\text{odd}}$ :
\begin{align}\label{jjowsalt} 
    \langle JJO_3 \rangle_{\text{odd}} &= k_1 k_2 P_1^{(1)}P_2^{(1)}\widetilde{D}_{12}\langle O_1 O_2 O_3 \rangle
\end{align}
The seed correlator has a non-local divergence and upon appropriate renormalization it is given by :
\begin{align}\label{o1o2o3renorm}
    \langle O_1 O_2 O_3 \rangle^{\text{reno}} = c_1\frac{k_1+k_2}{k_1}
\end{align}
If one calculates the r.h.s. of \eqref{jjowsalt} using \eqref{o1o2o3renorm}, it can be easily checked that the answer does not match the known result for $\langle JJO_3 \rangle_{\text{odd}}$. In fact, the final correlator obtained this away goes to zero upon using Schouten identities. The correct way to go about it is to put in the full, unrenormalized seed correlator into \eqref{jjowsalt}, which is given by
\begin{align}
    \langle O_1 O_2 O_3 \rangle &= c_1 \frac{k_1+k_2}{k_1 \epsilon}+c_1\left(\frac{\text{log}(k_1+k_2+k_3)(k_1+k_2)-(k_1+k_2+k_3)}{k_1}\right)
\end{align}
This way, in the final answer, we can see that the divergences cancel upon using Schouten identities and we get the correct answer for $\langle JJO_3 \rangle_{\text{odd}}$. 
\subsection{$\langle TTO_{\Delta} \rangle_{\text{odd}}$}
In \cite{Jain:2021wyn} we computed $\langle TTO_1 \rangle_{\text{odd}}$ and $\langle TTO_2 \rangle_{\text{odd}}$ using weight shifting operators. Here, we use these operators to calculate the correlator for any scaling dimension of the scalar operator. Also, our earlier analysis used a definition of the $3$-point function that resulted in the correlator having a non-trivial Ward-Takahashi identity. Here, following \cite{Bzowski:2018fql} we redefine the $3$-point function so that the correlator is completely transverse as this simplifies the calculations.
The following sequence of operators reproduces $\langle TTO_{\Delta} \rangle_{\text{odd}}$
\begin{align}
    \langle TTO_{\Delta}\rangle_{\text{odd}} = (k_1 k_2)^3P_1^{(2)}P_2^{(2)}H_{12}\widetilde{D}_{12}\langle O_1 O_2 O_{\Delta} \rangle
\end{align}
where the seed correlator is given by :
\begin{align}
    \langle O_1 O_2 O_{\Delta} \rangle = c_1 I_{\frac{1}{2}\{-\frac{1}{2},\frac{1}{2},\Delta-\frac{3}{2}\}}
\end{align}
This can be easily checked to give the correct result for the cases of $\Delta=1,2,3,4$ by calculating the correlator and converting it to its form in spinor-helicity variables. For $\Delta=3+2n$, the seed correlator has a non-local divergence and one must also be careful about the subtlety mentioned in Section \ref{subtlety}.
The explicit momentum space expression for $\langle TTO_3 \rangle_{\text{odd}}$ is given by
\begin{align}
    \langle TTO_3 \rangle_{\text{odd}} = A_1 \epsilon^{k_1 k_2 z_1}(k_2 \cdot z_1)(k_3 \cdot z_2)^2 +A_1(k_1 \leftrightarrow k_2)\epsilon^{k_1 k_2 z_2}(k_3 \cdot z_2)(k_2 \cdot z_1)^2
\end{align}
where we have used Schouten identities and the degeneracy in \eqref{degen} to ensure that the divergences cancel and to simplify the expression. The form factor is given \footnote{Like for $\ev{JJO}$ in \eqref{JJOff} - this form factor suggests an apparent collinear divergence. However, one can easily check that the numerator of the correlator also vanishes in the collinear limit leaving the correlator finite. Hence the only pole is at $E\rightarrow 0$ as expected.} by
\begin{align}
    A_1 = c_1\frac{k_1^3 k_2^2(k_1+k_2+4k_3)}{(k_1+k_2+k_3)^4(k_1^2+k_2^2-k_3^2-2k_1k_2)^2}
\end{align}
After doing a change of basis, this can be shown to match the answer obtained from spinor-helicity variables and bulk gravity calculations. 

In general, we have
\begin{align}
    \langle TTO_{\Delta} \rangle = c_1k_1^2k_2^2 I_{\frac{9}{2},\{\frac{1}{2},\frac{1}{2},\Delta-\frac{3}{2}\}}\left[\epsilon^{k_1 z_1 z_2}(z_1 \cdot z_2)k_2-\epsilon^{k_2 z_1 z_2}(z_1 \cdot z_2)k_1\right]
\end{align}
\subsection{$\langle TJJ \rangle_{\text{odd}}$}
As we saw in \eqref{tjjoddwi} and \eqref{tjjoddwifnl}, $\langle TJJ \rangle_{\text{odd}}$ is completely transverse with respect to all the momenta. We start with $\langle O_1 O_2 O_2\rangle$ as the seed correlator, which is given by
\begin{align}
     \langle O_1 O_2 O_2 \rangle &= \frac{c_1}{k_1} \text{log}\left(\frac{k_1+k_2+k_3}{\mu}\right)
\end{align}
where $\mu$ is the renormalization scale.
\par
The following sequence of operators reproduces $\langle TJJ \rangle_{\text{odd}}$
\begin{align}
    \langle T(k_1)J(k_2)J(k_3) \rangle_{\text{odd}} = k_1^3 (k_2 k_3)P_1^{(2)}P_2^{(1)}P_3^{(1)}H_{13}\widetilde{D}_{12}\langle O_1(k_1)O_2(k_2)O_2(k_3) \rangle+(2 \leftrightarrow 3)
\end{align}
where $P_i^{(s)}$ is a spin-$s$ projector. 
The explicit momentum space expression for the correlator is given in Appendix \ref{sdro}.

Although this expression looks very different from the expression obtained earlier in \eqref{jjta1a2}, it can be shown that they are the same upon using Schouten identities. This fact can easily be seen by converting both the momentum space expressions to spinor-helicity variables where they exactly match. 
\subsection{$\langle TTT \rangle_{\text{odd}}$}
We know from Section \ref{jsjsjsmomentumsp} that the non-homogeneous part of $\langle TTT \rangle_{\text{odd}}$ is a contact term and can therefore be ignored. In momentum space, this means that we only need to calculate the transverse part of the correlator. This is given by :
\begin{align}
    \langle TTT \rangle_{\text{odd}} = (k_1 k_2 k_3)^3P_1^{(2)}P_2^{(2)}P_3^{(2)}H_{13}H_{23}\widetilde{D}_{12}\langle O_1 O_2 O_2 \rangle +\text{cyclic perm.}
\end{align}
where we have added cyclic permutations to make the correlator manifestly symmetric. The explicit answer in momentum space obtained this way has many form-factors and is not very illuminating. However, one can use Schouten identities and the degeneracies in \eqref{degen} and \eqref{eid} to reduce the answer to just four form-factors, out of which two are independent. This matches the answer obtained using spinor-helicity variables in \eqref{TTTshal}. 
\subsection{Homogeneous part of general 3-point function using weight-shifting operators}

We can use weight-shifting operators $H_{ij}$ and $\widetilde{D}_{ij}$ to compute the homogeneous part of more general correlators. For example, consider $\langle J_s J_s J_s \rangle$ where $s$ is even. The parity-even and parity-odd parts of the correlator are given by : 
\begin{align}\label{gnss}
    \begin{split}
        \langle J_s J_s J_s \rangle_{\text{even}} &= (k_1 k_2 k_3)^{2s-1}P_1^{(s)}P_2^{(s)}P_3^{(s)}(H_{12}H_{13}H_{23})^{\frac{s}{2}}\langle O_2 O_2 O_2 \rangle\\[5 pt]
        \langle J_s J_s J_s \rangle_{\text{odd}} &= (k_1 k_2 k_3)^{2s-1}P_1^{(s)}P_2^{(s)}P_3^{(s)}H_{12}^{\frac{s-2}{2}}(H_{13}H_{23})^{\frac{s}{2}}\widetilde{D}_{12} \langle O_1 O_2 O_2 \rangle
    \end{split}
    \begin{split}
        (s=2n)
    \end{split}
\end{align}
The seed correlator $\langle O_2 O_2 O_2 \rangle$ is given by
    \begin{align}
    \langle O_2 O_2 O_2 \rangle &= c_1 \text{log}\left(\frac{k_1+k_2+k_3}{\mu}\right)
\end{align}
where $\mu$ is the renormalization scale.
Correlators involving scalar operator can also be written down as
\begin{align}
        \langle J_s J_s O_{\Delta} \rangle_{\text{even}} &= (k_1 k_2 )^{2s-1}P_1^{(s)}P_2^{(s)}H_{12}^{s}\langle O_2 O_2 O_{\Delta} \rangle\\[5 pt]
        \langle J_s J_s O_{\Delta} \rangle_{\text{odd}} &= (k_1 k_2)^{2s-1}P_1^{(s)}P_2^{(s)}H_{12}^{s-1}\widetilde{D}_{12} \langle O_1 O_2 O_{\Delta} \rangle
\end{align}
\par
When $s$ is odd in \eqref{gnss}, we require additional weight-shifting operators apart from $H_{ij}$ and $\widetilde{D}_{ij}$. This is analogous to the way we write correlators in the next section using a finite number of conformal invariant structures.
\par
Computation of the non-homogeneous part requires us to take linear combinations of different sequences of weight-shifting operators such that the Ward-Takahashi identity is saturated. See, for example, the computation of $\langle TTT \rangle_{\text{even}}$ in \cite{Baumann:2020dch}. Compared to the homogeneous part, this is much harder to do for a general correlator as we do not know the Ward-Takahashi identity for a general spin-$s$ current. 


\section{CFT correlators in terms of momentum space invariants}\label{INV}
The aim of this section is to write down CFT correlators derived in previous sections in terms of a few conformal invariant momentum space structures. Let us define  
\begin{align}
    &Q_{12} = \frac{1}{E^{2}}\left[2\left(\vec{z}_{1} \cdot \vec{k}_{2}\right)\left(\vec{z}_{2} \cdot \vec{k}_{1}\right)+E\left(E-2 k_{3}\right) \vec{z}_{1} \cdot \vec{z}_{2}\right]\label{Q12defn}\\
    &S_{12} = \frac{1}{E^{2}}\left[k_{2} \epsilon^{k_{1} z_{1} z_{2}}-k_{1} \epsilon^{k_{2} z_{1} z_{2}}\right]\\
   &P_{123}=\frac{1}{E^{3}}\left[2\left(\vec{z}_{1} \cdot \vec{k}_{2}\right)\left(\vec{z}_{2} \cdot \vec{k}_{3}\right)\left(\vec{z}_{3} \cdot \vec{k}_{1}\right)+E\left(k_{3}\left(\vec{z}_{1} \cdot \vec{z}_{2}\right)\left(\vec{z}_{3} \cdot \vec{k}_{1}\right)+\text { cyclic }\right)\right]\\
   &R_{123} = \frac{1}{E^3}\left[\left\{(\vec{k}_1 \cdot \vec{z}_3)\left(\epsilon^{k_3 z_1 z_2}k_1-\epsilon^{k_1 z_1 z_2}k_3\right)+(\vec{k}_3 \cdot \vec{z}_2)\left(\epsilon^{k_1 z_1 z_3}k_2-\epsilon^{k_2 z_1 z_3}k_1\right)\right.\right.\nonumber\\[5 pt]
&\hspace{1.5cm}\left.\left.-(\vec{z}_2 \cdot \vec{z}_3)\epsilon^{k_1 k_2 z_1}E+\frac{k_1}{2} \epsilon^{z_1 z_2 z_3}E(E-2k_1)\right\}+\text{cyclic perm}\right]
\end{align}
These can be used as building blocks for writing down momentum space 3-point conserved correlators since they arise naturally in the expressions for such correlators \footnote{That they are conformal invariants follows from \eqref{Homgeneric} - \eqref{sssinv}. If we put $s=1$, in each case a structure is equal to a particular correlator which is of course, by definition, conformally invariant.}. There are some interesting relations among the above defined quantities. For example
\begin{align}\label{invariantstructurerelations}
   & S_{ij}^2 = Q_{ij}^2,~~~~R_{ijk}^2= P_{ijk}^2,~~~P_{123}^2 = Q_{12} Q_{23} Q_{31},~~S_{ij}S_{jk} = Q_{ij}Q_{jk}\nonumber\\[5 pt]
   & P_{123} R_{123} = S_{12} Q_{23} Q_{31} + \text{cyclic perm.}
\end{align} up-to degeneracies.

\subsection*{Homogeneous contribution}
From the summary in \ref{spinsintermsofspin1},
 we may now write the momentum space three-point functions in a compact manner using the above invariants. Let us note that we are concerned only with correlators which satisfy triangle inequality. To do this we divide the correlator into two different classes.
 \subsection*{$s_1+s_2+s_3=2n\;\;(n \in \mathbb{Z})$}
 For this class of correlators we only require $Q_{ij}$ and $S_{ij}.$ 
 Consider $\langle J_{s_1}J_{s_2}J_{s_3} \rangle$ such that $s_1 \geq s_2 \geq s_3$, $s_1 \leq s_2+s_3$.  Then, we have
\begin{align}
    \langle J_{s_1} J_{s_2}J_{s_3} \rangle_{\text{even}} &= k_1^{s_1-1}k_2^{s_2-1}k_3^{s_3-1}Q_{12}^{\frac{1}{2}(s_1+s_2-s_3)}Q_{23}^{\frac{1}{2}(s_2+s_3-s_1)}Q_{13}^{\frac{1}{2}(s_1+s_3-s_2)}\nonumber\\[5 pt]
   \langle J_{s_1} J_{s_2}J_{s_3} \rangle_{\text{odd}} &= k_1^{s_1-1}k_2^{s_2-1}k_3^{s_3-1}S_{12}Q_{12}^{\frac{1}{2}(s_1+s_2-s_3-2)}Q_{23}^{\frac{1}{2}(s_2+s_3-s_1)}Q_{13}^{\frac{1}{2}(s_1+s_3-s_2)}\nonumber\\[5 pt]
   &+\text{cyclic perm.}
\end{align}
Correlators involving scalar operators can also be written this way,
\begin{align}\label{Homgeneric}
   &\langle J_sJ_sO_2\rangle_{\text{even}, \bf{h}} = b_{12}^{s-1}Q^s_{12}\\
   &\langle J_sJ_sO_2\rangle_{\text{odd}, \bf{h}} = b_{12}^{s-1}S_{12}Q^{s-1}_{12}
\end{align}
where $b_{ij}=k_i k_j$ and $c_{123}=k_1 k_2 k_3$.  The 3-point function involving $\Delta=1$ is obtained by shadow transforming \eqref{Homgeneric}. One can also write the correlator involving a generic scalar operator dimension $\delta$ but we do not reproduce this here.
\par

 \subsection*{$s_1+s_2+s_3=2n+1\;\;(n \in \mathbb{Z})$}
 We require $P_{123}$ and $R_{123}$ as well when the sum of the spins is odd. For example, when $s$ is odd, we have
 \begin{align}
 \langle J_s J_s J_s\rangle_{\text{even}, \bf{h}} = c^{s-1}_{123} P^s_{123}
  \end{align}
  \begin{align}\label{sssinv}
   \langle J_s J_s J_s\rangle_{\text{odd}, \bf{h}} = c^{s-1}_{123} R_{123} P^{s-1}_{123}
   \end{align}
   When $s$ is odd, these are the only structures using which $\langle J_s J_s J_s \rangle$ can be written. One can use \eqref{invariantstructurerelations} to substitute even powers of $P_{123}$ in terms of $Q_{ij}$'s. Other correlators with $s_1+s_2+s_3=\text{odd}$ can also be considered similarily.
\subsection*{Non-Homogeneous contribution}
We have discussed homogeneous contribution so far. The story for the non-homogeneous contribution is more complicated. We do not have a generic form of the WT identity to evaluate three point functions involving operators of arbitrary spin. For example, if we consider the solutions for $\langle J_s O_2 O_2 \rangle$ as given in \eqref{jso2o2momentum}, there is no discernible underlying structure to these expressions. The numerator becomes increasingly complicated as we consider higher values of $s$ and one cannot write these as the power of some simple structure.
We can similarly identify some structures based on the answers for $\langle J J J \rangle$, $\langle J J T \rangle$ and $\langle TTT \rangle$ however those relations are not illuminating  as in the case of the homogeneous part, (see \eqref{Homgeneric}) and we do not present them here.

\section{Some interesting observations}\label{sio}
In this section we collect a few interesting observations about the correlators discussed so far. For the purposes of this discussion, it will be useful to write the correlators as in \eqref{jssph1}.

\subsection{Contact terms}\label{contact}
To properly understand correlators in momentum space it is very important to understand the contact terms which arises in both 
parity-odd and parity-even cases. For example  $\langle JJJ\rangle$ correlation function has a contact term which is parity odd and is given by \eqref{JJJhn1}. Fourier transforming this to position space will give us a term of the form
\begin{equation}
    \langle J^{a}_{\mu}J^{b}_{\nu}J^{c}_{\rho}\rangle_{\text{contact}}\propto c'_{J} f^{abc} \epsilon_{\mu\nu\rho} \delta^3(x_1-x_2) \delta^3(x_2-x_3).
\end{equation}
Another example of a correlation function where both parity even and parity odd part has contact term, let us consider $\langle TTT\rangle$. The parity even contact term is given by \cite{Farrow:2018yni}
\begin{align}
    \langle TTT\rangle_{\text{even}} \propto c_{T}\left(k_1^3+k_2^3+k_3^3\right)z_1\cdot z_2 z_2\cdot z_3 z_3\cdot z_1
\end{align}
which when converted to position space gives contact term of the form
\begin{equation}
   \langle TTT\rangle_{\text{contact}} \propto c_{T} \left(f(x_1)\delta^3(x_2-x_3)+f(x_2)\delta^3(x_3-x_1)+f(x_3)\delta^3(x_1-x_2)\right).
\end{equation}
Parity odd contact term is given\footnote{Let us note that, we have neglected the precise functional dependence. We have just indicated the form of the delta function that arises.} by \eqref{tttnewbasis} which becomes
\begin{equation}
     \langle TTT\rangle_{\text{contact}} \propto c'_{T} \epsilon_{z_1 z_2 z_3}\delta^3(x_1-x_2) \delta^3(x_2-x_3)+ \cdots
\end{equation} where $z$ are polarization tensors. Once again we have not mentioned exact form of the contact term.
Interestingly, for both parity even and parity odd part the contact term arises in the non-homogeneous contribution.  One way to understand parity odd case is to  look at \eqref{nhpiece}. The right hand side of this equation for parity odd case is always a contact term for the correlator we have considered. For example, the transverse Ward identity for $\langle JJJ \rangle $ takes the form \eqref{WTjjj}
\begin{align}
\label{WTjjj1}
\begin{split}
k_{1\mu}\langle J^{\mu a}(k_1) J^{\nu b}(k_2) J^{\rho c}(k_2) \rangle &= f^{adc}\langle J^{\rho d}(k_2)J^{\nu b}(-k_2) \rangle-f^{abd}\langle J^{\nu d}(k_3)J^{\rho c}(-k_3) \rangle\nonumber \\[5 pt]
&= f^{abc}\epsilon^{\nu\rho k_1}
\end{split}
\end{align}
which is a contact term. One can check the same explicitly for  $\langle TTT \rangle$ as well as other correlators computed in previous sections. 
We expect on general grounds that $\langle J_{s_1}J_{s_2}J_{s_3}\rangle_{\text{odd},\bf{nh}}$ is a contact term. To conclude, we observe that \\ 

\noindent {\bf{A.}} Contribution to the contact term comes from non-homogeneous part of both parity-even and parity-odd correlator. For parity-even it was observed  in $\langle TTT\rangle$ only. \\

\noindent {\bf{B.}} Parity-odd non-homogeneous piece of the CFT correlator is always a contact term.
\begin{equation}
    \langle J_{s_1} J_{s_2} J_{s_3} \rangle_{\bf{nh},\text{odd}} = \rm{contact~ term.}
\end{equation}
However parity-even non-homogeneous piece can be nontrivial as is discussed in previous sections. The reader is referred to appendix \ref{contact} for a detailed discussion regarding whether the contact parity-odd terms can be set to zero by field redefinitions.

\subsection{Relation between parity-even and parity-odd solutions}
If we look at the correlator in momentum space, see section \ref{shmomentumspaceresults}, there seem to exist no clear relationship between parity odd and parity even part of the correlator. However, as is seen section \ref{CCSHV} and \ref{sectionSHresults}, there exist a remarkable relationship between them in spinor-helicity variables, namely
\begin{equation}\label{pepore}
   \langle J_{s_1} J_{s_2} J_{s_3} \rangle_{\bf{h},\text{odd}}  =  \langle J_{s_1} J_{s_2} J_{s_3} \rangle_{\bf{h},\text{even}}~~~~ ~~~(\rm{In~spinor~helicity~variables})
\end{equation} up-to some signs and factors of $i.$ 
Let us explain this in terms of some concrete equations. To start, let us consider the anstaz
\begin{align}
 \langle J_{s_1}^{-} J_{s_2}^{-} J_{s_3}^{-} \rangle =(F_{1}(k_1,k_2,k_3)+ i F_{2}(k_1,k_2,k_3)) \langle 1 2\rangle^{s_1 +s_2 -s_3}  \langle 2 3\rangle^{-s_1 +s_2 +s_3}  \langle 3 1\rangle^{s_1 -s_2 +s_3} \nonumber\\
  \langle J_{s_1}^{-} J_{s_2}^{-} J_{s_3}^{+} \rangle =(G_{1}(k_1,k_2,k_3)+ i G_{2}(k_1,k_2,k_3)) \langle 1 2\rangle^{s_1 +s_2 -s_3}  \langle 2 3\rangle^{-s_1 +s_2 +s_3}  \langle 3 1\rangle^{s_1 -s_2 +s_3} 
\end{align} 
where $F_1,G_1$ and $F_2,G_2$ are form-factors for the parity-even and parity-odd parts of the correlator. Both $F_1$ and $F_2$ satisfy the same non-homogeneous equation, see for example \eqref{Feqn1}, \eqref{Feqn1od}. However, the form factors $G_1$ and $G_2$ satisfy a different non-homogeneous equation,  see for example \eqref{Geqn1}, \eqref{Geqn1od} in appendix \ref{CWI}. This difference is coming due to the different contribution of WT identity to parity-even and parity-odd parts for $--+$ helicity component\footnote{For $---$ helicity the WT identity contributes the same for parity even and odd case.}. This implies non-homogeneous contribution to  parity-even and odd cases generally differ, whereas the homogeneous solution is always the same. 
\par
This relation becomes even more nontrivial in the cases where there is a divergence in the correlator. For example, for 
$ \langle J J O_{4} \rangle$ the solution of the conformal Ward identity is given by
\begin{align}
    \langle J^{-} J^{-} O_{4} \rangle_{\text{even}}=c_1(k_1k_2)I_{\frac 52,\{\frac 12,\frac 12,\frac 52\}} \\
    \langle J^{-} J^{-} O_{4} \rangle_{\text{odd}}=i\,c_2(k_1k_2)I_{\frac 52,\{\frac 12,\frac 12,\frac 52\}}
\end{align}
However, the triple-$K$ integral is divergent and one needs to regularise and renormalise the correlator.
To  do so  we go to momentum space (see section \ref{renor11}). The renormalization procedure for even and odd parts is also completely different and we required quite different kinds of counter-terms. However, converting back the renormalized  results in spinor-helicity variables, we remarkably obtained the same result again. This happens to all other correlators having divergences and it would be interesting to have a better understanding of this observation. 

\subsection{Manifest locality test}\label{MLT}
In the context of the cosmological bootstrap, a manifestly local test (MLT) was derived for wavefunctions of scalars of dimension 3 and gravitons in any manifestly local, unitary theory \cite{Jazayeri:2021fvk}. MLT imposes the following condition on such wavefunctions \cite{Jazayeri:2021fvk} :
\begin{equation}
\lim_{k_c\rightarrow 0}\frac{\partial}{\partial k_c}\psi_{n}(k_1,\ldots, k_c,\ldots, k_n)=0
\end{equation}
In the following we discuss how this analysis can be used for calculating CFT correlator $\langle TO_3O_3\rangle.$
Based on the symmetries of the correlator we write down the following ansatz for the correlator :
\begin{align}
\langle T_{\mu\nu}(k_1)O_3(k_2)O_3(k_3)\rangle&=\Pi_{\mu\nu\alpha\beta}(k_1)\,A_1(k_1,k_2,k_3)\,k_{2}^{\alpha}k_{2}^{\beta}
\end{align}
where we take the following ansatz for the form factor :
\begin{align}
A_1(k_1,k_2,k_3)&=\frac{1}{(k_1+k_2+k_3)^2}\left[c_1k_1^3+c_2k_1^2(k_2+k_3)+c_3k_1k_2k_3\right.\cr
&\hspace{2cm}\left.+c_4k_1(k_2+k_3)^2+c_5(k_2+k_3)k_2k_3+c_6(k_2+k_3)^3\right]
\end{align}
where the pole in $k_1+k_2+k_3=0$ can be argued on general grounds and the power of the pole is fixed by dilatation Ward identity.
We will now fix the coefficients that appear in the ansatz by imposing manifest locality. With respect to one of the scalar operators we have :
\begin{align}
\lim_{k_2\rightarrow 0}\frac{\partial}{\partial k_2}\langle T(k_1)O_3(k_2)O_3(k_3)\rangle=0
\end{align}
This gives the following relation between the coefficients :
\begin{align}
c_2=2c_1,\quad c_4=\frac{2c_1-c_3}{2},\quad c_6=-c_5,\quad c_3=2c_5
\end{align}
We can easily check that with these conditions, MLT with respect to the second scalar operator is also satisfied, i.e. : 
\begin{align}
\lim_{k_3\rightarrow 0}\frac{\partial}{\partial k_3}\langle T(k_1)O_3(k_2)O_3(k_3)\rangle=0.
\end{align}
Let us now impose manifest locality with respect to the stress-tensor operator :
\begin{align}
\lim_{k_1\rightarrow 0}\frac{\partial}{\partial k_1}\langle T(k_1)O_3(k_2)O_3(k_3)\rangle=0
\end{align}
This gives the following constraint $c_5=-c_1$.
We now substitute the coefficients back into the ansatz to get :
\begin{align}\label{MLTA}
A(k_1,k_2,k_3)=\frac{c_1}{(k_1+k_2+k_3)^2}\left[ k_1^3+k_2^3+k_3^3+2(k_1^2+k_2k_3)(k_2+k_3)+2k_1(k_2^2+k_2k_3+k_3^2)\right]
\end{align}
Notice that form factor in \eqref{MLTA} matches explicitly with form factor presented in \eqref{TO3O3}.
We hope to come back to this in future for a better understanding of other  3-point functions.
%
%
%

\subsection{A comparison between position and momentum space invariants}
It is interesting to compare momentum space invariants discussed in section \ref{INV} and position space invariants  introduced in \cite{Giombi:2011rz, Costa:2011mg}.
 To illustrate this, let us consider $\langle JJT \rangle$ even part. This is given by 
\begin{equation}\label{conformlinvp}
 \langle T(x_1) J(x_2) J(x_3) \rangle_{\text{even}} =\frac{1}{|x_{12}|| x_{23}||x_{31}|}\left(a_1 P_1^2 Q_1^2 + a_2 P_2^2 P_3^2 + a_3 Q_1^2 Q_2 Q_3 + a_4 P_1 P_2 P_3 Q_1\right),  
\end{equation}
We refer the reader to \cite{Giombi:2011rz} for details about the notation. We see that there are 4 structures. Demanding conservation equation for currents, we get two relation $a_2 = -4 a_1, ~a_3 =-\frac{5}{2} a_1$ which leaves two independent  structures
\begin{equation}\label{consconjt}
 \langle T(x_1) J(x_2) J(x_3) \rangle_{\text{even}} =\frac{1}{|x_{12}|| x_{23}||x_{31}|}\left[a_1 \left(P_1^2 Q_1^2 -4 P_2^2 P_3^2 - \frac{5}{2} Q_1^2 Q_2 Q_3\right) + a_4 P_1 P_2 P_3 Q_1\right]  
\end{equation}
 Furthermore, using WT identity we get a relation between $a_4, a_1$ and the two-point function coefficient $c_j.$ Eliminating $a_4$ we obtain
\begin{align}\label{JJTpos1}
\langle T(x_1) J(x_2) J(x_3) \rangle_{\text{even}} =   \frac{1}{|x_{12}|| x_{23}||x_{31}|}&\Big[a_1 \left(P_1^2 Q_1^2 -4 P_2^2 P_3^2 - \frac{5}{2} Q_1^2 Q_2 Q_3-2 P_1 P_2 P_3 Q_1\right) \nonumber\\
&+\frac{3}{8} c_j P_1 P_2 P_3 Q_1\Big]  
\end{align} where $c_j$ appears in two point function of $J_{\mu}.$ Let us emphasize that, \eqref{conformlinvp} is built out of conformal invariant structures whereas \eqref{JJTpos1} is built out of conformally invariant {\it conserved} structures\footnote{It is quite difficult to build conformally invariant conserved structures directly without first writing conformal invariants and then demanding conservation. Free theory generating functions defined in \cite{Giombi:2011rz} might be of help, however it will be difficult to separate out the homogeneous and non-homogeneous contributions. However, in momentum space we directly get the conformal invariant conserved structures and getting conformal invariant structures without the WT identity constraint would be a more challenging task.}. In \eqref{JJTpos1}, we can identify the term proportional to $a_1$ as homogeneous and the term proportional to $c_j$ as the non-homogeneous contribution. Let us note that, for a generic correlator involving arbitrary spin-$s$ currents, in general it is quite complicated to arrive at the analogue of \eqref{JJTpos1} starting from more the readily obtainable expression \eqref{conformlinvp}. Moreover, finding the non-homogeneous term in position space is equally complicated. However, in momentum space we naturally obtain the analogue of \eqref{JJTpos1} directly. In other words, in momentum space we naturally divide the answer into homogeneous and non-homogeneous contributions and the conformal invariant conserved structure is naturally built in. 

\section{Summary and Future directions}\label{fdir}
To summarize, we have systematically solved for 3-point CFT correlators involving higher spin conserved currents and scalar operators in three dimensions. Spinor-helicity formalism simplifies considerably the CWI based analysis of correlators. It solves the problems associated with degeneracy which makes direct computation in momentum space difficult.  In these variables, we found that the homogeneous part of the correlator gets an identical contribution from parity-even and parity-odd parts. We were also able to write down momentum space correlators in terms of conserved conformally invariant structures. For some correlators (for example $\langle TTO_{4}\rangle $) which are divergent in momentum space, a careful renormalization analysis is required. However, in spinor-helicity variables, we observed that it turns out we directly get the finite part of the correlator which does not require any renormalization. We also verified some of the results using weight-shifting operators. 
Below we discuss some future directions.

In this paper we focused exclusively on 3-point correlators of scalars and conserved currents with spin. If one considers spinning operators which are not conserved (no WT identity) then the approach has to be adjusted accordingly.  Some preliminary momentum space results in this direction were obtained in \cite{Isono:2019ihz}. This analysis is important as it is a useful first step for constructing general 4-point spinning conformal blocks in momentum space. The spinor-helicity formalism,  extended as in \cite{Arkani-Hamed:2017jhn} to account for scattering of massive particles in 4d, should be useful for this purpose. 

It would be interesting to utilise the spinor-helicity formalism in the analysis of 4-point functions. The momentum space CWI approach for 4-point CFT correlators has been used in \cite{Coriano:2019nkw, Maglio:2019grh, Coriano:2020ccb, Coriano:2020ees}. We have seen that the spinor-helicity expressions for 3-point correlators are much simpler compared to the momentum space ones and it is natural to expect a similar simplicity in the analysis of higher point correlators.

For correlators involving higher spin conserved currents the non-homogeneous part of the correlator requires the knowledge of Ward identity. It would be nice to find out if a general structure exists for the Ward identity which would then help us in getting the non-homogeneous part of correlation functions comprising operators with arbitrary spin. It would also be interesting to generalize our results to cases which break the triangle inequality and to cases where the current operators are not exactly conserved \cite{Maldacena:2012sf}. 

We observed in this paper the equivalence of parity-even and parity-odd parts of the 3-point correlator when expressed in spinor-helicity variables, and how this continues to hold in examples where, due to divergences, regularising and renormalising the correlator is required. It would be interesting to understand on general grounds why this is the case even though the counter-terms required in both cases are entirely different.

It would also be very interesting to understand the MLT condition \cite{Jazayeri:2021fvk} discussed in Section \ref{MLT} starting from basic CFT principles. In Section \ref{MLT}, we used MLT condition to compute the non-homogeneous contribution to $\langle TOO \rangle$ correlator. It would be interesting to understand how to use MLT conditions to calculate the non-homogeneous contribution to a generic 3-point CFT correlator. It would also be interesting to understand how this condition can be used to constrain 4-point correlators.

\section*{Acknowledgments}
The work of SJ and RRJ is supported by the Ramanujan Fellowship. AM would like to acknowledge the support of CSIR-UGC.
(JRF) fellowship (09/936(0212)/2019-EMR-I). The work of AS is supported by the KVPY scholarship. We would like to thank S. Mukhi and N. Prabhakar for valuable discussions. We are also thankful to the referee of the paper for pointing out various minor errors and inaccuracies and for suggestions which led to a number of improvements in the presentation.
We acknowledge our debt to the people of India for their steady support of research in basic sciences.

\section*{Appendices}
\appendix
\section{Spinor-helicity notation}\label{shn}
In this appendix we will quickly summarise the spinor-helicity variables for 3d CFTs. For more details see \cite{Mata:2012bx, Farrow:2018yni} . We first embed the Euclidean 3-momentum $\vec k$ into a null momentum vector $k_\mu$ in 3+1 dimensions :
\begin{align}
k_\mu=(k,\vec k)
\end{align}
such that $k=|\vec k|$. Given the 4-momentum we express it in spinor notation as :
\begin{align}
k_{\alpha\dot\alpha}=k_\mu\sigma^\mu_{\dot\alpha\alpha}=\lambda_\alpha\widetilde\lambda_{\dot\alpha}
\end{align}
where $\alpha$ and $\dot\alpha$ are $SL(2,\mathbb C)$ transform under inequivalent (conjugate) representations of $SL(2,\mathbb C)$. However, in 3 dimensions one has an identification between the dotted and undotted indices. To see this let us consider the vector $\tau^\mu=(1,0,0,0)$. In spinor-helicity variables :
\begin{align}
\tau_{\alpha\dot\alpha}=\tau_\mu(\sigma^{\mu})_{\alpha\dot\alpha}=-\mathbb I_{\alpha\dot\alpha}
\end{align}
We can now convert dotted indices to undotted indices using the following tensor :
\begin{align}
\tau^{\dot\alpha}_{\alpha}=-\epsilon^{\dot\alpha\dot\beta}\mathbb I_{\dot\beta\alpha}
\end{align}
We also introduce the barred spinors as follows :
\begin{align}
\bar\lambda_\alpha\equiv\widetilde\lambda_{\dot\alpha}\tau^{\dot\alpha}_\alpha
\end{align}
We then have the following relations between the 3-momentum and the spinors.
\begin{align}
&\lambda_{\alpha} \bar{\lambda}_{\beta}=k_{i}\left(\hat{\sigma}^{i}\right)_{\alpha \beta}+k \epsilon_{\alpha \beta}\\[5 pt]
&k^{i}=\frac{1}{2}\left(\sigma^{i}\right)_{\beta}^{\alpha} \lambda_{\alpha} \bar{\lambda}^{\beta}
\end{align} 
Since $\epsilon_{\alpha \beta}$ is an $SL(2, \mathbb{C})$ invariant, we can use it to define dot products between spinors. 
\begin{align}
\begin{split}
\langle ij \rangle &= \epsilon^{\alpha\beta}\lambda^i_{\alpha}\lambda^j_{\beta}\\[5 pt]
\langle \overline{ij} \rangle&= \epsilon^{\alpha\beta}\bar{\lambda}^i_{\alpha}\bar{\lambda}^j_{\beta}\\[5 pt]
\langle i\bar{j} \rangle &= \epsilon^{\alpha\beta}\lambda^i_{\alpha}\bar{\lambda}^j_{\beta}
\end{split}
\end{align}
It can be also be used to raise and lower indices on the spinors for which we will use the following convention.
\begin{align}
\lambda_{\beta} = \epsilon_{\alpha\beta}\lambda^{\alpha}
\end{align}
The reader is referred to appendix B of \cite{Farrow:2018yni} or appendix C in \cite{Baumann:2020dch} which contains a set of useful relations between spinor brackets that will be used throughout the main text. Finally, we also define the following polarization vectors which when dotted with the momentum space expression of a correlator, gives its expression in spinor-helicity variables.
\begin{align}
z_{\alpha\beta}^-= \frac{\lambda_{\alpha}\lambda_{\beta}}{2k}\quad\quad\quad\quad z_{\alpha\beta}^+= \frac{\bar{\lambda}_{\alpha}\bar{\lambda}_{\beta}}{2k}
\end{align}


\section{Homogeneous $\&$ non-homogeneous vs transverse $\&$ longitudinal contributions}
\label{hnhvstl}
While computing momentum space correlation functions one often splits the correlator into its transverse and longitudinal parts \cite{Bzowski:2013sza}.
In this paper we find it more useful to split correlators into their homogeneous and non-homogeneous parts as defined in sec. \ref{hnh}. In this appendix we emphasise and illustrate through examples that the transverse and homogeneous parts of a correlator are not identical, and also that the longitudinal and non-homogeneous parts are not identical.  In particular, we will show that while the homogeneous part of a correlator is always transverse, the non-homogeneous part in general contains both transverse and longitudinal contributions and is proportional to 2-point function coefficients.

As an example consider $\langle TOO\rangle$. The correlator is given by \cite{Bzowski:2013sza}
\begin{align}
\langle TOO\rangle =\langle TOO\rangle_{\text{transverse}}+\langle TOO\rangle_{\text{longitudinal}}
\end{align}
where the transverse part is given by
\begin{align}
\langle TOO\rangle_{\text{transverse}}=\Pi^{\mu_1\nu_1}_{\alpha_1\beta_1}(k_1)A_1k_2^{\alpha_1}k_2^{\beta_1}
\end{align}
 For example when the scalar operator $O$  has scaling dimension $\Delta=1$  the form factor is given by \cite{Bzowski:2013sza}
\begin{align}
A_1=c_O\frac{2k_1+k_2+k_3}{k_2k_3(k_1+k_2+k_3)^2}.
\end{align}
The form-factor is proportional to the coefficient of the scalar two-point function $c_O$
$$\langle O(k)O(-k)\rangle_{\Delta=1}= c_O \frac{1}{k}.$$
The longitudinal part of the correlator  for $\Delta=1$ is
\begin{align}
\langle TOO\rangle_{\text{longitudinal}}=\left[k_2^{\alpha}\mathcal I^{\mu_1\nu_1}_{\alpha}(k_1)-\frac{1}{2}\pi^{\mu_1\nu_1}(k_1)\right]c_O\frac{1}{k_2}+k_2\leftrightarrow k_3
\end{align}
where 
\begin{align}
\mathcal I^{\mu\nu}_{\alpha}(k)=\frac{1}{k^2}\left[2p^{(\mu}\delta^{\nu)}_{\alpha}-\frac{k_{\alpha}}{2}\left(\delta^{\mu\nu}+\frac{k^{\mu}k^{\nu}}{k^2}\right)\right].
\end{align}
We see that the full correlator is proportional to the two-point function coefficient $c_O$.  Thus in our terminology the full answer is non-homogeneous and there is no homogeneous contribution to it. To summarize we have
\begin{align}
\langle TOO\rangle &=\langle TOO\rangle_{\text{transverse}}+\langle TOO\rangle_{\text{longitudinal}}\nonumber\\
&=\langle TOO\rangle_{\textbf{nh}}
\end{align}

Let us now consider the case of $\langle TTT\rangle$. The full answer in the terminology of \cite{Bzowski:2013sza} is given by
 \begin{align}
 \langle TTT\rangle = \langle TTT\rangle _{\text{transverse}}+ \langle TTT\rangle _{\text{longitudinal}}
 \end{align}
 which can as well be split into homogeneous and non-homogeneous pieces as follows
 \begin{align}
 \langle TTT\rangle &= \langle TTT\rangle _{\text{transverse}}+ \langle TTT\rangle _{\text{longitudinal}}\cr
 &=\langle TTT\rangle _{\text{transverse},\bf{h}}+\langle TTT\rangle _{\text{transverse},\bf{nh}}+\langle TTT\rangle _{\text{longitudinal}}\cr
 &=\langle TTT\rangle _{\bf{h}}+\langle TTT\rangle _{\bf{nh}}
 \end{align}
 where we made the following identification
 \begin{align}
 \langle TTT\rangle _{\bf{h}}&=\langle TTT\rangle _{\text{transverse},\bf{h}}\cr
 \langle TTT\rangle _{\bf{nh}}&=\langle TTT\rangle _{\text{transverse},\bf{nh}}+\langle TTT\rangle _{\text{longitudinal}}
 \end{align}
 We now give explicit identification of the homogeneous and non-homogeneous contribution.  To simplify the discussion, we make use of  transverse, null polarization vectors that are contracted with the free indices of the correlator. The longitudinal term drops out and what remains are the transverse pieces. For convenience we reproduce it here \cite{Farrow:2018yni,Jain:2021qcl}
  \begin{align}\label{ttttransapp}
\langle TTT\rangle_{\text{even}}&=\frac{C_1c_{123}}{E^6}\mathcal M_{W^3}
+2C_{TT}\left(\frac{c_{123}}{E^2}+\frac{b_{123}}{E}-E\right)\mathcal M_{EG}
\end{align}
 where $C_{TT}$ is defined by the two-point function
 \begin{align}
\langle T(k)T(-k)\rangle=C_{TT}(z_1\cdot z_2)^{2}k^3
 \end{align}
In the transverse correlator \eqref{ttttransapp}, the term proportional to $C_{TT}$ is non-homogeneous and the rest of it (the term proportional to $C_1$) is homogeneous.  To summarize, the term that is dependent on the two-point function coefficient (fixed by secondary conformal Ward identity in the language of \cite{Bzowski:2013sza}) is the non-homogeneous contribution. From the $dS_4$ perspective the interpretation is  that the term getting contribution from $W^3$ (term proportional to $C_1$) is homogeneous and the term getting contribution from Einstein-gravity $\sqrt{g} R$ (term proportional to $C_{TT}$) is non-homogeneous.

To conclude, the non-homogeneous part of the correlator can contain both transverse as well as longitudinal parts. From the $dS_4$ perspective as well, the origins of the homogeneous and non-homogeneous contributions are distinct.

\section{Details of solutions of CWIs for various correlators}\label{CWI}
In this appendix we provide  details of the calculations related to solving conformal Ward identities (CWIs) in spinor-helicity variables. 

\subsection{$\langle J_s O_{\Delta}O_{\Delta} \rangle$}
The details of the conformal Ward identities for this case were already given in Section \ref{jsododshsection}. Here we consider a few examples. The $s=1$ and the $s=2$ cases have already been computed in \cite{Bzowski:2013sza}. 

\subsection*{Example - Spin one current: $\langle J_{\mu}O_{\Delta}O_{\Delta} \rangle$ }
Setting $s=1$ in \eqref{Jsoofnl} we obtain :
\begin{equation}\begin{split}
  \langle J^- O_{\Delta}O_{\Delta} \rangle &=  c_O I_{\frac{3}{2}\{-\frac{1}{2},\Delta-\frac{3}{2},\Delta-\frac{3}{2}\}}\langle 12 \rangle \langle\bar{2}1\rangle\\
  &=  c_O I_{\frac{3}{2}\{-\frac{1}{2},\Delta-\frac{3}{2},\Delta-\frac{3}{2}\}} \frac{\langle 12 \rangle \langle 13 \rangle}{\langle 23 \rangle}(k_2+k_3-k_1)
\end{split}\end{equation}
We see that the correlator gets a minus sign under a $(2 \leftrightarrow 3)$ exchange. Therefore, this correlator is non-zero only when all the three operators have non-abelian indices. The non-abelian indices add an extra factor of $f^{abc}$ to the correlator which results in a plus sign under a $(2 \leftrightarrow 3)$ exchange. This result holds for any $\langle J_s O_{\Delta} O_{\Delta} \rangle$ whenever $s$ is odd. 
For the specific case of $\Delta=2$, the correlator is given by
\begin{align}
\langle J^- O_2 O_2 \rangle = c_O \frac{1}{k_1 E}\langle 12 \rangle \langle \bar{2}1 \rangle
\end{align}
The correlator is divergent for $\Delta \ge 3$ and needs to be renormalized for higher scaling dimensions.

\subsection*{Example - Spin two current: $\langle T_{\mu\nu}O_{\Delta}O_{\Delta} \rangle$ } 
Setting $s=2$ and $\Delta=2$ in \eqref{Jsoofnl} we obtain :
\begin{align}
\langle T^- O_2 O_2 \rangle = c_O \frac{E+k_1}{k_1^2 E^2}\langle 12 \rangle^2 \langle \bar{2}1 \rangle^2
\end{align}
Setting $s=2$ and $\Delta=3$ in \eqref{Jsoofnl} we obtain :
\begin{align}
\langle T^- O_3 O_3 \rangle = c_O \frac{k_1^2(E+k_2+k_3)+(E+k_1)(k_2^2+k_2k_3+k_3^2)}{k_1^2 E^2}\langle 12 \rangle^2 \langle \bar{2}1 \rangle^2
\end{align}
For $\Delta >3$, the correlator is divergent and needs to renormalized. 

\subsection{$\ev{J_sJ_s O_{\Delta}}$}
\label{appJsJsOdetails}
From the action of the special conformal generator on the scalar operator and conserved spin-$s$ currents \eqref{Kkappageneralscalar} and \eqref{Kkappaspinning}, we get the following : 
\begin{align}
\label{kkappaJJO}
&\widetilde{K}^{\kappa}\left\langle \frac{J^{s-}}{k_1^{s-1}}\frac{J^{s-}}{k_2^{s-1}}\frac{O_{\Delta}}{k_3^{\Delta-2}} \right\rangle = 2\bigg[\frac{z_1^{-\kappa}}{k_1^{s+1}k_2^{s-1} k_3^{\Delta-2}} \langle k_1\cdot J_s(k_1)J_s^-(k_2) O(k_3) \rangle\bigg. \nonumber\\[5 pt]
&\bigg.+\frac{z_2^{-\kappa}}{k_1^{s-1}k_2^{s+1} k_3^{\Delta-2}} \langle J_s^-(k_1) k_2\cdot J_s(k_2) O(k_3) \rangle+\frac{k_3^{\kappa}(\Delta-2)(\Delta-1)}{k_1^{s-1} k_2^{s-1} k_3^{\Delta}}\langle J_s^-J_s^-O_{\Delta}\rangle\bigg]
\end{align}
Making use of the trivial transverse Ward identity \eqref{Wtjsjso}, the first and the second terms on the RHS of the above equation drop out and we obtain :
\begin{align}\label{jsjskko}
\widetilde{K}^{\kappa}\left\langle \frac{J^{s-}}{k_1^{s-1}}\frac{J^{s-}}{k_2^{s-1}}\frac{O_{\Delta}}{k_3^{\Delta-2}} \right\rangle = \frac{k_3^{\kappa}}{k_1^{s-1} k_2^{s-1} k_3^{\Delta}}(\Delta-2)(\Delta-1)\langle J_s^-J_s^-O_{\Delta} \rangle
\end{align}
Contracting \eqref{jsjskko} with  $k_1z_1^{-\kappa}$ and with $k_2z_2^{-\kappa}$ we get the following equations for the parity even part of the correlator \eqref{ssoansatzsh} \footnote{ $z_1^{-\kappa}=\frac{(\sigma^{\kappa})^{\alpha\beta}\lambda_{1\alpha}\lambda_{1\beta}}{2k_1}$ } :
\begin{align}\label{ssoeqn1ae}
\left(\frac{\partial^2 {F}_1}{\partial k_2^2}-\frac{\partial^2 {F}_1}{\partial k_3^2}\right)&= -\frac{{F}_1}{k_3^{2}}(\Delta-1)(\Delta-2)\nonumber\\
\left(\frac{\partial^2{F}_1}{\partial k_3^2}-\frac{\partial^2 {F}_1}{\partial k_1^2}\right)&= -\frac{{F}_1}{k_3^{2}}(\Delta-1)(\Delta-2)\\[5 pt]
\frac{(k_2+k_3-k_1)}{4}\left(\frac{\partial^2 {G}_1}{\partial k_2^2}-\frac{\partial^2 {G}_1}{\partial k_3^2}\right)+s\frac{\partial {G}_1}{\partial k_2} &=-\frac{{G}_1}{k_3^{2}}(\Delta-1)(\Delta-2)(k_2+k_3-k_1) \nonumber\\[5 pt]
\frac{(k_1+k_3-k_2)}{4}\left(\frac{\partial^2 {G}_1}{\partial k_3^2}-\frac{\partial^2 {G}_1}{\partial k_1^2}\right)-s\frac{\partial {G}_1}{\partial k_1} &= -\frac{{G}_1}{k_3^{2}}(\Delta-1)(\Delta-2)(k_1+k_3-k_2)\label{ssoeqnGeven}
\end{align}
From the form of the ansatz for the correlator in \eqref{ssoansatzsh} and since the conformal Ward identity takes the form in \eqref{jsjskko}, the equations satisfied by the odd parts $F_2$ and $G_2$ of the correlator \eqref{ssoansatzsh} are identical to those for the even parts $F_1$ and $G_1$ respectively.

We note that the equation for $F_1$ (and $F_2$) is independent of the spin $s$.
The dependence on the spin comes through the dilation Ward identity and is given by :
\begin{equation}\label{dwrd1s}
\left(\sum_{i=1}^3 k_i \frac{\partial {F}_{1}}{\partial k_i}\right)-(\Delta-2(s+1)){F}_{1}=0
\end{equation}
The same equation is satisfied by $F_2$ as well.
The equations \eqref{ssoeqnGeven} for $G_1$ (and $G_2$) do not have a non-trivial solution. Solving \eqref{ssoeqn1ae} and \eqref{dwrd1s} we obtain the result in \eqref{ssoformfactor}.
\subsection*{Examples}
In the following we consider a few examples of the correlator $\langle J_sJ_sO_{\Delta}\rangle$ for specific values of $s$ and $\Delta.$
\subsection*{Spin one current: $\langle J_\mu J_\nu O_\Delta\rangle$}
Setting $s=1$ in the expression for the generic correlator \eqref{jjofinalanswersh} we obtain :
\begin{align}
\label{jjosphappendix}
\begin{split}
\langle J^- J^+ O_{\Delta} \rangle &= 0\\
\langle J^-J^- O_{\Delta} \rangle&=\langle J^-J^- O_{\Delta} \rangle_{\text{even}}+\langle J^-J^- O_{\Delta} \rangle_{\text{odd}}=\left(c_1+ i c_2\right)\,I_{\frac{5}{2}\{\frac{1}{2},\frac{1}{2},\Delta-\frac{3}{2}\}} \langle 12 \rangle^2\\[5 pt]
\langle J^+J^+ O_{\Delta} \rangle&=\langle J^+J^+ O_{\Delta} \rangle_{\text{even}}+\langle J^+J^+ O_{\Delta} \rangle_{\text{odd}}=\left(c_1 - i c_2\right)\,I_{\frac{5}{2}\{\frac{1}{2},\frac{1}{2},\Delta-\frac{3}{2}\}} \langle \bar 1\bar 2 \rangle^2\\[5 pt]
\end{split}
\end{align}
\subsubsection*{Example: $\Delta=1$}
When $\Delta=1$ we have :
\begin{align}
\begin{split}
&\langle J^-J^-O_1 \rangle_{\text{even}} = c_1 \frac{1}{k_3(k_1+k_2+k_3)^2}\langle 12 \rangle^2\\[5 pt]
&\langle J^-J^-O_1 \rangle_{\text{odd}} = ic'_1 \frac{1}{k_3(k_1+k_2+k_3)^2}\langle 12 \rangle^2
\end{split}
\begin{split}
&\langle J^-J^+O_1 \rangle_{\text{even}} =0\\[5 pt]
&\langle J^-J^+O_1 \rangle_{\text{odd}} =0
\end{split}
\end{align}
\subsubsection*{Example: $\Delta=2$}
When $\Delta=2$ we have :
\begin{align}
\begin{split}
&\langle J^-J^-O_2 \rangle_{\text{even}} = c_1 \frac{1}{(k_1+k_2+k_3)^2}\langle 12 \rangle^2\\[5 pt]
&\langle J^-J^-O_2 \rangle_{\text{odd}} = ic'_1 \frac{1}{(k_1+k_2+k_3)^2}\langle 12 \rangle^2
\end{split}
\begin{split}
&\langle J^-J^+O_2 \rangle_{\text{even}} =0\\[5 pt]
&\langle J^-J^+O_2 \rangle_{\text{odd}} =0
\end{split}
\end{align}
\subsubsection*{Example: $\Delta=3$}
When $\Delta=3$ we have :
\begin{align}
\begin{split}
&\langle J^-J^-O_3 \rangle_{\text{even}} = c_1 \frac{k_1+k_2+2k_3}{(k_1+k_2+k_3)^2}\langle 12 \rangle^2\\[5 pt]
&\langle J^-J^-O_3 \rangle_{\text{odd}} = ic'_1 \frac{k_1+k_2+2k_3}{(k_1+k_2+k_3)^2}\langle 12 \rangle^2
\end{split}
\begin{split}
&\langle J^-J^+O_3 \rangle_{\text{even}} =0\\[5 pt]
&\langle J^-J^+O_3 \rangle_{\text{odd}} =0
\end{split}
\end{align}
We see that the solution for $\Delta=1$ is just the shadow transform of the $\Delta=2$ solution. In Section \ref{shmomentumspaceresults} we convert this answer to momentum space and check that it matches the known answer previously computed in \cite{Jain:2021wyn}. 

\subsection*{Spin Two current : $\langle TTO_\Delta\rangle$}
Setting $s=2$ in the expression for the generic correlator \eqref{jjofinalanswersh} we obtain :
\begin{align}
\label{ttodeltasph}
\begin{split}
\langle T^- T^+ O_{\Delta} \rangle &= 0\\
\langle T^-T^- O_{\Delta} \rangle&=\langle T^-T^- O_{\Delta} \rangle_{\text{even}}+\langle T^-T^- O_{\Delta} \rangle_{\text{odd}}=\left(c_1+ i c_2\right)\,k_1 k_2 I_{\frac{9}{2}\{\frac{1}{2},\frac{1}{2},\Delta-\frac{3}{2}\}} \langle 12 \rangle^4\\[5 pt]
\langle T^+T^+ O_{\Delta} \rangle&=\langle T^+ T^+ O_{\Delta} \rangle_{\text{even}}+\langle T^+ T^+ O_{\Delta} \rangle_{\text{odd}}=\left(c_1 - i c_2\right)\,k_1 k_2 I_{\frac{9}{2}\{\frac{1}{2},\frac{1}{2},\Delta-\frac{3}{2}\}} \langle \bar 1\bar 2 \rangle^4\\[5 pt]
\end{split}
\end{align}
\subsection*{Example: $\Delta=1$}
When $\Delta=1$ we have :
\begin{align}
\begin{split}
&\langle T^-T^-O_1 \rangle_{\text{even}} = c_1 k_1 k_2\frac{1}{k_3(k_1+k_2+k_3)^4}\langle 12 \rangle^4\\[5 pt]
&\langle T^-T^-O_1\rangle_{\text{odd}} = ic'_1 k_1k _2\frac{1}{k_3(k_1+k_2+k_3)^4}\langle 12 \rangle^4
\end{split}
\begin{split}
&\langle T^-T^+O_1 \rangle_{\text{even}} =0\\[5 pt]
&\langle T^-T^+O_1 \rangle_{\text{odd}} =0
\end{split}
\end{align}
\subsection*{Example: $\Delta=2$}
When $\Delta=2$ we have :
\begin{align}
\begin{split}
&\langle T^-T^-O_2 \rangle_{\text{even}} = c_1 k_1 k_2\frac{1}{(k_1+k_2+k_3)^4}\langle 12 \rangle^4\\[5 pt]
&\langle T^-T^-O_2\rangle_{\text{odd}} = ic'_1 k_1k _2\frac{1}{(k_1+k_2+k_3)^4}\langle 12 \rangle^4
\end{split}
\begin{split}
&\langle T^-T^+O_2 \rangle_{\text{even}} =0\\[5 pt]
&\langle T^-T^+O_2 \rangle_{\text{odd}} =0
\end{split}
\end{align}
\subsection*{Example: $\Delta=3$}
When $\Delta=3$ we have :
\begin{align}
\begin{split}
&\langle T^-T^-O_3 \rangle_{\text{even}} = c_1 k_1 k_2\frac{k_1+k_2+4k_3}{(k_1+k_2+k_3)^4}\langle 12 \rangle^4\\[5 pt]
&\langle T^-T^-O_3 \rangle_{\text{odd}} = ic'_1 k_1 k_2\frac{k_1+k_2+4k_3}{(k_1+k_2+k_3)^4}\langle 12 \rangle^4
\end{split}
\begin{split}
&\langle T^-T^+O_3 \rangle_{\text{even}} =0\\[5 pt]
&\langle T^-T^+O_3 \rangle_{\text{odd}} =0
\end{split}
\end{align}
Again, we see that the $\Delta=1$ solution and the $\Delta=2$ solution are just shadow transforms of each other. For $\Delta \ge 6,$ the triple-$K$ integrals show a divergence and the correlators need to be renormalized.

\subsection*{Higher spin example}
Let us now discuss a few correlators involving higher spin conserved currents.
When the scalar operator $O_\Delta$ has scaling dimension $\Delta=3$ and the conserved current operator $J_s$ has spin $s=3$, we have from \eqref{jjofinalanswersh} :
\begin{align}
\label{jjo3}
\langle J^{3-} J^{3-} O_{3} \rangle& =(c_1+i\, c_2)  (k_1 k_2)^{2}I_{\frac{13}{2}\{\frac{1}{2},\frac{1}{2},\frac{3}{2}\}}\nonumber\\[5pt]
&=(c_1+i\, c_2)  (k_1 k_2)^2\frac{E+5k_3}{E^6}\langle 12\rangle^6
\end{align}
When the scalar operator $O_\Delta$ has scaling dimension $\Delta=3$ and the conserved current operator $J_s$ has spin $s=4$, we have from \eqref{jjofinalanswersh} :
\begin{align}
\label{jjo4}
\langle J^{4-} J^{4-} O_{3} \rangle &=(c_1+i\, c_2)  (k_1 k_2)^{4}I_{\frac{17}{2}\{\frac{1}{2},\frac{1}{2},\frac{3}{2}\}}\langle 12\rangle^8\nonumber\\[5pt]
&=(c_1+i\, c_2)(k_1 k_2)^4\frac{E+7k_3}{E^8}\langle 12\rangle^8
\end{align}
We can also get the parity even part of the above two results using weight-shifting and spin-raising operators in momentum space \cite{Baumann:2019oyu, Jain:2021wyn} and then converting the answer into spinor-helicity variables :
\begin{align}
\langle J^3 J^3 O_3 \rangle = (k_1 k_2)^2P^{(3)}_1 P^{(3)}_2 H_{12}^3 \langle O_2 O_2 O_3 \rangle\\[5 pt]
\langle J^4 J^4 O_3 \rangle = (k_1 k_2)^3P^{(4)}_1 P^{(4)}_2 H_{12}^4 \langle O_2 O_2 O_3 \rangle
\end{align}
where $P^{(s)}_{i}$ are spin-$s$ projectors transverse to $k_i$ and $H_{12}$ is a bilocal operator that raises the spin of the operators at insertions 1 and 2.
It can be verified that the answers obtained this way match the answers in \eqref{jjo3} and \eqref{jjo4}.

\subsection{$\ev{JJJ}$}
\label{CWIjjj}
The ansatz for the correlator is given in \eqref{jjjevenansatzab}. We will analyze the parity-odd and the parity-even parts separately here as they have different WT identities. 
\subsubsection*{$\langle JJJ \rangle_{\text{even}}$}\label{jjjevenansatz}
\begin{align}
\langle J^-(k_1) J^-(k_2)J^-(k_3) \rangle_{\text{even}} &=F_1(k_1, k_2, k_3)\langle 12 \rangle \langle 23 \rangle \langle 31 \rangle\\
\langle J^-(k_1) J^-(k_2)J^+(k_3) \rangle_{\text{even}} &=G_1(k_1, k_2, k_3)\langle 12 \rangle \langle 2\bar{3} \rangle \langle \bar{3}1 \rangle
\end{align}
The action of the conformal generator is given by :
\begin{align}\label{Kkappajjj}
\widetilde{K}^{\kappa} \langle J^- J^- J^- \rangle &= 2\left(z_1^{-\kappa}\frac{k_{1\mu}}{k_1^2}\langle J^{\mu} J^- J^- \rangle+z_2^{-\kappa}\frac{k_{2\mu}}{k_2^2}\langle J^- J^{\mu} J^- \rangle+z_3^{-\kappa}\frac{k_{3\mu}}{k_3^2}\langle J^- J^- J^{\mu} \rangle\right)\nonumber\\[5 pt]
\widetilde{K}^{\kappa} \langle J^- J^- J^+ \rangle &=2\left(z_1^{-\kappa}\frac{k_{1\mu}}{k_1^2}\langle J^{\mu} J^- J^+\rangle+z_2^{-\kappa}\frac{k_{2\mu}}{k_2^2}\langle J^- J^{\mu} J^+ \rangle+z_3^{+\kappa}\frac{k_{3\mu}}{k_3^2}\langle J^- J^- J^{\mu} \rangle\right)
\end{align}
The transverse Ward identities of $\langle JJJ\rangle$ \cite{Bzowski:2017poo} are non-trivial :
\begin{align}\label{WTJJJ}
    \frac{k_{1\mu}}{k_1^2}\langle J^{\mu} J^- J^- \rangle_{\text{even}}&= c_J\frac{1}{k_1^2 k_2 k_3}\langle 23 \rangle^2(k_3-k_2)\notag \\[5 pt]
    \frac{k_{1\mu}}{k_1^2}\langle J^{\mu} J^- J^+\rangle_{\text{even}}&= c_J\frac{1}{k_1^2 k_2 k_3}\langle 2\bar{3} \rangle^2(k_3-k_2)
    \end{align}
Using \eqref{WTJJJ} in the R.H.S. of \eqref{Kkappajjj} we obtain :
\begin{align}
\widetilde{K}^{\kappa} \langle J^- J^- J^- \rangle_{\text{even}} &= z_1^{-\kappa} c_J\frac{\langle 23 \rangle^2}{k_1^2 k_2 k_3}(k_2-k_3)+\text{cyclic perm.}\\[5 pt]
 \widetilde{K}^{\kappa} \langle J^- J^- J^+ \rangle_{\text{even}} &= z_1^{-\kappa}c_J\frac{\langle 2\bar{3} \rangle^2}{k_1^2 k_2 k_3}(k_2-k_3)+\text{cyclic perm.}
\end{align}
Expanding out the left hand side and dotting with $(\sigma^{\kappa})_{\alpha}^{\beta}(\lambda_2^{\alpha}\lambda_{3\beta}+\lambda_{2\beta}\lambda_3^{\alpha})$ gives us the following equations for the form factors :
\begin{align}
2\left(\frac{\partial F_1}{\partial k_2}-\frac{\partial F_1}{\partial k_3}\right)+k_2\left(\frac{\partial^2 F_1}{\partial k_2^2}-\frac{\partial^2 F_1}{\partial k_1^2}\right)+k_3\left(\frac{\partial^2 F_1}{\partial k_1^2}-\frac{\partial^2 F_1}{\partial k_3^2}\right)&=2c_J\frac{(k_3-k_2)}{k_1^3 k_2 k_3}\label{Feqn1}\\[5 pt]
2\left(\frac{\partial G_1}{\partial k_2}+\frac{\partial G_1}{\partial k_3}\right)+k_2\left(\frac{\partial^2 G_1}{\partial k_2^2}-\frac{\partial^2 G_1}{\partial k_1^2}\right)-k_3\left(\frac{\partial^2 G_1}{\partial k_1^2}-\frac{\partial^2 G_1}{\partial k_3^2}\right)&=2 c_J\frac{(k_3-k_2)}{k_1^3 k_2 k_3}\label{Geqn1}
\end{align}
Similarly, dotting with $(\sigma^{\kappa})_{\alpha}^{\beta}(\lambda_1^{\alpha}\lambda_{3\beta}+\lambda_{1\beta}\lambda_3^{\alpha})$ gives :
\begin{align}
2\left(\frac{\partial F_1}{\partial k_1}-\frac{\partial F_1}{\partial k_3}\right)-k_1\left(\frac{\partial^2 F_1}{\partial k_2^2}-\frac{\partial^2 F_1}{\partial k_1^2}\right)+k_3\left(\frac{\partial^2 F_1}{\partial k_2^2}-\frac{\partial^2 F_1}{\partial k_3^2}\right)&=2c_J\frac{(k_3-k_1)}{k_1 k_2^3 k_3}\label{Feqn2}\\[5 pt]
2\left(\frac{\partial G_1}{\partial k_1}+\frac{\partial G_1}{\partial k_3}\right)-k_1\left(\frac{\partial^2 G_1}{\partial k_2^2}-\frac{\partial^2 G_1}{\partial k_1^2}\right)-k_3\left(\frac{\partial^2 G_1}{\partial k_2^2}-\frac{\partial^2 G_1}{\partial k_3^2}\right)&=2c_J\frac{(k_3-k_1)}{k_1 k_2^3 k_3}\label{Geqn2}
\end{align}
The dilatation Ward identity is given by
\begin{align}
\begin{split}
\left(\sum_{i=1}^3 k_i \frac{\partial F_1}{\partial k_i}\right)+3F_1=0,
\end{split}
\begin{split}
\left(\sum_{i=1}^3 k_i \frac{\partial G_1}{\partial k_i}\right)+3G_1=0
\end{split}
\end{align}
Solving these equations we obtain $F_1(k_1, k_2, k_3)$ and $G_1(k_1, k_2, k_3)$ in \eqref{F1G1F2G2JJJ}.
\subsubsection*{$\langle JJJ \rangle_{\text{odd}}$}
We now turn our attention to the odd part of the correlator. The ansatz is given by
\begin{align}\label{jjjoddansatz}
    \langle J^-(k_1) J^-(k_2)J^-(k_3) \rangle_{\text{odd}} &=iF_2(k_1, k_2, k_3)\langle 12 \rangle \langle 23 \rangle \langle 31 \rangle\\
\langle J^-(k_1) J^-(k_2)J^+(k_3) \rangle_{\text{odd}} &=iG_2(k_1, k_2, k_3)\langle 12 \rangle \langle 2\bar{3} \rangle \langle \bar{3}1 \rangle
\end{align}
The transverse WT identity in this case is given by :
\begin{align}\label{WTJJJodd}
   \frac{k_{1\mu}}{k_1^2}\langle J^{\mu} J^- J^- \rangle_{\text{odd}}&= c'_J\frac{1}{k_1^2 k_2 k_3}\langle 23 \rangle^2(k_3-k_2)\notag \\[5 pt]
    \frac{k_{1\mu}}{k_1^2}\langle J^{\mu} J^- J^+\rangle_{\text{odd}}&= c'_J\frac{1}{k_1^2 k_2 k_3}\langle 2\bar{3} \rangle^2(k_3+k_2)
\end{align}
Substituting \eqref{WTJJJodd} into the right hand side of the conformal identity \eqref{Kkappajjj}, we get :
\begin{align}
\begin{split}
\widetilde{K}^{\kappa} \langle J^- J^- J^- \rangle_{\text{odd}} &= z_1^{-\kappa}i\,c_{J}'\frac{\langle 23 \rangle^2}{k_1^2 k_2 k_3}(k_2-k_3)+\text{cyclic perm.}\\[5 pt]
\widetilde{K}^{\kappa} \langle J^- J^- J^+ \rangle_{\text{odd}} &= z_1^{-\kappa}i\,c_{J}'\frac{\langle 2\bar{3} \rangle^2}{k_1^2 k_2 k_3}(k_2+k_3)+\text{cyclic perm.}
\end{split}
\end{align}
Following the same procedure as in the parity-even case, we get :
\begin{align}
2\left(\frac{\partial F_2}{\partial k_2}-\frac{\partial F_2}{\partial k_3}\right)+k_2\left(\frac{\partial^2 F_2}{\partial k_2^2}-\frac{\partial^2 F_2}{\partial k_1^2}\right)+k_3\left(\frac{\partial^2 F_2}{\partial k_1^2}-\frac{\partial^2 F_2}{\partial k_3^2}\right)&=2c_J'\frac{(k_3-k_2)}{k_1^3 k_2 k_3}\label{Feqn1od}\\[5 pt]
2\left(\frac{\partial G_2}{\partial k_2}+\frac{\partial G_2}{\partial k_3}\right)+k_2\left(\frac{\partial^2 G_2}{\partial k_2^2}-\frac{\partial^2 G_2}{\partial k_1^2}\right)-k_3\left(\frac{\partial^2 G_2}{\partial k_1^2}-\frac{\partial^2 G_2}{\partial k_3^2}\right)&=2 c_J'\frac{(k_3+k_2)}{k_1^3 k_2 k_3}\label{Geqn1od}
\end{align}
and 
\begin{align}
2\left(\frac{\partial F_2}{\partial k_1}-\frac{\partial F_2}{\partial k_3}\right)-k_1\left(\frac{\partial^2 F_2}{\partial k_2^2}-\frac{\partial^2 F_2}{\partial k_1^2}\right)+k_3\left(\frac{\partial^2 F_2}{\partial k_2^2}-\frac{\partial^2 F_2}{\partial k_3^2}\right)&=2c_J'\frac{(k_3-k_1)}{k_1 k_2^3 k_3}\label{Feqn2od}\\[5 pt]
2\left(\frac{\partial G_2}{\partial k_1}+\frac{\partial G_2}{\partial k_3}\right)-k_1\left(\frac{\partial^2 G_2}{\partial k_2^2}-\frac{\partial^2 G_2}{\partial k_1^2}\right)-k_3\left(\frac{\partial^2 G_2}{\partial k_2^2}-\frac{\partial^2 G_2}{\partial k_3^2}\right)&=2c_J'\frac{(k_3+k_1)}{k_1 k_2^3 k_3}\label{Geqn2od}
\end{align}
Let us note that \eqref{Feqn1od}, \eqref{Feqn2od} are exactly identical to \eqref{Feqn1}, \eqref{Feqn2}, whereas comparing \eqref{Geqn1od}, \eqref{Geqn2od} with \eqref{Geqn1}, \eqref{Geqn2}, we see that the r.h.s. of the equations are different. Solving these equations we obtain $F_2(k_1, k_2, k_3)$ and $G_2(k_1, k_2, k_3)$ in \eqref{F1G1F2G2JJJ}.



\subsection{$\ev{TTT}$}
\label{CWItttapp}
The even part of this correlator was obtained earlier in \cite{Baumann:2020dch, Bzowski:2013sza}. We focus on obtaining the odd part. 
\subsubsection*{$\langle TTT \rangle_{\text{odd}}$}
We start with the following ansatz for $\langle TTT \rangle_{\text{odd}}$ : 
\begin{align}
\left\langle \frac{T^-}{k_1}\frac{T^-}{k_2} \frac{T^-}{k_3} \right\rangle_{\text{odd}} &=i\,F(k_1,k _2, k_3)\langle 12 \rangle^2 \langle 23 \rangle^2 \langle 31 \rangle^2\\[6 pt]
\left\langle \frac{T^-}{k_1} \frac{T^-}{k_2} \frac{T^+}{k_3} \right\rangle_{\text{odd}} &=i\,G(k_1,k _2, k_3)\langle 12 \rangle^2 \langle 2\bar{3} \rangle^2 \langle \bar{3}1 \rangle^2
\end{align}
The action of the conformal generator is given by :
\begin{align}\label{Kkappattt}
\widetilde{K}^{\kappa}\left\langle \frac{T^-}{k_1}\frac{T^-}{k_2}\frac{T^-}{k_3} \right\rangle &= 12 z^-_{1\kappa} \frac{k_{(1\mu}z_{1\nu)}^-}{k_1^3}\left\langle T^{\mu\nu} \frac{T^-}{k_2} \frac{T^-}{k_3} \right\rangle+12 z^-_{2\kappa} \frac{k_{(2\mu}z_{2\nu)}^-}{k_2^3}\left\langle \frac{T^-}{k_1} T^{\mu\nu} \frac{T^-}{k_3} \right\rangle\notag\\[5 pt]
&+12 z^-_{3\kappa} \frac{k_{(3\mu}z_{3\nu)}^-}{k_3^3}\left\langle \frac{T^-}{k_1} \frac{T^-}{k_2} T^{\mu\nu} \right\rangle \notag\\[5 pt]
\widetilde{K}^{\kappa}\left\langle \frac{T^-}{k_1}\frac{T^-}{k_2}\frac{T^+}{k_3} \right\rangle &= 12 z^-_{1\kappa} \frac{k_{(1\mu}z_{1\nu)}^-}{k_1^3}\left\langle T^{\mu\nu} \frac{T^-}{k_2} \frac{T^+}{k_3} \right\rangle+12 z^-_{2\kappa} \frac{k_{(2\mu}z_{2\nu)}^-}{k_2^3}\left\langle \frac{T^-}{k_1} T^{\mu\nu} \frac{T^+}{k_3} \right\rangle\notag\\[5 pt]
&+12 z^+_{3\kappa} \frac{k_{(3\mu}z_{3\nu)}^+}{k_3^3}\left\langle \frac{T^-}{k_1} \frac{T^-}{k_2} T^{\mu\nu} \right\rangle
\end{align}
Using \eqref{wtidttt} we find for parity odd contribution
\begin{align}\label{WTTT}
    \frac{k_{(1\mu}z_{1\nu)}^-}{k_1^3}\left\langle T^{\mu\nu} \frac{T^-}{k_2} \frac{T^-}{k_3} \right\rangle&= E \frac{\langle 12 \rangle \langle 23 \rangle^3\langle 31 \rangle}{k_1^4k_2^3k_3^3}(k_3^3-k_2^3)\notag \\[5 pt]
      \frac{k_{(1\mu}z_{1\nu)}^-}{k_1^3}\left\langle T^{\mu\nu} \frac{T^-}{k_2} \frac{T^+}{k_3} \right\rangle&= (E-2k_3) \frac{\langle 12 \rangle \langle 2\bar{3} \rangle^3\langle \bar{3}1 \rangle}{k_1^4k_2^3k_3^3}(k_3^3+k_2^3)
\end{align}
The action of $\widetilde{K}^{\kappa}$ on the ansatz, after dotting with $b_{\kappa}=(\sigma_{\kappa})_{\alpha}^{\beta}(\lambda_2^{\alpha}\lambda_{3\beta}+\lambda_{2\beta}\lambda_3^{\alpha})$, becomes
\begin{align}
4\left(\frac{\partial F}{\partial k_2}-\frac{\partial F}{\partial k_3}\right)+k_3\left(\frac{\partial^2 F}{\partial k_1^2}-\frac{\partial^2 F}{\partial k_3^2}\right)-k_2\left(\frac{\partial^2 F}{\partial k_1^2}-\frac{\partial^2 F}{\partial k_2^2}\right)&= c_T'\frac{E (k_2^3-k_3^3)}{k_1^2(k_1 k_2 k_3)^3}\\[5 pt]
4\left(\frac{\partial G}{\partial k_2}+\frac{\partial G}{\partial k_3}\right)-k_3\left(\frac{\partial^2 G}{\partial k_1^2}-\frac{\partial^2 G}{\partial k_3^2}\right)-k_2\left(\frac{\partial^2 G}{\partial k_1^2}-\frac{\partial^2 G}{\partial k_2^2}\right)&= c_T'\frac{(E-2k_3) (k_2^3+k_3^3)}{k_1^2(k_1 k_2 k_3)^3}
\end{align}
The dilatation Ward identity is given by
\begin{align}
\left(\sum_{i=1}^3 k_i \frac{\partial F}{\partial k_i}\right)+6F=0,\quad
\left(\sum_{i=1}^3 k_i \frac{\partial G}{\partial k_i}\right)+6G=0
\end{align}
The solutions for $F$ and $G$ are then given by :
\begin{align}
\label{tttsh}
F(k_1, k_2, k_3) &= \frac{c_1'}{E^6}+c_T'\frac{E^3-E\,b_{123}-c_{123}}{c_{123}^3}\\[5 pt]
G(k_1, k_2, k_3) &= c_T'\frac{(E-2k_3)^3-(E-2k_3)(b_{123}-2k_3\,a_{12})+c_{123}}{c_{123}^3}
\end{align}
where  $a_{12}=k_1+k_2$, $b_{123}=k_1k_2+k_2k_3+k_1k_3$ and $c_{123}=k_1\,k_2\,k_3$.


\subsection{$\ev{TJJ}$}
We once again focus on only the odd part of the correlator. 
Since we have shown that the transverse WT identities are trivial in  \eqref{tjjoddwi}and \eqref{tjjoddwifnl}, the action of $\widetilde{K}^{\kappa}$ on the ansatz \eqref{tjjansatz} becomes :
\begin{align}
\widetilde{K}^{\kappa}\left\langle \frac{T^-}{k_1}J^-J^- \right\rangle_{\text{odd}}&=0\nonumber\\
\widetilde{K}^{\kappa}\left\langle \frac{T^-}{k_1}J^-J^+ \right\rangle_{\text{odd}}&=0.
\end{align}
Expanding out the l.h.s. and dotting with an appropriate $b_{\kappa}=(\sigma_{\kappa})_{\alpha}^{\beta}(\lambda_2^{\alpha}\lambda_{3\beta}+\lambda_{2\beta}\lambda_3^{\alpha})$, we get
\begin{align}
&k_3\left(\frac{\partial^2 F}{\partial k_1^2}-\frac{\partial^2 F}{\partial k_3^2}\right)-k_2\left(\frac{\partial^2 F}{\partial k_1^2}-\frac{\partial^2 F}{\partial k_2^2}\right)+2\left(\frac{\partial F}{\partial k_2}-\frac{\partial F}{\partial k_3}\right)=0 \\[5 pt]
&k_3\left(\frac{\partial^2 G}{\partial k_1^2}-\frac{\partial^2 G}{\partial k_3^2}\right)-k_2\left(\frac{\partial^2 G}{\partial k_1^2}-\frac{\partial^2 G}{\partial k_2^2}\right)+2\left(\frac{\partial G}{\partial k_2}+\frac{\partial G}{\partial k_3}-2\frac{\partial G}{\partial k_1}\right)=0
\end{align}
The solutions to these are given by \eqref{tjjoddformfactorssh}.


\subsection{$\ev{J_{s_1} J_s J_s}$}
Dotting \eqref{Kkappajs1jsjs} with $b_{\kappa} = (\sigma^{\kappa})\lambda_{1\alpha}\lambda_1^{\;\;\beta}$, we get :
\begin{align}\label{js1jsjseqn1}
&(-k_1+k_2+k_3)\left(\frac{\partial^2 F}{\partial k_2^2}-\frac{\partial^2 F}{\partial k_3^2}\right)+2(2s-s_1)\left(\frac{\partial F}{\partial k_2}-\frac{\partial F}{\partial k_3}\right)=0\notag\\[5 pt]
&(-k_1+k_2-k_3)\left(\frac{\partial^2 H}{\partial k_2^2}-\frac{\partial^2 H}{\partial k_3^2}\right)+2(2s-s_1)\left(\frac{\partial H}{\partial k_2}+\frac{\partial H}{\partial k_3}\right)=0
\end{align}
Similarly, dotting \eqref{Kkappajs1jsjs} with $b_{\kappa} = (\sigma^{\kappa})\lambda_{2\alpha}\lambda_2^{\;\;\beta}$, we get :
\begin{align}\label{js1jsjseqn2}
&(k_1-k_2+k_3)\left(\frac{\partial^2 F}{\partial k_3^2}-\frac{\partial^2 F}{\partial k_1^2}\right)+2s_1\left(\frac{\partial F}{\partial k_3}-\frac{\partial F}{\partial k_1}\right)=0\notag\\[5 pt]
&(k_3-k_2-k_1)\left(\frac{\partial^2 G}{\partial k_3^2}-\frac{\partial^2 G}{\partial k_1^2}\right)+2s_1\left(\frac{\partial G}{\partial k_3}+\frac{\partial G}{\partial k_1}\right)=0
\end{align}
The dilatation Ward identity is given by
\begin{align}
&\left(\sum_{i=1}^3 k_i \frac{\partial F}{\partial k_i}\right)+(2s+s_1)F=0\notag\\[5 pt]
&\left(\sum_{i=1}^3 k_i \frac{\partial G}{\partial k_i}\right)+(2s+s_1)G=0\notag\\[5 pt]
&\left(\sum_{i=1}^3 k_i \frac{\partial H}{\partial k_i}\right)+(2s+s_1)H=0
\end{align}
We have considered only one equation for $G$ and $H$ as these by themselves imply that there is no homogeneous solution for the two form factors. The solutions for $F$, $G$ and $H$ are then given by \eqref{s1ssJ}.


\section{Identities involving Triple-$K$ integrals}\label{idtrk}
In this section we obtain non-trivial identities involving triple-$K$ integrals by matching our results obtained for the correlator in spinor-helicity variables to the results obtained for the same in momentum space after converting to spinor-helicity variables. 

Let us first consider the correlator $\langle JJO_\Delta\rangle$. We will work in a convenient regularisation scheme in which we set $u=v_1=v_2=0$ and $v_3\ne 0$. The momentum space expression for the correlator after converting to spinor-helicity variables takes the following form :
\begin{align}
\langle J^-J^-O\rangle=-\frac{2A_2+A_1\left[(k_1-k_2)^2-k_3^2\right]}{4k_1k_2}\langle 12\rangle^2
\end{align}
where \cite{Bzowski:2018fql} :
\begin{align}
A_1&=c_1I_{\frac 52,\{\frac 12,\frac 12,\Delta-\frac 32+v_3\epsilon\}}\cr
A_2&=c_1I_{\frac 32,\{\frac 12,\frac 12,\Delta-\frac 12+v_3\epsilon\}}+c_1\frac {\Delta}{2}(1-\Delta)\,I_{\frac 12,\{\frac 12,\frac 12,\Delta-\frac 32+v_3\epsilon\}}
\end{align}
Comparing with our results for the same correlator obtained by solving the conformal Ward identities directly in spinor-helicity variables \eqref{jjosphappendix} we get the following identity involving triple$-K$ integrals which we have verified to $O(1)$ in the regulator the following relation :
\begin{align}
-\frac{2A_2+A_1\left[(k_1-k_2)^2-k_3^2\right]}{4k_1k_2}=c_1I_{\frac 52,\{\frac 12,\frac 12,\Delta-\frac 32+v_3\epsilon\}}
\end{align}
Let us now consider the correlator $\langle TTO_\Delta\rangle$. The momentum space expression for the correlator after converting to spinor-helicity variables takes the following form :

\begin{align}
\langle T^-T^-O\rangle=\frac{4A_3+\left[(k_1-k_2)^2-k_3^2\right]\left[2A_2+A_1((k_1-k_2)^2-k_3^2)\right]}{16k_1^2k_2^2}\langle 12\rangle^4
\end{align}
We will continue to work in the scheme where $u=v_1=v_2=0$ and only $v_3$ is non-zero and in this scheme the form factors are given by \cite{Bzowski:2018fql} :
\begin{align}
A_1&=c_1\,I_{\frac 92,\{\frac 32,\frac 32,\Delta-\frac 32+v_3\epsilon\}}\cr
A_2&=4c_1\,I_{\frac 72,\{\frac 32,\frac 32,\Delta-\frac 12+v_3\epsilon\}}+c_2\,I_{\frac 52,\{\frac 32,\frac 32,\Delta-\frac 32+v_3\epsilon\}}\cr
A_3&=2c_1\,I_{\frac 52,\{\frac 32,\frac 32,\Delta+\frac 12+v_3\epsilon\}}+c_2\,I_{\frac 32,\{\frac 32,\frac 32,\Delta-\frac 12+v_3\epsilon\}}+c_3\,I_{\frac 12,\{\frac 32,\frac 32,\Delta-\frac 32+v_3\epsilon\}}
\end{align}
where
\begin{align}
c_2&=c_1(1-\Delta-v_3\epsilon)(\Delta+2+v_3\epsilon)\cr
c_3&=\frac{c_1}{4}(\Delta-3+v_3\epsilon)(\Delta-1+v_3\epsilon)(\Delta+v_3\epsilon)(\Delta+2+v_3\epsilon)
\end{align}
Matching with our answers obtained by solving conformal Ward identities in spinor-helicity variables \eqref{ttodeltasph} we obtain the following identity for triple-$K$ integrals we have verified to $O(1)$ in the regulator :
\begin{align}
&{4A_3+\left[(k_1-k_2)^2-k_3^2\right]\left[2A_2+A_1((k_1-k_2)^2-k_3^2)\right]}=16c_1k_1^3\,k_2^3\,I_{\frac 92,\{\frac 12,\frac 12,\Delta-\frac 32+v_3\epsilon\}}
\end{align}

\section{Higher-spin momentum space correlators}\label{higherspin-correlator}
In this section we summarise the momentum space expression for the parity-even and parity-odd homogeneous parts of higher spin correlators using the results of section \ref{shmomentumspaceresults}, see also Appendix D of \cite{Jain:2021qcl}.

For $\ev{J_s J_s O_2}$ we have
\begin{align}\label{spinsintermsofspin1}
\langle J_sJ_sO_2\rangle_{\text{even},\bf{h}}
&=(k_1 k_2)^{s-1}\left[\frac{1}{E^2} \left\{2(\vec{z}_1\cdot \vec{k}_2)( \vec{z}_2\cdot \vec{k}_1) +E (E-2k_3)\vec{z}_1\cdot \vec{z}_2 \right\}\right]^s\cr
\langle J_sJ_sO_2\rangle_{\text{odd},\bf{h}}
&= (k_1 k_2)^{s-1}\frac{1}{E^{2s}}\left[k_2\,\epsilon^{ k_1 z_1 z_2}- k_1\,\epsilon^{k_2 z_1  z_2}\right]\cr
&\hspace{.5cm}\times \left[2(\vec{z}_1\cdot \vec{k}_2)( \vec{z}_2\cdot \vec{k}_1) +E (E-2k_3)\vec{z}_1\cdot \vec{z}_2 \right]^{s-1}
\end{align}
while for $\ev{J_s J_s O_3}$ we get
\begin{align}\label{spinsintermsofspin1b}
\langle J_sJ_sO_3\rangle_{\text{even},\bf{h}}
&=(k_1 k_2)^{s-1}(E+(2s-1)k_3)\left[\frac{1}{E^2} \left\{2(\vec{z}_1\cdot \vec{k}_2)( \vec{z}_2\cdot \vec{k}_1) +E (E-2k_3)\vec{z}_1\cdot \vec{z}_2 \right\}\right]^s\cr
\langle J_sJ_sO_3\rangle_{\text{odd},\bf{h}}
&= (k_1 k_2)^{s-1}\frac{(E+(2s-1)k_3)}{E^{2s}}\left[k_2\,\epsilon^{ k_1 z_1 z_2}- k_1\,\epsilon^{k_2 z_1  z_2}\right]\cr
&\hspace{.5cm}\times \left[2(\vec{z}_1\cdot \vec{k}_2)( \vec{z}_2\cdot \vec{k}_1) +E(E-2k_3)\vec{z}_1\cdot \vec{z}_2 \right]^{s-1}
\end{align}
The homogeneous part of the $J_s$ 3-point correlator is
\begin{align}\label{spinsintermsofspin1c}
\langle J_sJ_sJ_s\rangle_{\text{even},\bf{h}}
&=(k_1 k_2 k_3)^{s-1}\left[\frac{1}{E^3} \Big\{2\,(\vec{z}_1\cdot \vec{k}_2) \, (\vec{z}_2\cdot \vec{k}_3) \, (\vec{z}_3\cdot \vec{k}_1)+E \{k_3\, (\vec{z}_1\cdot \vec{z}_2) \, (\vec{z}_3\cdot \vec{k}_1)+ \text{cyclic}\}\Big\}\right]^s\cr
\langle J_sJ_sJ_s\rangle_{\text{odd},\bf{h}}
&=(k_1 k_2 k_3)^{s-1}\frac{1}{E^3}\left[\left\{(\vec{k}_1 \cdot \vec{z}_3)\left(\epsilon^{k_3 z_1 z_2}k_1-\epsilon^{k_1 z_1 z_2}k_3\right)+(\vec{k}_3 \cdot \vec{z}_2)\left(\epsilon^{k_1 z_1 z_3}k_2-\epsilon^{k_2 z_1 z_3}k_1\right)\right.\right.\nonumber\\[5 pt]
&\hspace{1.5cm}\left.\left.-(\vec{z}_2 \cdot \vec{z}_3)\epsilon^{k_1 k_2 z_1}E+\frac{k_1}{2} \epsilon^{z_1 z_2 z_3}E(E-2k_1)\right\}+\text{cyclic perm}\right]\cr
&\hspace{.5cm}\times\left[\frac{1}{E^3} \Big\{2\,(\vec{z}_1\cdot \vec{k}_2) \, (\vec{z}_2\cdot \vec{k}_3) \, (\vec{z}_3\cdot \vec{k}_1)+E \{k_3\, (\vec{z}_1\cdot \vec{z}_2) \, (\vec{z}_3\cdot \vec{k}_1)+ \text{cyclic}\}\Big\}\right]^{s-1}\nonumber\\[5 pt]
\end{align}
whereas for $\ev{J_{2s} J_s J_s }$ we have
\begin{align}\label{spinsintermsofspin1d}
 \langle J_{2s} J_s J_s \rangle_{\text{odd},\mathbf{h}}&= \frac{k_1^{2s-1} (k_2 k_3)^{s-1}}{E^{4s}} \bigg[\bigg((k_3 \cdot z_2)(k_2 \cdot z_1)-\frac{1}{2}E(E-2k_3)(z_1 \cdot z_2)\bigg)\left( k_1\epsilon^{z_{1} z_{3} k_{3}}-  k_3  \epsilon^{z_{1} z_{3} k_{1}}\right)\bigg]\cr
 &\hspace{-0.5cm}\times\bigg[\bigg((k_3 \cdot z_2)(k_2 \cdot z_1)\bigg.\bigg.\bigg.\bigg.-\frac{1}{2}E(E-2k_3)(z_1 \cdot z_2)\bigg)\left((k_1 \cdot z_3)(k_2 \cdot z_1)-\frac{1}{2}E(E-2k_2)(z_1 \cdot z_3)\right)\bigg]^{s-1}\nonumber\\[5 pt]
     \langle J_{2s} J_s J_s \rangle_{\text{even},\mathbf{h}}&=\frac{k_1^{2s-1} (k_2 k_3)^{s-1}}{E^{4s}}\bigg[\bigg((k_3 \cdot z_2)(k_2 \cdot z_1)-\frac{1}{2}E(E-2k_3)(z_1 \cdot z_2)\bigg)\bigg.\cr 
     &\hspace{3.5cm}\times\bigg.\left((k_1 \cdot z_3)(k_2 \cdot z_1)-\frac{1}{2}E(E-2k_2)(z_1 \cdot z_3)\right)\bigg]^{s} 
\end{align}

\section{Weight-shifting operators}\label{sdro}
We need the following spin and dimension raising operators \cite{Baumann:2019oyu,Baumann:2020dch,Jain:2021wyn},
\begin{align}\label{wsoe}
    H_{12}=&2 \left(z_1\cdot K_{12}z_2\cdot K_{12}-2 z_1\cdot z_2 W_{12}^{--}\right),\nonumber\\
    \tilde{D}_{12}=&-\frac{1}{2}\Big[\epsilon(z_{1}z_{2}K^-_{12})(\Delta_1-d-k_1\cdot\frac{\partial}{\partial k_1}) +\frac{K_{12}^-K^+_{12}}{2}\,\epsilon(k_{1}z_{1}z_{2})+\epsilon(k_{1}K_{12}^-z_{1})(z_2\cdot \frac{\partial}{\partial k_2})\nonumber \\
    &+\epsilon(k_{1}z_{2}K^-_{12})(z_1\cdot\frac{\partial}{\partial k_1})\Big].
\end{align}

where expressions for $K_{12}, W_{12}^{--}$ can be found in the above mentioned references.
The following sequence of operators reproduces $\langle TJJ \rangle_{\text{odd}}$
\begin{align}
    \langle T(k_1)J(k_2)J(k_3) \rangle_{\text{odd}} = P_1^{(2)}P_2^{(1)}P_3^{(1)}H_{13}\widetilde{D}_{12}\langle O_1(k_1)O_2(k_2)O_2(k_3) \rangle+(2 \leftrightarrow 3)
\end{align}
where $P_i^{(s)}$ is a spin$-s$ projector. 
The explicit momentum space expression for the correlator is given by
\begin{align}
\begin{split}
    \langle TJJ \rangle_{\text{odd}} &= \bigg[A_1\epsilon^{k_1 k_2 z_1}(k_2 \cdot z_1)(k_3 \cdot z_2)(k_1 \cdot z_3)+A_2 \epsilon^{k_1 k_2 z_1}(z_2 \cdot z_3)(k_2 \cdot z_1)\bigg.\\
    &+A_3 \epsilon^{k_1 z_1 z_2}(k_2 \cdot z_1)(k_1 \cdot z_3)+A_4 \epsilon^{k_2 z_1 z_2}(k_2 \cdot z_1)(k_1 \cdot z_3)\\
    &+A_5 \epsilon^{k_1 z_1 z_2}(z_1 \cdot z_3)+A_6 \epsilon^{k_2 z_1 z_2}(z_1 \cdot z_3)\\
    &+\bigg.A_7\epsilon^{k_1 k_2 z_1}(z_1 \cdot z_2)(k_1 \cdot z_3)+A_8 \epsilon^{z_1 z_2 z_3}(k_2 \cdot z_1)\bigg]+(2 \leftrightarrow 3)
    \end{split}
\end{align}
where the form factors are given by
\begin{align}
    \begin{split}
    &A_1 = 12\frac{5k_1^2+4k_1(k_2+k_3)+(k_2+k_3)^2}{k_1^2(k_1+k_2+k_3)^4}\\[5 pt]
    &A_2 = 4\frac{k_1 +k_2 +3k_3}{(k_1+k_2+k_3)^3}\\[5 pt]
    &A_3 = \frac{15k_1^3+13k_1^2(k_2+k_3)+9k_1(k_2+k_3)^2+3(k_2+k_3)^3}{k_1^2(k_1+k_2+k_3)^3}\\[5 pt]
    &A_4 = \frac{k_1+k_2+3k_3}{(k_1+k_2+k_3)^3}\\[5 pt]
    &A_5 = \frac{-3k_1^4+2k_1^3(5k_2-3k_3)+4k_1^2k_2(2k_2-k_3)+6k_1(k_2-k_3)^2(k_2+k_3)+3(k_2^2-k_3^2)^2}{2k_1^2(k_1+k_2+k_3)^2}\\[5 pt]
    &A_6= 4\frac{k_2(k_1+k_2+2k_3}{(k_1+k_2+k_3)^2}\\[5 pt]
    &A_7 = \frac{-3k_1^3-3(k_2-3k_3)(k_2+k_3)^2+k_1^2(-9k_2+23k_3)-9k_1(k_2^2-2k_2k_3-3k_3^2))}{k_1^2(k_1+k_2+k_3)^3}\\[5 pt]
    &A_8 = -2\frac{3k_1^2+2k_1(k_2+k_3)+(k_2+k_3)^2}{(k_1+k_2+k_3)^2}
    \end{split}
\end{align}
Although this expression looks very different from the expression obtained earlier in \eqref{jjta1a2}, they are actually the same up to some Schouten identities. This can easily be seen by converting both of them to spinor-helicity variables where they match exactly.

\section{Parity-odd contact terms and field redefinitions}\label{contact}
Here we address the following question: is it possible to set the non-homogeneous part of the parity-odd contribution, which is argued to be a contact term always, to zero using some field re-definition or a suitable definition of the correlation function? 
 \\ 
We will argue below that it is not always possible to set the non-homogeneous parity-odd contribution to zero. It should be emphasised that the parity-odd contribution to the non-homogeneous piece is also proportional to the parity-odd two-point function coefficient. Our main argument below is that the parity-odd contribution to the Ward identity is non-zero in general and hence this will contribute nontrivially to the non-homogeneous parity-odd three-point function. To augment our CFT argument, we also perform some simple $dS_4$ computations and identify interaction terms in the lagrangian which give rise to non-homogeneous parity-odd contributions. This serves as an additional check that the parity-odd non-homogeneous contribution cannot be set to zero. To illustrate these points let us consider a few examples. 
 
 \section*{$\langle JJJ\rangle $}
 For the case of $\langle JJJ\rangle $, irrespective of whether the functional derivatives involved in the definition of the correlator act on the measure factors or not, the correlator takes the same form \cite{Bzowski:2017poo}. Following \cite{Bzowski:2017poo} we have
 \begin{align}
 \langle JJJ\rangle_{\text{here}}=\langle JJJ\rangle_{\text{there}}
 \end{align}
 where ``here" refers to the correlator as per the definition in \cite{Bzowski:2017poo} (where the functional derivatives do not act on the measure factors)  and ``there" refers to the correlator as defined in \cite{Bzowski:2013sza} (where the functional derivatives do act on the measure factors).  The transverse Ward identity for the correlator is \cite{Bzowski:2013sza} 
 \begin{align}
\label{WTjjj}
\begin{split}
k_{1\mu}\langle J^{\mu a}(k_1) J^{\nu b}(k_2) J^{\rho c}(k_2) \rangle &= \left(f^{adc}\langle J^{\rho d}(k_2)J^{\nu b}(-k_2) \rangle-f^{abd}\langle J^{\nu d}(k_3)J^{\rho c}(-k_3)\rangle\right)\\[5 pt]
&\hspace{-1cm}+\bigg[\left(\frac{k_2^{\nu}}{ k_2^2}f^{abd}k_{2\alpha}\langle J^{\alpha d}(k_3)J^{\rho c}(-k_3)\rangle\right)+\left((k_2, \nu) \leftrightarrow (k_3, \rho)\right)\bigg]\cr
&\hspace{-1cm}=f^{adc}\langle J^{\rho d}(k_2)J^{\nu b}(-k_2)\rangle_{\text{odd}}-f^{abd}\langle J^{\nu d}(k_3)J^{\rho c}(-k_3)\rangle_{\text{odd}}\cr
&\hspace{-1cm}=-f^{abc}\left(\epsilon^{k_2 \nu \rho}+\epsilon^{k_3 \nu \rho}\right)\cr
&\hspace{-1cm}\ne 0
\end{split}
\end{align}
This leads to the following parity-odd non-homogeneous contribution to the correlator 
 \begin{align}
 \langle J(z_1, k_1)J(z_2, k_2)J(z_3, k_3)\rangle_{\text{odd},\bf{nh}} = c_{JJ}\epsilon^{z_1z_2z_3}\label{nhpo}
 \end{align}
 This is a contact term and can be explained from the $dS_4$ perspective using the following interaction term 
 \begin{align}
 \int F^a_{\mu\nu} F^a_{\rho\sigma} ~dx^{\mu}\wedge dx^{\nu}\wedge dx^{\rho} \wedge dx^{\sigma}\label{gs}
 \end{align}
 where $F^a_{\mu\nu} = \partial_{\mu}A^a_{\nu}-\partial_{\nu}A^a_{\mu}+i f^{abc}A^b_{\mu}A^c_{\nu}$. The three-point tree-level amplitude due to this interaction is given by
 \begin{align}
 \mathcal{M}_{F\widetilde{F}} &= \epsilon(k_1 z_1 z_2 z_3)+\text{cyclic terms}\cr
 & = E\epsilon^{z_1z_2z_3}
 \end{align}
Following the prescription of \cite{Maldacena:2011nz} the correlator is given by
 \begin{align}
 \langle JJJ\rangle_{\text{odd},\textbf{nh}}=\frac{1}{E} \mathcal{M}_{F\widetilde{F}} 
 \end{align}
where the prefactor $\frac 1E$ can be understood to come from the associated conformal time integral 
 \begin{align}
 \text{Im}\left[i\int_{0}^{\infty}d\eta e^{-i\eta E}\right] \propto \frac{1}{E}
 \end{align}
 Thus we see that the non-homogeneous part of the correlator \eqref{nhpo} is reproduced by the interaction $F\widetilde F$ in $dS_4$. Such a contact term cannot be erased by any field re-definition. The $dS_4$ interaction \eqref{gs} also contributes to the two-point function $\langle JJ\rangle$ as $\epsilon^{z_1z_2k_1}$ which is consistent with the contribution of $F\widetilde F$ to the non-homogeneous part of the three-point function.
 
 \section*{$\langle TJJ\rangle $}
The transverse Ward identities for $\langle TJJ\rangle $ takes the form 
 \begin{align}\label{tjjoddwi}
\begin{split}
k_{1\mu}\langle T^{\mu\nu}(k_1)J^{\rho}(k_2)J^{\sigma}(k_3) \rangle &= k_{3\mu}\delta^{\nu\sigma} \langle J^{\mu}(k_1+k_3)J^{\rho}(k_2) \rangle+k_{2\mu}\delta^{\nu\rho}\langle J^{\mu}(k_1+k_2)J^{\sigma}(k_3) \rangle\\[5 pt]
&\hspace{.5cm}-k_{3\nu}\langle J^{\sigma}(k_1+k_3)J^{\rho}(k_2) \rangle-k_{2\nu}\langle J^{\rho}(k_1+k_2) J^{\sigma}(k_3) \rangle\\
k_{2\rho}\langle T^{\mu\nu}(k_1)J^{\rho}(k_2)J^{\sigma}(k_3) \rangle &=0.
\end{split}
\end{align}
In the parity-odd case $\langle J^{\alpha}(k_1+k_2) J^{\beta}(k_3) \rangle = -c_J' \epsilon^{\alpha \beta k_3}$ and upon using Schouten identities the Ward identity w.r.t $k_{1\mu}$ also becomes trivial
\begin{align}\label{tjjoddwifnl}
k_{1\mu}\langle T^{\mu\nu}(k_1)J^{\rho}(k_2)J^{\sigma}(k_3) \rangle_{\text{odd}} &= 0
\end{align}
This immediately implies that the parity-odd part of the non-homogeneous term is zero 
\begin{equation}
\label{tjjcftnhodd}
   \langle T^{\mu\nu}(k_1)J^{\rho}(k_2)J^{\sigma}(k_3) \rangle_{\textbf{nh},\text{odd}} =0.
\end{equation}
Let us now check that the $dS_4$ computation would also lead to a vanishing contribution. There is no interaction term that one can have from the gravity side that contributes to the non-homogeneous parity odd part of $\langle TJJ\rangle.$  
 The only term that could possibly have contributed to this correlator is \eqref{gs}. However since this term is  independent of the metric the contribution from it to $\langle TJJ\rangle$ is zero, 
which is consistent with the CFT computation in \eqref{tjjcftnhodd}.

  \section*{$\langle TTT\rangle $}
 Depending on the definition of stress-tensor the transverse Ward identity of $\langle TTT\rangle$ can take different forms. If the functional derivatives involved in the definition of the correlator acts on the measure factors, it takes the form \cite{Bzowski:2013sza} 
\begin{align}
k_1^{\mu_1}\langle T_{\mu_1\nu_1}(k_1)T_{\mu_2\nu_2}(k_2)T_{\mu_3\nu_3}(k_3)\rangle&=2p_1^{\mu_1}\left\langle \frac{\delta T_{\mu_1\nu_1}}{\delta g^{\mu_3\nu_3}}(k_1,k_3)T_{\mu_2\nu_2}(k_2)\right\rangle+2p_1^{\mu_1}\left\langle \frac{\delta T_{\mu_1\nu_1}}{\delta g^{\mu_2\nu_2}}(k_1,k_2)T_{\mu_3\nu_3}(k_3)\right\rangle\cr
&\hspace{.5cm}+2p_{1(\mu_3}\langle T_{\nu_3)\nu_1}(k_2)T_{\mu_2\nu_2}(-k_2)\rangle+2p_{1(\mu_2}\langle T_{\nu_2)\nu_1}(k_3)T_{\mu_3\nu_3}(-k_3)\rangle\cr
&\hspace{.5cm}+\delta_{\mu_3\nu_3}k_3^\alpha\langle T_{\alpha\nu_1}(k_2)T_{\mu_2\nu_2}(-k_2)\rangle+\delta_{\mu_2\nu_2}p_2^\alpha\langle T_{\alpha\nu_1}(k_3)T_{\mu_3\nu_3}(-k_3)\rangle\cr
&\hspace{.5cm}-p_{3\nu_1}\langle T_{\mu_2\nu_2}(k_2)T_{\mu_3\nu_3}(-k_2)\rangle-p_{2\nu_1}\langle T_{\mu_2\nu_2}(k_3)T_{\mu_3\nu_3}(-k_3)\rangle
\end{align}
With a suitable definition of the correlator where the measure factors are not acted upon by the functional derivatives the transverse identity takes the following form \cite{Bzowski:2017poo}
\begin{align}
k_1^{\mu_1}\langle T_{\mu_1\nu_1}(k_1)T_{\mu_2\nu_2}(k_2)T_{\mu_3\nu_3}(k_3)\rangle&=2p_{1(\mu_3}\langle T_{\nu_3)\nu_1}(k_2)T_{\mu_2\nu_2}(-k_2)\rangle+2p_{1(\mu_2}\langle T_{\nu_2)\nu_1}(k_3)T_{\mu_3\nu_3}(-k_3)\rangle\cr
&\hspace{.5cm}-p_{3\nu_1}\langle T_{\mu_2\nu_2}(k_2)T_{\mu_3\nu_3}(-k_2)\rangle-p_{2\nu_1}\langle T_{\mu_2\nu_2}(k_3)T_{\mu_3\nu_3}(-k_3)\rangle
\end{align}
In addition to the above one can consider field re-definition of the sources and get an alternate definition of the stress-tensor such as changing the metric as $\gamma_{ij}=c^{-1}e^{c\hat\gamma}_{ij}$ \cite{Farrow:2018yni}. In this case the new stress-energy tensor takes the form \cite{Baumann:2020dch, Farrow:2018yni}
\begin{align}
\hat T^{ij}=T^{ij}+\frac c2\gamma^{ik}T_{k}^j+\frac c2\gamma^{jk}T^{i}_k
\end{align}
Using this field re-definition, we obtain the following Ward identity \cite{Baumann:2020dch}
\begin{align}
\begin{split}
z_{1\mu}k_{1\nu}&\langle T^{\mu\nu}(k_1) T(k_2) T(k_3) \rangle \\
&= -(z_1 \cdot k_2) z_{2\mu}z_{2\nu}\langle T^{\mu\nu}(k_1 + k_2) T(k_3) \rangle +2(z_1 \cdot z_2)k_{2\mu}z_{2\nu}\langle T^{\mu\nu}(k_1+k_2) T(k_3) \rangle\\[5 pt]
&\hspace{.3cm}-(z_1 \cdot k_3)z_{3\mu} z_{3\nu}\langle  T^{\mu\nu}(k_1+k_3) T(k_2) \rangle + 2(z_1 \cdot z_3)k_{3\mu}z_{3\nu}\langle T^{\mu\nu}(k_1+k_3) T(k_2) \rangle\\[5 pt]
&\hspace{.3cm}-\frac{c}{2}(k_1 \cdot z_2)z_{1\mu}z_{2\nu}\langle T^{\mu\nu}(k_1+k_2) T(k_3) \rangle-\frac{c}{2}(z_1 \cdot z_2)k_{1\mu}z_{2\nu}\langle T^{\mu\nu}(k_1+k_2) T(k_3) \rangle\\[5 pt]
&\hspace{.3cm}-\frac{c}{2}(k_1 \cdot z_3)z_{1\mu}z_{3\nu}\langle T^{\mu\nu}(k_1+k_3) T(k_2) \rangle-\frac{c}{2}(z_1 \cdot z_3)k_{1\mu}z_{3\nu}\langle T^{\mu\nu}(k_1+k_3) T(k_2) \rangle
\end{split}
\end{align}
where $T(k)\equiv z_{\mu}z_{\nu}T^{\mu\nu}(k)$.
Plugging in the parity-odd contribution to $\langle TT\rangle$,
\begin{align}\label{abcs}
    \langle T^{\mu\nu}(k)T^{\rho\sigma}(-k) \rangle_{\text{odd}} &= c'_{T} \Delta^{\mu\nu\rho\sigma}(k)k^2
    \end{align}
    where 
\begin{align}
    \Delta^{\mu\nu\rho\sigma}(k) &=\epsilon^{\mu \rho k} \pi^{\nu \sigma}(k)+\epsilon^{\mu \sigma k} \pi^{\nu \rho}(k)+\epsilon^{\nu \sigma k} \pi^{\mu \rho}(k)+\epsilon^{\nu \rho k} \pi^{\mu \sigma}(k)\\[5 pt]
    \pi^{\mu\nu}(k) &= \delta^{\mu\nu}-\frac{k^{\mu}k^{\nu}}{k^2}
\end{align}
it is easy to check using Schouten identities and degeneracy that for no value of $c$ can the rhs of the Ward identity be set to zero, i.e. 
\begin{align}
k_{1\mu_1}&\langle T^{\mu_1\nu_1}(k_1) T^{\mu_2\nu_2}(k_2) T^{\mu_3\nu_3}(k_3) \rangle_{\text{odd}}\ne 0
\end{align}
This will in turn result in a non-zero contribution to the non-homogeneous parity-odd part, which takes the form 
\begin{align}\label{tttnewbasis}
\begin{split}
\langle TTT \rangle_{\bf{nh},\text{odd}} &= \bigg[B_1 \epsilon^{k_1 z_1 z_2} (z_1 \cdot z_2) (k_1 \cdot z_3)^2 - B_1 (k_1 \leftrightarrow k_2)\epsilon^{k_2 z_1 z_2} (z_1 \cdot z_2) (k_1 \cdot z_3)^2\bigg.\\[5 pt]
&\bigg.+B_2 \epsilon^{k_1 z_1 z_2} (z_1 \cdot z_3) (z_2 \cdot z_3)-B_2(k_1 \leftrightarrow k_2)\epsilon^{k_2 z_1 z_2} (z_1 \cdot z_3) (z_2 \cdot z_3)\bigg] + \text{cyclic perm.}
\end{split}
\end{align}
where
\begin{align}
\begin{split}
B_1=c'_T \frac{1}{24},\quad\quad
B_2 = c'_T \frac{1}{12}\left(k_1^2 + \frac{7}{4} k_2^2+ \frac{7}{4} k_3^2\right)
\end{split}
\end{align}
We again see that the non-homogeneous contribution is proportional to the parity-odd two-point function coefficient $c'_{T}$. From the $dS_4$ perspective this contribution to the correlator can be understood to arise from the $W\widetilde W$ interaction. The $W\widetilde W$  interaction also reproduces the  parity-odd two-point function of the stress tensor \eqref{abcs}.

\providecommand{\href}[2]{#2}\begingroup\raggedright
\bibliography{SH-v1}
\bibliographystyle{JHEP}
\endgroup

\end{document}